\documentclass{ieeeaccess}
\usepackage{cite}
\usepackage{amsmath,amssymb,amsfonts}
\usepackage{algorithmic}
\usepackage{graphicx}
\usepackage{balance}
\usepackage{textcomp}

\usepackage{url}
\usepackage{physics}
\usepackage{balance}
\usepackage{bm}
\usepackage{colortbl}
\usepackage{lipsum}      
\usepackage{changepage}
\usepackage{caption}
\usepackage{color}
\newtheorem{theorem}{Theorem}
\newtheorem{proposition}{Proposition}
\newtheorem{definition}{Definition}
\newcommand{\eqdef}{\mathrel{\mathop:}=}

\def\BibTeX{{\rm B\kern-.05em{\sc i\kern-.025em b}\kern-.08em
    T\kern-.1667em\lower.7ex\hbox{E}\kern-.125emX}}
\begin{document}
	\history{\textbf{Received March 19, 2023, accepted April 17, 2023, date of current version April 21, 2023.}}
	\doi{mm.yyyy/ACCESS.2023.DOI}
	
	\title{Making Sense of Meaning: A Survey on Metrics for Semantic and Goal-Oriented Communication}
	\author{\uppercase{Tilahun~M.~Getu}\authorrefmark{1,}\authorrefmark{2}, \IEEEmembership{Member, IEEE},
		\uppercase{Georges~Kaddoum}\authorrefmark{2,}\authorrefmark{3}, \IEEEmembership{Senior Member, IEEE}, and \uppercase{Mehdi~Bennis}\authorrefmark{4}, \IEEEmembership{Fellow, IEEE}
	}
	\address[1]{Communications Technology Laboratory (CTL), National Institute of Standards and Technology (NIST), Gaithersburg, MD 20899, USA}
	\address[2]{Electrical Engineering Department, \'Ecole de Technologie Sup\'erieure (ETS), Montr\'eal, QC H3C 1K3, Canada}
	\address[3]{Cyber Security Systems and Applied AI Research Center, Lebanese American University, Beirut, Lebanon}
	\address[4]{Centre for Wireless Communications, University of Oulu, 90570 Oulu, Finland}
	\tfootnote{The first author acknowledges the U.S. Department of Commerce and NIST for funding this work, and Dr. Hamid Gharavi (\textit{IEEE Life Fellow} of NIST, MD, USA) for funding and leadership support.}
	
	\markboth
	{Getu \headeretal: Making Sense of Meaning: A Survey on Metrics for Semantic and Goal-Oriented Communication}
	{Getu  \headeretal: Making Sense of Meaning: A Survey on Metrics for Semantic and Goal-Oriented Communication}
	
	
	\corresp{Corresponding author: Tilahun M. Getu (e-mail: tilahun-melkamu.getu.1@ ens.etsmtl.ca).}
	
	\begin{abstract}
		Semantic communication (SemCom) aims to convey the meaning behind a transmitted message by transmitting only semantically-relevant information. This semantic-centric design helps to minimize power usage, bandwidth consumption, and transmission delay. SemCom and goal-oriented SemCom (or effectiveness-level SemCom) are therefore promising enablers of 6G and developing rapidly. Despite the surge in their swift development, the design, analysis, optimization, and realization of robust and intelligent SemCom as well as goal-oriented SemCom are fraught with many fundamental challenges. One of the challenges is that the lack of unified/universal metrics of SemCom and goal-oriented SemCom can stifle research progress on their respective algorithmic, theoretical, and implementation frontiers. Consequently, this survey paper documents the existing metrics -- scattered in many references -- of wireless SemCom, optical SemCom, quantum SemCom, and goal-oriented wireless SemCom. By doing so, this paper aims to inspire the design, analysis, and optimization of a wide variety of SemCom and goal-oriented SemCom systems. This article also stimulates the development of unified/universal performance assessment metrics of SemCom and goal-oriented SemCom, as the existing metrics are purely statistical and hardly applicable to reasoning-type tasks that constitute the heart of 6G and beyond.
	\end{abstract}
	
	\begin{keywords}
	6G, wireless SemCom, optical SemCom, quantum SemCom, goal-oriented  wireless SemCom, metrics of SemCom and goal-oriented SemCom. 
	\end{keywords}
	
	\titlepgskip=-15pt
	
	\maketitle

\section{Introduction}
\label{sec: introduction}
\subsection{Motivation}
\label{subsec: motivation}
Following the global rollout of fifth-generation (5G) wireless communication system applications and services, researchers in academia, industry, and national laboratories have been developing visions \cite{Saad_6G_Vision_20,Letaief_Edge_AI_Vision'22,Alwis_Survey_GG_Networks'21,Akyildiz_6G_and_Beyond'20,Alsabah_6G_Wireless_Commun_Network'21,Dang_Alouini_6G_20,You_Towards_6G'21,Road_towards_6G'21,Roadmap_6G_Privacy_and_Security'21,6G_Ecosystem'21,Survey_6G_Networks'21,Lu_6G_survey_20,Yaacoub_PIEEE_20,zhao2019survey_IRS_19,Chowdhury_6G_2020,Viswanathan_6G_2020,Bariah_6G_2020,Tataria_6G_Wire_Systems'21,Fettweis_6G_Per_Tactile_Internet'21,Uusitalo_6G_Hexa-X'21,De_Lima_6G'21,Khan_6G_20,Rappaport_wires_commun_above_100GHz,6G_mailbox_theory'21,Chen_6G_2020,Shaping_Future_6G_Networks'22} regarding the next generation of wireless communication systems -- commonly known as the sixth-generation (6G). 6G is driven -- as envisaged in the last four years -- by multiple widely envisioned applications as varied as wireless brain-computer interactions, multi-sensory extended reality (XR) applications, blockchain and distributed Ledger technologies, and connected robotic and autonomous systems \cite{Saad_6G_Vision_20}; haptic communication, massive Internet of things (IoT) \cite{IoT_Connectivity_in_6G'21}, integrated smart city, and automation and manufacturing \cite{Spec_stu_6G_19}; the \textit{internet of no things} (metaverse) \cite{InoT_Maier'20,Edge_Enabled_Metaverse'22}; industrial IoT \cite{Gui_6G_20}, internet of robots \cite{Chen_6G_2020}, flying vehicles \cite{Bariah_6G_2020}, and wireless data centers \cite{Bariah_6G_2020,Wireless_DCN_18}; accurate indoor positioning, new communication terminals, high-quality communication services onboard aircraft, worldwide connectivity, integrated networking, communications that support industry verticals \cite{RWH_6G_VTM_19}, holographic communication, tactile communication, and human bond communication \cite{Dang_Alouini_6G_20}; Smart Grid 2.0, Industry 5.0, personalized body area networks, Healthcare 5.0; and the internet of industrial smart things and the internet of healthcare \cite{Alwis_Survey_GG_Networks'21}.

To make the aforementioned 6G applications a reality, many researchers propose to use a wide variety of 6G enabling technologies \cite{Saad_6G_Vision_20,Letaief_Edge_AI_Vision'22,Alwis_Survey_GG_Networks'21,You_Towards_6G'21,Survey_6G_Networks'21,Shaping_Future_6G_Networks'22} at the infrastructure, spectrum, and algorithm/protocol level \cite{Scoring_Terabit_per_second_goal'20,6G_BBC_20_White_Paper}. Despite the variety of proposals, realizing 6G -- as many researchers are presently contemplating -- demands not only evolutionary developments but also a revolutionary paradigm shift \cite{Saad_6G_Vision_20}. The revolutionary paradigm shift -- in particular -- must tackle the following fundamental challenges of 6G: 
\begin{itemize}
	\item Guaranteeing ultra-high data rate for most users.
	\item Ensuring an ultra-reliability and low latency for the bulk of users.
	\item Managing ultra-heterogeneity
	\item Taming ultra-high complexity in 6G networks.
	\item Addressing ultra-high mobility
	\item Accommodating users' needs or perspectives (see \cite{KDHB_18}).
	\item Designing with respect to (w.r.t.) various key performance indicators (KPIs).
	\item Attaining high energy efficiency.
	\item Realizing energy-efficient artificial intelligence (AI).
	\item Ensuring security, privacy, and trust across the 6G network.
	\item Attaining \textit{full intelligence and autonomy}. 
	\item Dealing with the technological uncertainty \cite{Latvaaho2019KeyDAC} of 6G technology enablers.
\end{itemize}
Addressing the itemized fundamental challenges would translate to overcoming numerous interdisciplinary, multidisciplinary, and transdisciplinary (IMT) challenges. 

To mitigate the astronomical IMT challenges of 6G, the  design of 6G systems and networks must be holistically geared towards minimizing power usage, bandwidth consumption, and transmission delay by minimizing the transmission of semantically irrelevant information. This semantic-centric information transmission calls for the efficient transmission of \textit{semantics} by a semantic transmitter followed by their reliable recovery by a semantic receiver. This type of communication paradigm is now widely regarded as semantic communication (SemCom). SemCom -- which was first put forward by Weaver around 1949 \cite{Shannon_Weaver_Math_Theory_Commun'49} -- is a communication paradigm aimed at conveying the transmitter's intended meaning. SemCom targets the transmission of only the \textit{semantic information}\footnote{Since \textit{semantics} is built upon syntax and studies signs and their relationship to the world \cite{Gunduz_Beyond_Transmitting_Bits'22}, the fundamental concept of semantic information relies on the information ecosystem, which is a complete process of \textit{information-knowledge-intelligence conversion} \cite{hong_Theory_Semantic_Info'17,Zhong_Theory_Sem_Information_Book_Chapter}. See \cite[Fig.1]{hong_Theory_Semantic_Info'17} for more information. Meanwhile, semantic information can be represented using knowledge graphs (KGs) \cite{Knowledge_Graphs_Survey'22}, deep neural networks (DNNs), topos \cite{Topos_and_Stacks'21}, and \textit{quantum corollas} \cite{Tetlow_Toward_Semantic_Info_Theory'22}.} relevant to the communication goal in order to minimize the divergence between the intended meaning of the transmitted messages and the meaning of the messages ultimately recovered \cite{Tong_FL_ASC'21}, reducing data traffic considerably \cite{Xie_DL-based_SemCom'21}. SemCom involves the transmission of less data than the traditional communications techniques do \cite{Tong_FL_ASC'21} because only the semantic information that is pertinent to accurate interpretation at the destination is transmitted. In this respect, SemCom makes it possible to utilize the available network capacity more effectively \cite{Kalfa_Toward_GO_Semantic_Signal_Processing'21}. A network's capacity can certainly be utilized effectively by avoiding the bit-by-bit reconstruction of the transmitted information at the receiver. Moreover, SemCom aims to incorporate the purpose of transmission when doing so to simplify the data to be transmitted and avoid transmitting redundant information \cite{Zhou_WiCom_Letters'22}.

SemCom epitomizes the \textquotedblleft provisioning of the right and significant piece of information to the right point of computation (or actuation) at the right point in time'' \cite{SemCom_Net_Systems'21}. This philosophy is of paramount importance for networked control systems in which a system designer has to deal with not only the transmission of relevant semantic information but also the effectiveness of the transmitted semantic information to effectively execute a desired goal/action. As for the desired goal/action, a SemCom in which the efficiency/effectiveness of semantic transmission is explicitly defined and targeted can be qualified as a goal-oriented SemCom \cite{Zhang_Goal-Oriented_Commun'22}.\footnote{Goal-oriented communication and task-oriented communication -- that are based on semantic information -- are discussed throughout this paper under the heading \textquotedblleft goal-oriented wireless SemCom''. However, the authors of \cite{Sem_Empowered_Commun'22} underscore that goal communication is much broader than SemCom. Per Weaver’s vision, they classify SemCom as \textit{semantic level-SemCom} and \textit{effectiveness level-SemCom}.} Goal-oriented SemCom is a subset of SemCom that provides a pragmatic view of SemCom wherein the receiver is interested in the significance (semantics) and the effectiveness of the source's transmitted message to accomplish a certain goal \cite{Zhang_Goal-Oriented_Commun'22}. Therefore, goal-oriented SemCom targets the extraction and transmission of only task-relevant information so that the transmitted source signal can be substantially compressed, communication efficiency is improved, and low end-to-end latency can be achieved \cite{Xie_Robust_IB'22}. 

The state-of-the-art on SemCom and goal-oriented SemCom features many proposals concerning SemCom \cite{Sem_Empowered_Commun'22,Chaccour_Building_NG_SemCom_Networks'22,Qin_Sem_Com_Principles_Apps'22,Luo_SemCom_Overview'22,Rethinking_modern_com_Lu_2022,Niu_Towards_SemCom'22,SemCom_for_6G_Future_Internet'22,Engineering_SemCom'22,Gunduz_Beyond_Transmitting_Bits'22,Zhang_a_New_Paradigm'22,Jiang_Wireless_Semantic_Transmission'22} and goal-oriented SemCom \cite{Sem_Empowered_Commun'22,SemCom_for_6G_Future_Internet'22,Gunduz_Beyond_Transmitting_Bits'22,Zhang_Goal-Oriented_Commun'22} techniques. Despite the numerous state-of-the-art techniques that exist for SemCom and goal-oriented SemCom, the design, analysis, optimization, and realization of systems that are based on SemCom and goal-oriented SemCom are fraught with various fundamental challenges. Among the challenges, one important fundamental challenge is the lack of \textit{unified/universal performance assessment metrics} -- of SemCom and goal-oriented SemCom -- that help facilitate research developments in SemCom and goal-oriented SemCom. To this end, a detailed discussion of the existing performance metrics of SemCom and goal-oriented SemCom -- either used or proposed in state-of-the-art works -- is therefore required to develop a unified/universal performance assessment metrics. To serve this purpose, this survey paper reports on the existing metrics -- from many distinct references -- of SemCom and goal-oriented SemCom while aiming to inspire the development of unified/universal performance assessment metrics of SemCom and goal-oriented SemCom. This translates to the following paper contributions.

\begin{table*}
	\centering
	\begin{tabular}{| l | l | l | l | l | l| }
		\hline
		Semantic metrics & Scope of & Scope of    & Scope of  & Scope of    & \textbf{Scope of}   \\   
		&  Ref. \cite{Sem_Empowered_Commun'22} & Ref. \cite{Qin_Sem_Com_Principles_Apps'22}  & Ref. \cite{Luo_SemCom_Overview'22}   & Ref. \cite{SemCom_for_6G_Future_Internet'22}  & \textbf{this paper}   \\   \hline
		Semantic metrics for text quality assessment & Partially & Partially  & Partially & Partially  &  \textbf{Completely} \\ \hline
		Semantic metrics for speech quality assessment & Partially & Partially & Partially  & Partially &\textbf{Completely}   \\ \hline
		Semantic metrics for image quality assessment & Partially & Partially & Partially & Partially  &  \textbf{Completely} \\ \hline
		Semantic metrics for video quality and 3D human sensing assessment & -- & -- & -- & --  &  \textbf{Completely} \\ \hline
		Age of information- and value of information-based semantic metrics & Partially & -- & --  & Almost completely   &  \textbf{Completely} \\ \hline
		Resource allocation semantic metrics & -- & -- &  -- & -- &   \textbf{Completely} \\ \hline
		Generic semantic metrics of SemCom & -- & -- &  -- &  -- & \textbf{Completely}  \\ \hline
		Semantic metrics of quantum SemCom & -- & -- &  -- & -- & \textbf{Almost }   \\
		&  &  &   &  &\textbf{completely}   \\ \hline
		Semantic metrics of goal-oriented wireless SemCom  & Partially & -- &  -- & Partially &  \textbf{Completely}  \\   [2mm]\hline
	\end{tabular}
	\caption{Scope of this survey paper w.r.t. related state-of-the-art SemCom survey papers that discuss metrics of SemCom -- Ref.: reference; \textquotedblleft --'' means the particular reference didn't discuss the semantic metric listed on a given row.}
	\label{table: Scope_of_our_survey}
\end{table*}

\subsection{Contributions}
\label{subsec: contributions}
The key contributions of this survey paper -- a product of multidisciplinary research -- are enumerated below.
\begin{enumerate}
	\item We discuss existing as well as emerging developments of SemCom in multiple domains including \textit{wireless SemCom}, \textit{optical SemCom}, and \textit{quantum SemCom}.
	
	\item We discuss existing as well as emerging developments in goal-oriented wireless SemCom.
	
	\item We detail the numerous semantic metrics that are used for text, speech, and image quality assessment.
	
	\item We present the semantic metrics that are deployed for video quality and three-dimensional (3D) human sensing assessment.
	
	\item We provide an overview of age of information- and value of information-based semantic metrics.
	
	\item We outline resource allocation semantic metrics.
	
	\item We present generic semantic metrics of SemCom.
	
	\item We discuss semantic metrics of quantum SemCom.
	
	\item We delineate semantic metrics of goal-oriented wireless SemCom.
	
\end{enumerate}

The scope of our enumerated contributions w.r.t. the contributions of related state-of-the-art  SemCom papers that also discuss semantic metrics are put in perspective by Table \ref{table: Scope_of_our_survey}. Considering the fact that the various metrics of SemCom and goal-oriented SemCom are scattered in different references that disseminate them in different times, we discuss most of the corresponding metrics in this paper with the aim of inspiring the development of unified performance assessment metrics for SemCom and goal-oriented SemCom.

\begin{figure*}[t!]
	\centering
	\includegraphics[scale=0.80]{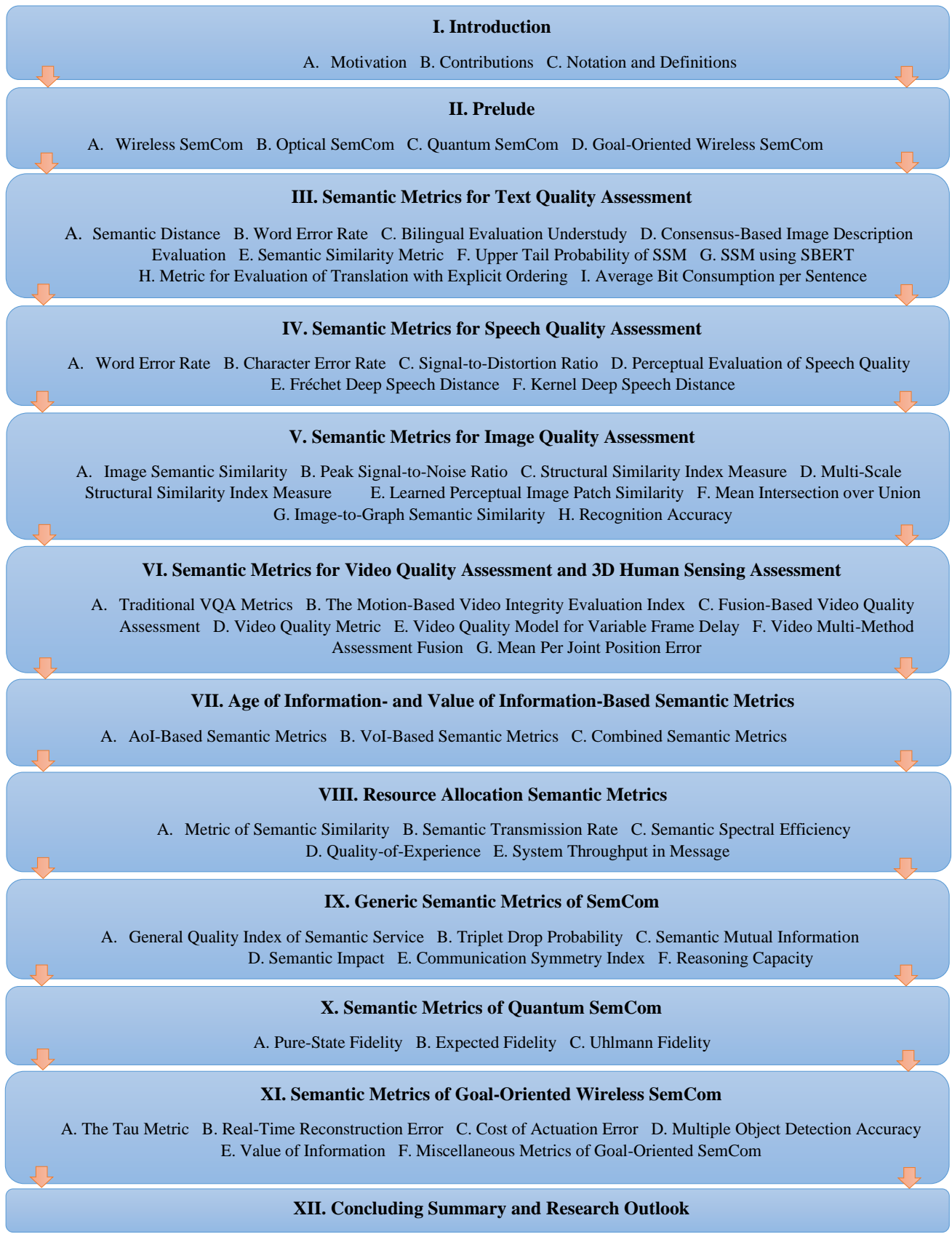}  \vspace{2mm} 
	\caption{The organization and structure of this survey paper.}
	\label{fig: Paper_structure_IEEE_Access_20230421}
\end{figure*}

The rest of this paper is organized as follows. Section \ref{sec: prelude}~presents this paper's prelude. Sections \ref{sec: metrics_for_text}, \ref{sec: metrics_for_speech}, and \ref{sec: metrics_for_image}~detail the semantic metrics that are used for text, speech, and image quality assessment, respectively. Section \ref{sec: metrics_for_video_and_3D_human_sensing}~reports on the semantic metrics that are deployed for video quality and 3D human sensing assessment. Section \ref{sec: AoI_and_VoI_based_semantic_metrics}~provides an overview of age of information- and value of information-based semantic metrics. Section \ref{sec: resource_allocation_semantic_metrics}~outlines the resource allocation semantic metrics. Sections \ref{sec: generic_semantic_metrics_of_SemCom} and \ref{sec: semantic_metrics_of_QSemCom}~present generic semantic metrics of wireless SemCom and semantic metrics of quantum SemCom, respectively. Section \ref{sec: Goal_oriented_SemCom_semantic_metrics}~summarizes the semantic metrics of goal-oriented wireless SemCom. Finally, Section \ref{sec: conc_summary_and_research_outlook}~contains the concluding summary and research outlook. Meanwhile, the organization and structure of this survey paper are depicted in Fig. \ref{fig: Paper_structure_IEEE_Access_20230421}.

\subsection{Notation and Definitions}
\label{subsec: notation} 
Scalars, vectors, and matrices are represented by italic letters, bold lowercase letters, and bold uppercase letters, respectively. Sets, datasets, skeletons, deep networks, and the Hilbert space are denoted by calligraphic letters. Calligraphic letters that are bold represent tensors. Random variables (RVs) and multivariate RVs (or random vectors) are represented by uppercase letters and bold lowercase letters, respectively. $\mathbb{N}$, $\mathbb{R}(\mathbb{C})$, $\mathbb{R}^+$, $\mathbb{R}^n(\mathbb{C}^n)$, and $\mathbb{R}^{m\times n}$ denote the set of natural numbers, the set of real(complex) numbers, the set of non-negative real numbers, the set of $n$-dimensional vectors of real(complex) numbers, and the set of $m\times n$ matrices of real numbers, respectively. $\eqdef$ denotes an equality by definition. For $n\in\mathbb{N}$, we let $[n] \eqdef \{1, 2, \ldots, n\}$. $\min$, $\max$, $\jmath$, $\| \cdot\|_1 $, and $\bm{I}_n$ denote minimum, maximum, $\sqrt{-1}$, the Schatten-1 norm, and an $n \times n$ identity matrix, respectively. $\otimes$, $\| \cdot \|$ (or $\| \cdot \|_2$), $(\cdot)^*$, $(\cdot)^T$, and $(\cdot)^H$ stand for tensor product, Euclidean norm, complex conjugate, transpose, and Hermitian, respectively.

$\textnormal{tr}(\cdot)$, $\mathbb{E}\{\cdot\}$, $\mathbb{E}_X\{\cdot\}$, $\mathbb{P}(\cdot)$, and $\mathbb{P}(A|B)$ denote trace (of a matrix), expectation, expectation w.r.t. an RV $X$, probability, and the probability of event $A$ conditioned on event $B$, respectively. $\Gamma(\cdot)$, $\Gamma(\cdot, \cdot)$, and $\mathbb{I}\{\cdot\}$ represent the gamma function, the upper incomplete gamma function, and an indicator function that returns 1 if the argument is true and 0 otherwise, respectively. For $z\in\mathbb{C}$ that $z=x+\jmath y$, its magnitude is denoted by $|z|$ and defined as $|z|\eqdef \sqrt{x^2+y^2}$. For a real vector $\bm{a}\in \mathbb{R}^n$, its $i$-th element is denoted by $(\bm{a})_i$ for all $i\in[n]$. For two real vectors $\bm{a}, \bm{b} \in \mathbb{R}^{1\times n}$, their dot product is denoted by $\bm{a} \cdot \bm{b}$ and defined as $\bm{a} \cdot \bm{b} \eqdef \sum_{i=1}^n (\bm{a})_i(\bm{b})_i$. For two vectors $\bm{c}, \bm{d} \in \mathbb{R}^{m}$, their element-wise product is denoted by $\bm{c} \odot \bm{d}$. For a three-way tensor $\bm{\mathcal{Y}}\in\mathbb{R}^{H \times W \times C}$, its element vector w.r.t. the given $h$-th and $w$-th dimension -- for $h\in[H]$ and $w\in[W]$ -- is denoted by $(\bm{\mathcal{Y}})_{h,w}\in \mathbb{R}^{C}$.

$\ket{\psi}$ is the Dirac's ket notation for a column vector such that $\ket{\psi_A} \eqdef \begin{bmatrix}
	a_0  & a_1 & \ldots & a_{N-1}
\end{bmatrix}^T$ \cite{Van_Meter_QNetworking'14}. $\bra{\psi}$ is the Dirac's bra notation corresponding to $\ket{\psi}$ and defined as the complex conjugate transpose (Hermitian) of $\ket{\psi}$: i.e., $\bra{\psi_A} \eqdef (\ket{\psi_A})^H =\begin{bmatrix}
	a_0^*  & a_1^* & \ldots & a_{N-1}^*
\end{bmatrix}$ \cite{Van_Meter_QNetworking'14}. For $\bra{\psi_A}$ and the quantum state $\ket{\psi_B} \eqdef \begin{bmatrix}
	b_0  & b_1 & \ldots & b_{N-1}
\end{bmatrix}^T$, $\bra{\psi_A}\ket{\psi_B}$ is the \textit{inner product (dot product)} -- of the two vectors $\psi_A$ and $\psi_B$ -- and defined as \cite[eq. (2.3)]{Van_Meter_QNetworking'14}
\begin{equation}
	\label{dot_product_defn}
	\bra{\psi_A}\ket{\psi_B}   \eqdef \sum_{i=0}^{N-1} a_i^{*}b_i.
\end{equation}
For $\ket{\psi_A}$ defined in above and $\bra{\psi_B} (\equiv \big( \ket{\psi_B} \big)^H)$, $\ket{\psi_A} \bra{\psi_B}$ is their \textit{outer product} and defined as \cite[eq. (2.4)]{Van_Meter_QNetworking'14}
\begin{equation}
	\ket{\psi_A} \bra{\psi_B} \eqdef 
	\begin{bmatrix}
		a_0 b_0^{*}  & \ldots  & a_0 b_{N-1}^{*} \\
		\vdots  & \ddots   & \vdots    \\
		a_{N-1} b_0^{*}  & \ldots  & a_{N-1} b_{N-1}^{*}
	\end{bmatrix}.
\end{equation}

In light of this bra-ket notation, a (noiseless) quantum bit (\textit{qubit}) $\ket{\psi}$ -- a basic unit of quantum information -- is a vector in a two-dimensional complex vector space (two-dimensional Hilbert space) and expressed as \cite[eq. (1.1)]{Nielsen_Chuang_QC'10}
\begin{equation}
	\label{ket_phi_defn}
	\ket{\psi} \eqdef \alpha \ket{0}  +   \beta \ket{1}, 
\end{equation}
where $\ket{0}$ and $\ket{1}$ are the special states known as \textit{computational basis states} that form an orthonormal basis for the vector space, and $\alpha, \beta \in \mathbb{C}$ such that $|\alpha|^2+|\beta|^2=1$ \cite{Nielsen_Chuang_QC'10}. The complex coefficients $\alpha$ and $\beta$ are \textit{probability amplitudes}; these amplitudes are not themselves probabilities but allow us to calculate probabilities \cite{Wilde_QIT'2017}. Per (\ref{ket_phi_defn}), the qubit $\ket{\psi}$ is a \textit{linear superposition}\footnote{A quantum mechanical \cite{Quantum_Mech_for_Scien_Engineers'08} equivalent of a bit, a qubit can be in state of 0, state of 1, and a superposition of state 0 and state 1 \cite{Nielsen_Chuang_QC'10}. A qubit can be physically materialized as a quantum mechanical system based on \textit{nuclear spin}, \textit{electron spin}, \textit{ion trap}, \textit{quantum dot}, \textit{optical cavity}, and \textit{microwave cavity} \cite[Ch. 7]{Nielsen_Chuang_QC'10}.} of two quantum states (i.e., $\ket{0}$ and $\ket{1}$), which underscores the fact that a qubit can be in one of the infinitely\footnote{Despite its infinitely many possible quantum states, a qubit cannot be examined, and \textit{quantum mechanics} (see \cite{Quantum_Mech_for_Scien_Engineers'08}) asserts that we can obtain only very limited information about $\ket{\psi}$ \cite{Nielsen_Chuang_QC'10}. To this end, when we measure $\ket{\psi}$, its inherent superposition will collapse and we get the result 0 or 1 with probability $|\alpha|^2$ or probability $|\beta|^2$ (under the probability constraint $|\alpha|^2+|\beta|^2=1$), respectively \cite{Nielsen_Chuang_QC'10}.} many quantum states that are possible \cite{Nielsen_Chuang_QC'10}. This is explained by the fact that measuring a qubit makes the wave function collapse, pushing the quantum state into just one term of the superposition \cite{Van_Meter_QNetworking'14}.

A generalized version of qubit -- called \textit{qudit}\footnote{Compared to qubit, qudit offers a larger state space to store and process information \cite{Wang_Qudits_HD_QC_2020,Cozzolino_HD_QC'19}. Hence, qudit can simplify the experimental setup, reduce the circuit complexity, and enhance algorithm efficiency \cite{Wang_Qudits_HD_QC_2020}. Generally, qudits offer many advantages over qubits, including higher information and communication capacity, greater noise resilience, enhanced robustness to \textit{quantum cloning} (see \cite{Wootters_Q_no_cloning'82}), greater violation of local theories, and benefits when it comes to communication complexity problems \cite{Cozzolino_HD_QC'19}.} -- is a multi-level computational unit alternative to the conventional \textit{2-level} qubit \cite{Wang_Qudits_HD_QC_2020}. More specifically, a qudit\footnote{As the basic computational element for quantum algorithms, qudit can replace qubit and the state of a qudit is altered by \textit{qudit gates} \cite{Wang_Qudits_HD_QC_2020}. Meanwhile, high-dimensional quantum states such as qudits can be generated with \textit{bulk optics} and \textit{integrated photonics} \cite{Cozzolino_HD_QC'19}. The following physical platforms have been used to implement qudit gates or qudit algorithms: the \textit{time and frequency bin of a photon}, \textit{ion trap}, \textit{nuclear magnetic resonance} (NMR), and \textit{molecular magnets} \cite{Wang_Qudits_HD_QC_2020}.} is a quantum version of $d$-ary digits whose state can be characterized by a vector in the $d$-dimensional Hilbert space $\mathcal{H}_d$ \cite{Wang_Qudits_HD_QC_2020}. $\mathcal{H}_d$ is spanned by a set of orthonormal basis vectors $\{\ket{0}, \ket{1}, \ket{2}, \ldots, \ket{d-1}\}$ \cite{Wang_Qudits_HD_QC_2020}. Using these basis vectors, the state of a qudit takes the general form \cite[eq. (1)]{Wang_Qudits_HD_QC_2020}  
\begin{subequations}
	\begin{align}
		\label{qudit_defn_1}
		\ket{\phi} &\eqdef \alpha_0 \ket{0}+\alpha_1 \ket{1}+ \alpha_2 \ket{2} + \ldots +  \alpha_{d-1} \ket{d-1}  \\
		\label{qudit_defn_2}
		&=\begin{bmatrix}
			\alpha_0 & \alpha_1 & \alpha_2 & \ldots & \alpha_{d-1}    
		\end{bmatrix}^T \in \mathbb{C}^d,
	\end{align}
\end{subequations}
where $\alpha_0,  \alpha_1, \alpha_2, \ldots,  \alpha_{d-1} \in \mathbb{C}$ and $\sum_{i=0}^{d-1} |\alpha_i|^2=1$ \cite{Wang_Qudits_HD_QC_2020}. The qudit $\ket{\phi}$ can also be expressed as a sum of pure states $\ket{\alpha_d}$ within a density matrix representation  given by \cite{Chehimi_Quantum_SemCom'22}
\begin{equation}
	\label{density_matrix_representation}
	\rho \eqdef   \sum_{d=0}^{d-1} p_d \ket{\alpha_d} \bra{\alpha_d},
\end{equation}
where $p_d$ is the selection probability pertaining to the $d$-th pure state.

We now proceed to this paper's prelude. 

\section{Prelude}
\label{sec: prelude}
A number of SemCom techniques inspired by the advancements in 6G research \cite{Tong_Zhu__6G'21,Shaping_Future_6G_Networks'22}; AI \cite{Russel_AI_Book_18,Rebooting_AI'19,Dietterich_Toward_Robust_AI'17}, machine learning (ML) \cite{Jordan_ML_Science'15,MUWCM19,Ghahramani'2015_Probabilistic_ML}, and deep learning (DL) \cite{YYBGH_15,DL_Revolution'18,IGYAC16} research; research on quantum computation \cite{Nielsen_Chuang_QC'10,Scherer_Math_for_QC'19,Preskill2018quantumcomputingin}, quantum communication \cite{Gisin_QCommun'07,Imre_Advanced_QC'12,Cariolaro2015QuantumC}, and quantum networking \cite{Van_Meter_QNetworking'14,Bassoli_QCNs'2021,Djordjevic_QC_QN_and_QS'22}; and research on optical communications have been proposed in not only the wireless domain -- hereinafter referred to \textit{wireless SemCom} -- but also in the optical and quantum  domains. These latter domains' respective SemCom paradigms are henceforth referred to as \textit{optical SemCom} and \textit{quantum SemCom}. Quantum SemCom, optical SemCom, and wireless SemCom are promising 6G enabling technologies that need much more development and discussion. Stimulating a comprehensive discussion toward rigorous theoretical/algorithmic developments of SemCom, we begin our discussion of the state-of-the-art developments of wireless SemCom.

\subsection{Wireless SemCom}
\label{subsec: Wireless_SemCom}
\begin{figure*}[t!]
	\centering
	\includegraphics[scale=0.52]{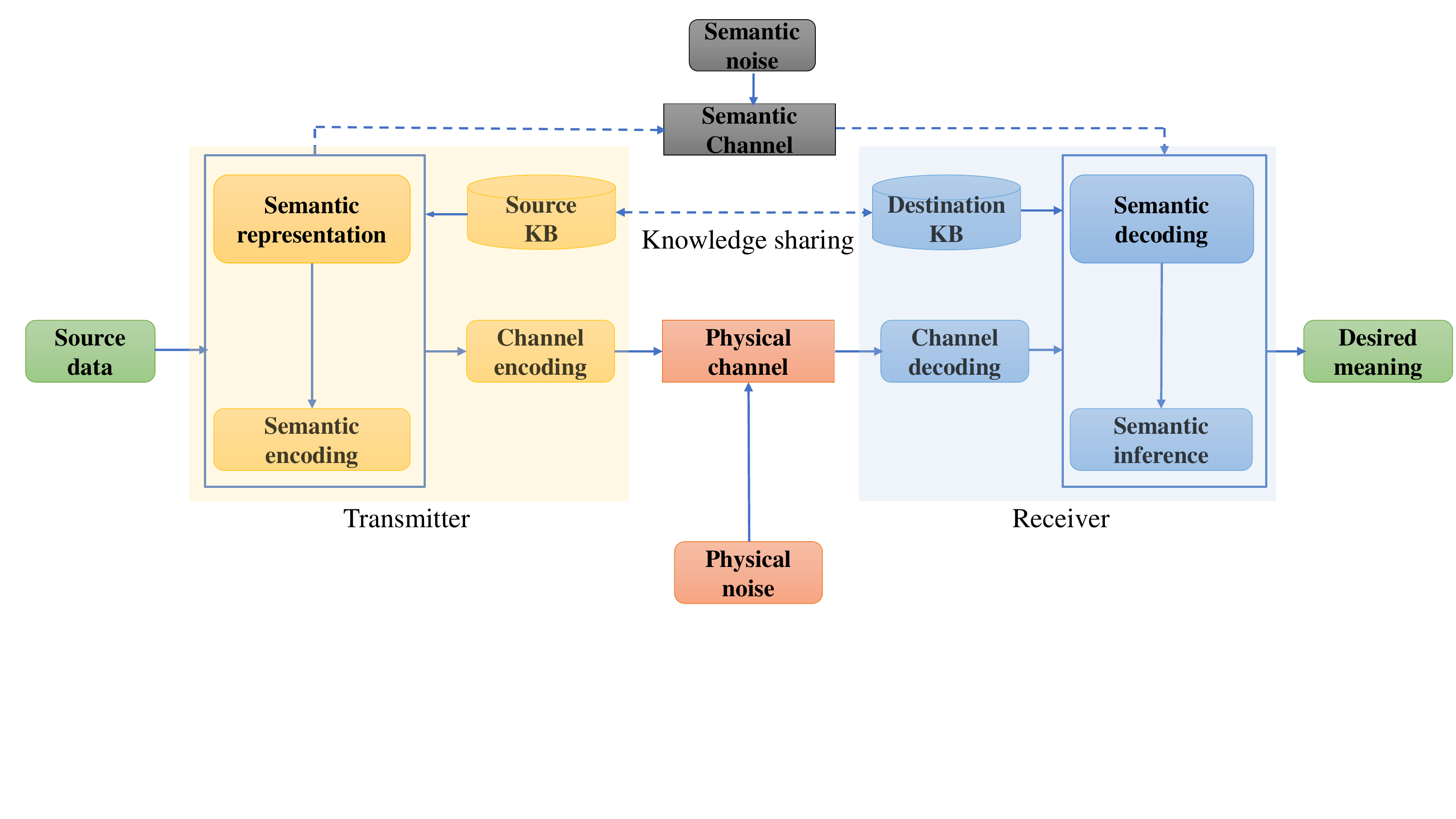}  \vspace{-2.1cm} 
	\caption{System model for SemCom -- modified from \cite[Fig. 6(b)]{SemCom_for_6G_Future_Internet'22}.}
	\label{fig: SemCom_oriented_system_model}
\end{figure*}

Aiming to convey a message's desired meaning (rather than supporting symbol-by-symbol reconstruction), wireless SemCom revolves around the extraction of semantic information that is transmitted -- by the semantic transmitter -- through a wireless communication channel and received by a semantic receiver that has been designed to faithfully recover the transmitted message's intended meaning. Hence, the first step in the wireless SemCom design is the extraction of semantic information to be transmitted from the source data/message to be transmitted. This semantic information extraction is accomplished using a semantic encoder -- by employing the source knowledge base (KB) -- which is often designed by training deep networks such as \textit{transformers} \cite{Universal_transformers'19,Liu_Swin_Transformer'21,Wang_Transformer_empowered_6G'22}. In many the state-of-the-art works, the semantic encoder's function comprises both semantic representation and semantic encoding as schematized in Fig. \ref{fig: SemCom_oriented_system_model}. The output of such a semantic encoder is then fed to a channel encoder, which is usually designed using a trained DNN, that comprises a trained end-to-end semantic transmitter. 

\begin{table}
	\centering
	\begin{tabular}{ | l | l |  }
		\hline
		Abbreviation & Definition    \\ \hline \hline 
		3D  &  Three-dimensional \\ \hline
		5G  & Fifth-generation   \\ \hline
		6G  & Sixth-generation   \\ \hline  
		3-SSIM & Three-component weighted SSIM   \\ \hline
		AI & Artificial intelligence  \\ \hline
		ANSI	& American National Standards Institute  \\ \hline
		AN-SNR &  Anti-noise SNR \\ \hline
		AoI  &  Age of information  \\ \hline
		AoII & Age of incorrect information  \\ \hline
		BER &  Bit error rate  \\ \hline
		BERT & Bidirectional encoder representations    \\ 
		&  from transformers \\ \hline
		BLEU & Bilingual evaluation understudy  \\ \hline
		BS  &  Base station \\ \hline
		CE & cross-entropy    \\ \hline
		CER & 	Character error rate  \\ \hline
		CIDEr & Consensus-based image description evaluation  \\ \hline
		CLUB  &  Contrastive log-ratio upper bound \\ \hline
		CNN & Convolutional neural network  \\ \hline
		CVQ & Continuous video quality  \\ \hline
		CW-SSIM  & Complex-wavelet SSIM \\ \hline
		DL	& Deep learning   \\ \hline
		DLM &  Detail loss metric \\ \hline
		DNNs&   Deep neural networks  \\ \hline
		DTMC &  Discrete-time Markov chain \\ \hline
		Abbreviation & Definition    \\ \hline \hline
		DVQ & Digital video quality  \\ \hline
		FDSD &  Fréchet deep speech distance \\ \hline
		FID & Fréchet inception distance  \\ \hline
		FR &  Full-reference \\ \hline
		FR-TV     & Full reference television  \\ \hline
		FSIM & Feature similarity index for image quality    \\ 
		&  assessment  \\ \hline
		FVQA  & Fusion-based video quality assessment \\ \hline
		GANs &  Generative adversarial networks \\ \hline
		HARQ & Hybrid automatic repeat request  \\ \hline
		HDTV & High Definition TV  \\ \hline
		HVS  &  Human visual system \\ \hline
		IFC & Information fidelity criterion  \\ \hline
		IM/DD	&    Intensity modulation / direct detection  \\\hline
		IMT & Interdisciplinary, multidisciplinary, and \\ 
		& transdisciplinary  \\ \hline
		ind  &  The indicator error  \\ \hline
		IoT & Internet of things  \\ \hline
		IQA &  Image quality assessment \\ \hline
		IS &  Inception score \\ \hline
		iSemCom &  Intelligent SemCom \\ \hline
		iSemCom-HetNet & An iSemCom-enabled heterogeneous     \\ 
		&  network \\ \hline
		ISS &  Image-to-graph semantic similarity \\ \hline
		ITU & International Telecommunication Union   \\ \hline
		IW-SSIM	&  Information content weighted SSIM \\ \hline
		KB & Knowledge base    \\ \hline
		KDSD &  Kernel deep speech distance \\ \hline
		KGs & Knowledge graphs  \\ \hline
		KID & Kernel inception distance  \\ \hline
		KPIs	&  Key performance indicators \\ \hline
		LPIPS &  Learned perceptual image patch similarity \\ \hline
		MAD &  Most apparent distortion \\ \hline 
		MCPD  &  Mean co-located pixel difference  \\ \hline
		MGA-based IQA &  Multi-scale geometric analysis-based IQA \\ \hline
		METEOR	& Metric for evaluation of translation with   \\ 
		&   explicit ordering  \\ \hline 
			MI & Mutual information   \\ \hline
		mIoU & Mean intersection over union  \\ \hline
		ML & Machine learning   \\ \hline
		MMF & Multi-metric fusion  \\ \hline
		MODA &  Multiple object detection accuracy  \\ \hline
	\end{tabular}   \\ [2mm]
	\caption{List of abbreviations and acronyms I.}
	\label{table: abbreviations_and_acronyms_I}
\end{table}

\begin{table}
	\centering
	\begin{tabular}{ | l | l |  }
		\hline
		Abbreviation & Definition    \\ \hline \hline
		MOS & 	Mean opinion score  \\ \hline
		MOVIE &  Motion-based video integrity evaluation \\ \hline
		MPJAE &  Mean per joint angle error \\ \hline
		MPJLE & Mean per joint localization error  \\ \hline
		MPJPE &  Mean per joint position error \\ \hline
		MSE & Mean squared error  \\ \hline
		MSS  &  Metric of semantic similarity \\ \hline
		MS-SSIM & Multi-scale structural similarity index measure   \\ \hline
		MSSIM & Mean SSIM  \\ \hline
		MUs & Mobile users  \\ \hline
		N-MODA  &   Normalized MODA  \\ \hline
		NQM & Noise quality measure  \\ \hline
		NTIA & National Telecommunications and Information   \\ 
		& Administration   \\ \hline
		NR &   No-reference \\ \hline
		OAM & Orbital angular momentum  \\ \hline
		OFC & Optical fiber communication   \\ \hline
		OFDMA & Orthogonal frequency division multiple access  \\ \hline
		PAM8 & Pulse-amplitude modulation 8    \\ \hline
		PAMS &  Perceptual analysis measurement system \\ \hline
		PDF  &  Probability distribution function \\ \hline
		PESQ & Perceptual evaluation of speech quality   \\ \hline
		PSNR & Peak signal-to-noise ratio  \\ \hline
		PSNR-HVS-M  &  Peak signal-to-noise ratio-human vision system  \\ 
		&   modified   \\ \hline
		PSQM & Perceptual speech quality measure  \\ \hline
		QAoI   &   Age of information at query  \\ \hline
		QC & Quantum computing   \\ \hline
		QCIF &  Quarter Common Intermediate Format \\ \hline
		QKD & Quantum key distribution  \\ \hline
		QML &  Quantum machine learning \\ \hline
		QoE &  Quality-of-experience  \\ \hline
		QRAM & Quantum random access memory \\ \hline
		QSC &  Quantum semantic communication \\ \hline
		rAoI &  relative age of information \\ \hline
		RR &  	Reduced-reference \\ \hline
		RHS  & Right-hand side  \\ \hline
		RVs & Random variables  \\ \hline 
		SBERT	&  Sentence-BERT  \\ \hline
		SemCom  & Semantic communication   \\ \hline
		SDR & Signal-to-distortion ratio  \\ \hline
		SINR &  Signal-to-interference-plus-noise ratio \\ \hline
		SMI & Semantic mutual information  \\ \hline
		SNR & Signal-to-noise ratio  \\ \hline
		sq  & The squared error  \\ \hline
		S-R & Semantic transmission rate  \\ \hline
		S-SE &   Semantic spectral efficiency \\ \hline
		SSIM	& Structural similarity index measure   \\ \hline
		SSM & Semantic similarity metric  \\ \hline
		ST &  Spatio-temporal  \\ \hline
		STM & System throughput in message  \\ \hline
		STAQ & Spatial–temporal assessment of quality  \\ \hline
		ST-MAD &  Spatiotemporal MAD \\ \hline
		SVM & Support vector machine  \\ \hline
		TDP &   Triplet drop probability\\ \hline
		threshold & The threshold error   \\ \hline
		VFD &  Variable frame delay \\ \hline
		VIF	&  Visual information fidelity \\ \hline
		VMAF  &  Video multi-method assessment fusion \\ \hline
		VoI & Value of information  \\ \hline 
		VQA &  Video quality assessment \\ \hline
		VQEG &  Video Quality Experts Group \\ \hline
		VQM & Video quality metric  \\ \hline
		VQM$\_$VFD &  Video quality model for variable frame delay \\ \hline
		VSNR &  Visual signal-to-noise ratio \\ \hline
		WER &  Word error rate \\ \hline
		w.r.t.  &  With respect to \\ \hline
		XR  & Extended reality   \\ \hline
	\end{tabular}   \\ [2mm]
	\caption{List of abbreviations and acronyms II.}
	\label{table: abbreviations_and_acronyms_II}
\end{table}

\newpage

The semantic transmitter's output is sent through a channel whose output is received by the semantic receiver. As shown in Fig. \ref{fig: SemCom_oriented_system_model}, the semantic receiver is built using a DL-based channel decoder followed by a deep network-based semantic decoder. The DL-based semantic decoder performs semantic decoding followed by semantic inference   -- using the destination KB as viewed in Fig. \ref{fig: SemCom_oriented_system_model} -- to faithfully recover the transmitted message's intended meaning. While the semantic receiver aims to determine the intended meaning, it can suffer greatly from \textit{semantic noise}\footnote{Semantic noise causes semantic information to be misunderstood by producing a misleading meaning between the transmitter's intended meaning and the receiver's recovered meaning \cite{Hu_Robust_SemCom_with_Masked_VQ-VAE'22}.} so long as there is a mismatch between the source KB and the destination KB. The destination KB, meanwhile, needs to be shared with the source KB in real-time for effective SemCom akin to productive human conversation, which requires common knowledge of the communicating parties' language and culture \cite{SemCom_for_6G_Future_Internet'22}.

Advancements in DL, in particular, and AI, in general, have spurred a surge in research contributions pertaining to the design and optimization of various DL-enabled wireless SemCom systems. Such SemCom systems constitute the state-of-the-art algorithmic research developments in wireless text SemCom \cite{SemCom_Game'18,Farsad_DL_JSCC'18,Xie_DL-based_SemCom'21,Xie_Lite_distributed_SemCom'21,Zhou_WiCom_Letters'22,Peng_Robust_DL-Based_SemCom'22,Yao_Semantic_Coding'22,Lu_RL-powered_SemCom'21,Rethinking_modern_com_Lu_2022,Luo_SemCom_with_relay'21,Jiang_Deep_Source-Channel_coding'22,Liu_Context-Based_SemCom'22}; wireless audio SemCom \cite{Weng_SemCom_Sys_Speech_Trans'21,SemCom_for_speech_signals'20,Weng_SemCom_Speech_Recognition'21,Han_Semantic-aware_Speech2Text_Transmission'22,Weng_DL-enabled_SemCom'22,Tong_FL_ASC'21}; wireless image SemCom \cite{Eirina_JSCC'19,Kurka_Deep_JSCC-f'20,Bandwidth_Agile_Image_Transmission'21,Zhang_Wireless_Information_Transmission_of_Image'22,Xu_Wireless_Image_Transmission'22,Pan_IM-SemCom'22,Yang_WITT'22,Dai_NLT_SCC'22,Lee_Joint_Transmission_Recognition_for_IoTs'19,Hu_Robust_SemCom'22,Huang_Toward_SemCom'23}; wireless video SemCom \cite{Jiang_Wireless_SemCom'22,Wang_Wireless_Deep_Video_Transmission,Tung_DeepWiVe'21,Huang_IS-SemCom'22}; wireless multimodal SemCom \cite{WANG_Multimodal_SemCom'23}; and wireless cross-modal SemCom \cite{Li_Cross-Modal_SemCom'22} pertaining to the efficient wireless transmission of text data, audio data, image data, video data, multimedia data, and multimedia and haptic data, respectively. All these wireless SemCom techniques have been demonstrated to outperform traditional/conventional wireless communication schemes, especially in low signal-to-noise ratio (SNR) regimes.

In addition to the aforementioned wireless SemCom techniques, the rapidly evolving state-of-the-art research landscape of SemCom also encompasses numerous SemCom techniques and trends such as cognitive SemCom \cite{Cognitive_SemCom_Systems'22}; implicit SemCom \cite{Xiao_Reasoning_on_the_Air'22}; adaptive SemCom \cite{Dai_Adaptive_SemCom'22}; context-based SemCom \cite{zhang_Context-Based_SemCom'22,Liu_Context-Based_SemCom'22}; digital SemCom \cite{Learning_Based_Digital_SemCom'22,Fu_VQ_SemCom'22}; SemCom with conceptual spaces \cite{SemCom_with_Conceptual_Spaces'22}; inverse SemCom \cite{Du_RIS-aided_Encoding'22}; one-to-many SemCom \cite{Hu_One-to-Many_SemCom'22}; cooperative SemCom \cite{Xu_SemCom_for_IoV'22}; strategic SemCom \cite{Xiao_RD_Theory_Strategic_SemCom'22}; and encrypted SemCom \cite{Luo_Encrypted_SemCom'22}. These wireless SemCom techniques have also been corroborated to outperform traditional wireless communication techniques in low SNR regimes. For further details, meanwhile, the reader is referred to the vision papers \cite{Wang_Transformer_empowered_6G'22,SemCom_Game'18}, and \cite{Qiao_What_is_SemCom'21,Yang_SemCom_meets_Edge_Intelligence'22,Strinati_Beyond_Shannon'20,Zhang_Wisdom_Evolutionary_6G'21,Beck_SemCom_Info_Bottleneck_View'22,Phopovski_SE_Filtering'19,Shi_to_Semantic_Fidelity'21,Shi_From_SemCom_to_Sematic-aware_Networking'20,Commun_Beyond_Transmitting_Bits'22,Dong_Semantic_Cognitive_Intell'22,Zhao_Semantic-Native_Communication'22,Seo_SemCom_Protocols'22,Pokhrel_UBT_SemCom'22} and the tutorial/survey papers \cite{Gunduz_Beyond_Transmitting_Bits'22,Sem_Empowered_Commun'22}, and \cite{Chaccour_Building_NG_SemCom_Networks'22,Qin_Sem_Com_Principles_Apps'22,Luo_SemCom_Overview'22,Rethinking_modern_com_Lu_2022,Niu_Towards_SemCom'22,SemCom_for_6G_Future_Internet'22,Engineering_SemCom'22,Zhang_a_New_Paradigm'22,Jiang_Wireless_Semantic_Transmission'22} on state-of-the-art developments in wireless SemCom. 

Inspired by some of the aforementioned wireless SemCom techniques, there are also some SemCom proposals and experimental demonstrations in the domain of optical communications. Thus, we continue with the techniques of optical SemCom.

\subsection{Optical SemCom}
\label{subsec: Optical_SemCom}
The authors of \cite{Yu_Optical_SemCom'22} design and experimentally demonstrate an optical SemCom system in which DL is exploited to extract semantic information from the source and the generated semantic symbols are then directly transmitted through an optical fiber. This optical SemCom system produce higher information compression and achieve more stable performance, particularly in the low received optical power regime, while enhancing the robustness against optical link impairments \cite{Yu_Optical_SemCom'22}.

As part of their proposed optical SemCom system, the authors of \cite{Yu_Optical_SemCom'22} experimentally substantiate the semantic transmission of text and images through an intensity modulation / direct detection (IM/DD)-based optical fiber link. For text transmission, the authors of \cite{Yu_Optical_SemCom'22} design the language attention network to restore the meaning of sentences while minimizing semantic errors. For image transmission, on the other hand, they design the dual-attention residual network to extract rich semantic features from images while keeping semantic errors to a minimum. Moreover, to make semantic decoding robust against evident optical link impairments, they deploy a convolutional neural network (CNN) in the semantic decoding network and perform joint optimization.

For the purpose of comparison, the authors of \cite{Yu_Optical_SemCom'22} carry out experiments on traditional IM/DD pulse-amplitude modulation 8 (PAM8) and PAM4 optical fiber communication (OFC) systems. For these systems, the results reported by the authors of \cite{Yu_Optical_SemCom'22} corroborate that their proposed optical SemCom system achieves higher information compression and is more robust to Gaussian noise as well as optical link impairments \cite{Yu_Optical_SemCom'22}. When the optical channel environment is harsh, the performance of traditional OFC systems drops off a \textquotedblleft cliff,'' whereas the optical SemCom system's performance remains stable \cite{Yu_Optical_SemCom'22}. These results attest to the proposed optical SemCom system's considerable advantages over traditional OFC systems, especially in the low received optical power and high optical link impairment regimes \cite{Yu_Optical_SemCom'22}.  

This optical SemCom system's significant advantages demonstrate the viability of SemCom for 6G and beyond in not only the wireless domain but also the optical domain. Apart from optical and wireless domains, SemCom is also proposed in the quantum domain, which we discuss below.

\subsection{Quantum SemCom}
\label{subsec: Quantum_SemCom}
At the crossroads of  SemCom \cite{Sem_Empowered_Commun'22,Chaccour_Building_NG_SemCom_Networks'22,Qin_Sem_Com_Principles_Apps'22,SemCom_for_6G_Future_Internet'22}; ML \cite{Jordan_ML_Science'15,MUWCM19,Ghahramani'2015_Probabilistic_ML}; quantum ML (QML) \cite{JBPPNWS_17,SSSMM19,NW_QDL_2019}; quantum computing \cite{Nielsen_Chuang_QC'10,Scherer_Math_for_QC'19,Preskill2018quantumcomputingin}; quantum communication \cite{Gisin_QCommun'07,Imre_Advanced_QC'12,Cariolaro2015QuantumC}; and quantum networking \cite{Van_Meter_QNetworking'14,Bassoli_QCNs'2021,Djordjevic_QC_QN_and_QS'22}, the authors of \cite{Chehimi_Quantum_SemCom'22} propose a SemCom system in the quantum domain dubbed \textit{quantum semantic communication} (QSC) \cite[Fig. 1]{Chehimi_Quantum_SemCom'22}. QSC is based on the premise that the $d$-dimensional quantum state -- per (\ref{qudit_defn_1}) -- can be viewed as equivalent to the concept of \textit{finite vocabulary} in the information-theoretic domain \cite{Chehimi_Quantum_SemCom'22}. Accordingly, the set of $d$ orthonormal basis vectors $\{\ket{0}, \ket{1}, \ket{2}, \ldots, \ket{d-1}\}$ spanning the Hilbert space $\mathcal{H}_d$ construct a common language\footnote{Since it is part of a common language, every superposition of the $d$ basis vectors corresponds to a unique contextual meaning \cite{Chehimi_Quantum_SemCom'22}.} -- \textit{vocabulary of contextual meanings} -- that can be employed to create a fitting semantic representation of the data \cite{Chehimi_Quantum_SemCom'22}. The raw data's semantic representation can be efficiently achieved by \textit{quantum embedding} using \textit{quantum feature maps} \cite{Schuld_PRL'2019}.

Using quantum feature maps \cite{Schuld_PRL'2019}, the authors of \cite{Chehimi_Quantum_SemCom'22} propose to encode a classical datum $x\in\mathcal{X}$ into quantum states $\ket{\psi(x)}$ in the $d$-dimensional\footnote{In this particular setting, it is assumed that the Hilbert space dimension $d$ is much greater than the dimension of the classical dataset $\mathcal{X}$ \cite{Chehimi_Quantum_SemCom'22}.} Hilbert space $\mathcal{H}_d$ using a quantum feature map $\psi: \mathcal{X} \rightarrow \mathcal{H}_d$ such that $x \rightarrow \ket{\psi(x)}$. This mapping can be achieved using $U_{\psi}(x)$ which is known as a \textit{feature-embedding circuit}\footnote{Other than circuit-based (gate-based) quantum computing (QC), which is a very popular approach to QC, various other approaches exist, including measurement-based QC \cite{Raussendorf_PRL_one-way_QC'01}, adiabatic QC \cite{Adiabatic_QC_SIAM'07}, and topological QC \cite{Freedman_Topological'02}.} \cite{Schuld_PRL'2019} (or \textit{quantum-embedding circuit} \cite{Chehimi_Quantum_SemCom'22}). $U_{\psi}(x)$ acts\footnote{From a quantum computing viewpoint, the quantum feature map given by $x \rightarrow \ket{\psi(x)}$ corresponds to a state preparation circuit $U_{\psi}(x)$ that acts on the ground state $\ket{0 \ldots 0}$ \cite{Schuld_PRL'2019}.} on the ground or vacuum state $\ket{0 \ldots 0}$ of the Hilbert space $\mathcal{H}_d$ as $U_{\psi}(x) \ket{0 \ldots 0} = \ket{\psi(x)}$ \cite{Schuld_PRL'2019}. This makes it possible to construct the classical datum's quantum-embedded semantic representations via \textit{semantic-embedded quantum states} \cite{Chehimi_Quantum_SemCom'22}. To transmit these states reliably, the authors of \cite{Chehimi_Quantum_SemCom'22} propose to process the semantic-embedded quantum states to be transmitted as follows \cite[Fig. 1]{Chehimi_Quantum_SemCom'22}:
\begin{enumerate}
	\item The semantically-embedded $d$-dimensional quantum states are stored in quantum random access memory (QRAM) \cite{Chehimi_Quantum_SemCom'22}.
	
	\item The speaker implements quantum clustering techniques to construct efficient representations of the quantum semantics \cite{Chehimi_Quantum_SemCom'22}.

	\item The qudits -- corresponding to the quantum semantics -- are generated using orbital angular momentum (OAM) \cite{Chaccour_Fellowshop_Com_Sensing'22} encoding \cite{Chehimi_Quantum_SemCom'22}.
	
	\item One of the generated \textit{entangled photons}\footnote{\textit{Quantum entanglement} -- which Albert Einstein famously referred to as \textquotedblleft spooky action at a distance'' \cite{Van_Meter_QNetworking'14} -- is the very striking (\textit{counter-intuitive}) quantum mechanical phenomenon that the states of two or more quantum subsystems are correlated in a manner that is not possible in classical systems \cite{Van_Meter_QNetworking'14}. Quantum entanglement is a peculiarly quantum mechanical resource that usually plays a prominent role in the applications of quantum computation, quantum information, quantum communication, and quantum networking \cite{Van_Meter_QNetworking'14,Nielsen_Chuang_QC'10}.} is transmitted to the listener over a quantum channel (optical fiber or free-space optical channel) to initiate the quantum entanglement link \cite{Chehimi_Quantum_SemCom'22}.
	
	\item The listener can then detect the transmitted entangled photon and store it in QRAM. Entanglement purification protocols (e.g, \cite{Pan_EntanglementPF_01}) can be subsequently applied whenever needed.
	\item The entanglement link between the speaker and the listener is established \cite{Chehimi_Quantum_SemCom'22}.
	
	\item The speaker maps each of the $K$ semantic-representing $d$-dimensional quantum states to one of its entangled photons \cite{Chehimi_Quantum_SemCom'22}.
	
	\item The quantum teleportation protocol is implemented to deliver the semantics to the listener \cite{Chehimi_Quantum_SemCom'22}.
	
	\item Lastly, the listener conducts quantum measurements (and applies some quantum gates) to retrieve the embedded semantics and recover the context from the raw data using quantum operations \cite{Chehimi_Quantum_SemCom'22}.
\end{enumerate}

\begin{figure*}[htb!]
	\centering
	\includegraphics[scale=0.51]{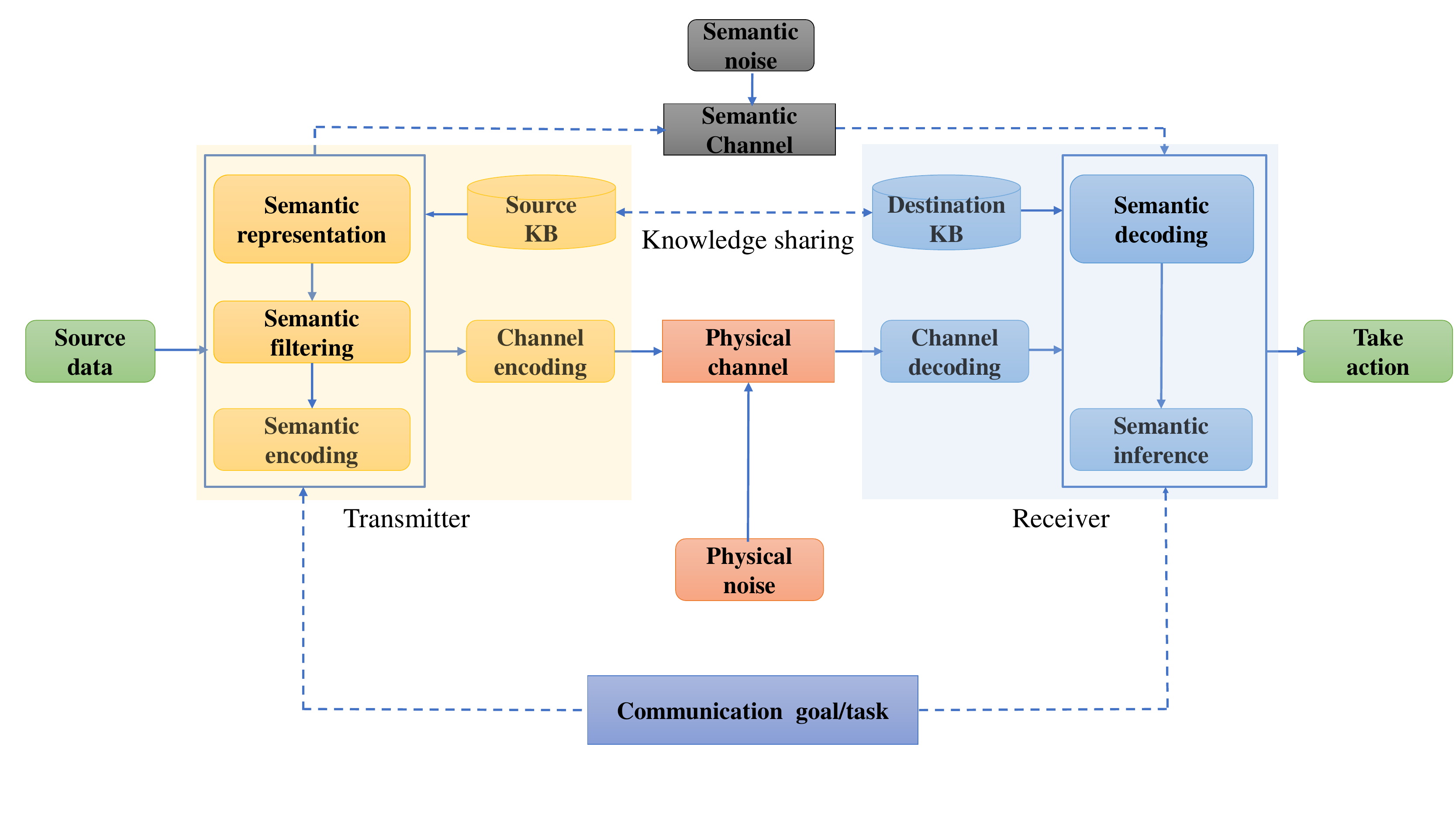} \vspace{-0.4cm}  
	\caption{System model for goal-oriented SemCom -- adapted from \cite[Fig. 6(c)]{SemCom_for_6G_Future_Internet'22}.}
	\label{fig: Goal_oriented_SemCom_system_model}
\end{figure*}

The itemized steps comprise the quantum SemCom technique dubbed QSC \cite[Fig. 1]{Chehimi_Quantum_SemCom'22}. Apart from QSC, the authors of \cite{Kaewpuang_Cooperative_Resource_Management'22} present a quantum SemCom system that is secured by quantum key distribution\footnote{As a secure communication paradigm, QKD utilizes a cryptographic protocol that incorporates components of quantum mechanics. The reader is referred to \cite{Gisin_QCommun'07,Van_Meter_QNetworking'14}, and \cite{Cao_QKD_Networks_Survey'22} for details about state-of-the-art QKD techniques and developments.} (QKD). Meanwhile, it is worth mentioning that quantum SemCom -- like optical SemCom, wireless SemCom, and other communication paradigms -- is not an end but a means to achieve specific goals \cite{Kountouris_Semantics_EmpoweredCF'21,Juba_Universal_SemCom'08}. This goal-oriented viewpoint justifies the need for goal-oriented wireless SemCom techniques, as discussed below.  

\subsection{Goal-Oriented Wireless SemCom}
\label{subsec: GO_SemCom}
Revolving around the effectiveness of communication using semantic information, goal-oriented SemCom enables interested communicating parties to achieve a joint communication goal/task \cite{SemCom_for_6G_Future_Internet'22,Juba_Universal_SemCom'08}. In view of a joint communication goal/task, Fig. \ref{fig: Goal_oriented_SemCom_system_model} shows a generic system model for goal-oriented SemCom, where the goal-oriented SemCom transmitter transforms the source data into semantically encoded information via the cascaded processing -- using the source KB w.r.t. a given communication goal/task -- of semantic representation, semantic filtering, and semantic encoding. The semantically encoded data is then fed into the channel encoder, whose output is transmitted through a wireless physical channel.

The output of the physical channel is received by the goal-oriented SemCom receiver's channel decoder. Acting on the channel decoder's output, the receiver aims to take a desired action -- regarding a communication goal/task and a destination KB (which is shared with the source KB in real-time) -- via a semantic decoding operation followed by semantic inference \cite{SemCom_for_6G_Future_Internet'22}. The inference module's output -- for example, in self-driving cars -- can incorporate action execution instructions such as acceleration and braking; responding to pedestrians, roadblocks, traffic signal changes; and the angle for the steering wheel and flashing the headlights \cite{SemCom_for_6G_Future_Internet'22}. Each of these goals would require application/goal-tailored semantic extraction at the receiver followed by semantic filtering which is, in turn, followed by semantic post-processing prior to the source signal transmission \cite{Kalfa_Toward_GO_Semantic_Signal_Processing'21}, as schematized in \cite[Figure 12]{Kalfa_Toward_GO_Semantic_Signal_Processing'21}.

Concerning wireless SemCom and goal-oriented wireless SemCom, the state-of-the-art also comprises many goal-oriented wireless SemCom developments. Major trends in these developments include task-oriented communication with digital modulation \cite{Xie_Robust_IB'22}; goal-oriented SemCom with AI tasks \cite{Yang_SemCom_with_AI_Tasks'21}; intent-based goal-oriented SemCom \cite{Thomas_NeuroSymb_AI_SemCom'22,Thomas_Neuro-Symbolic_Causal_Reasoning'22}; and multi-user goal-oriented SemCom \cite{Xie_Task-Oriented_MU-SemCom'22}. The reader is referred to the vision papers \cite{Kountouris_Semantics_EmpoweredCF'21,SemCom_Net_Systems'21,Strinati_Beyond_Shannon'20,Kalfa_Toward_GO_Semantic_Signal_Processing'21}, and \cite{Tung_Effective_Commun'21,Effective_Commun_for_6G'22,Sana_Learning_Semantics'21} and the tutorial/survey papers \cite{Sem_Empowered_Commun'22,SemCom_for_6G_Future_Internet'22,Gunduz_Beyond_Transmitting_Bits'22,Zhang_Goal-Oriented_Commun'22}, and \cite{Goal-oriented_SemCom_Thesis} on goal-oriented wireless SemCom.      

Prior to detailing the metrics of SemCom and goal-oriented SemCom (Sections \ref{sec: metrics_for_text} through \ref{sec: Goal_oriented_SemCom_semantic_metrics}), let us first look at the basic \textit{semantic unit (sut)} or \textit{semantic base (Seb)} \cite{Zhang_Wisdom_Evolutionary_6G'21}. Regarding the latter, the authors of \cite{Zhang_Wisdom_Evolutionary_6G'21} introduce the concept of \textit{Seb}\footnote{The authors of \cite{Zhang_Wisdom_Evolutionary_6G'21} believe that Seb will be an essential building block for a more comprehensive semantic information-processing framework that integrates SemCom and semantic computation. To this end, they recommend studying the Seb representation to enable unified/generalized semantic information extraction and representation for multimodal (syntactic) information \cite{Zhang_Wisdom_Evolutionary_6G'21}.} as a basic representation framework for semantic information much like \textit{bit} is the representation and measurement framework for information entropy. According to the authors of \cite{Zhang_Wisdom_Evolutionary_6G'21}, Seb provides a modularized and abstractive method to symbolize semantic information, which inspires SemCom to be more efficient \cite{Zhang_Wisdom_Evolutionary_6G'21}. As an alternative definition of the basic unit of semantic information, the authors of \cite{Resource_allocation_text_SemCom'22} advocate that semantic information can be measured by the sut, defined to designate the basic unit of semantic information. In light of sut and Seb, the design, analysis, and optimization of goal-oriented wireless SemCom systems, quantum SemCom systems, optical SemCom systems, and wireless SemCom systems hinge on adequate semantic metrics. Therefore, we continue below with state-of-the-art semantic metrics for text quality assessment.

\section{Semantic Metrics for Text Quality Assessment}
\label{sec: metrics_for_text}
To assess the quality of text, several semantic metrics have been developed over the years. Some of these metrics have been exploited since recently in the design, analysis, and optimization of state-of-the-art wireless text SemCom systems \cite{SemCom_Game'18,Farsad_DL_JSCC'18,Xie_DL-based_SemCom'21,Xie_Lite_distributed_SemCom'21,Zhou_WiCom_Letters'22,Peng_Robust_DL-Based_SemCom'22,Yao_Semantic_Coding'22,Lu_RL-powered_SemCom'21,Rethinking_modern_com_Lu_2022,Luo_SemCom_with_relay'21,Jiang_Deep_Source-Channel_coding'22,Liu_Context-Based_SemCom'22} and an optical text SemCom system \cite{Yu_Optical_SemCom'22}. Deployed in both optical and wireless text SemCom systems, semantic metrics such as \textit{semantic distance}, \textit{word error rate} (WER), \textit{bilingual evaluation understudy} (BLEU), \textit{consensus-based image description evaluation} (CIDEr), the \textit{semantic similarity metric} (SSM), \textit{the upper tail probability of SSM}, \textit{SSM using sentence-BERT}\footnote{BERT: bidirectional encoder representations from transformers \cite{Peters_BERT_Paper'18}.} (SSM using SBERT), the \textit{metric for evaluation of translation with explicit ordering} (METEOR), and \textit{average bit consumption per sentence} are commonly used by designers of wireless and optical text SemCom systems. The mentioned metrics are discussed below, beginning with semantic distance.

\subsection{Semantic Distance}
Semantic distance (semantic distortion) measures the semantic dissimilarity between two words \cite{Luo_SemCom_Overview'22,SemCom_Game'18}. More specifically, semantic distance quantifies the distortion between two words $w, \hat{w}$ on a semantic level and is defined as \cite[eq. (5)]{SemCom_Game'18}   
\begin{equation}
	\label{semantic_distance_defn}
	d(w,\hat{w} ) \eqdef 1 - \textnormal{sim}(w,\hat{w}),
\end{equation}
where $w, \hat{w}\in\mathcal{W}$ -- $\mathcal{W}$ being a finite set of all meaningful words -- and $\textnormal{sim}(w,\hat{w})\in[0,1]$ denotes the semantic similarity between $w$ and $\hat{w}$. Using (\ref{semantic_distance_defn}), we can determine the average semantic error, which is the average semantic distance in probability \cite{Luo_SemCom_Overview'22,SemCom_Game'18}. To define this probability formally, let the encoder of \cite[Fig. 2]{SemCom_Game'18} observe a word $w$ from a finite set $\mathcal{W}$ with a probability $\mathbb{P}(W = w)$. The encoder maps $w$ into a channel input $\bm{x} = [x_1, \ldots , x_n]\in\mathcal{X}^{(n)}$ using an encoding function $g: \mathcal{W} \to \mathcal{X}^{(n)}$, where $\mathcal{X}^{(n)} \subseteq \mathcal{X}^n$ and $\mathcal{X}$ is a finite alphabet, and $g\in\mathcal{G}$ given $\mathcal{G}$ is the set of all encoding functions. The channel input $\bm{x}$ is then transmitted through a noisy channel -- which is characterized by the conditional probability $p(\bm{Y} = \bm{y}|\bm{X} = \bm{x})$\footnote{As opposed to our notation, $\bm{Y}$ and $\bm{X}$ denote multivariate RVs in this particular case.} -- that produces channel output $\bm{y}=[y_1, \ldots, y_n] \in \mathcal{Y}^{(n)}$, where $\mathcal{Y}^{(n)} \subseteq \mathcal{Y}^n$ and $\mathcal{Y}$ is a finite alphabet. The channel output $\bm{y}$ is fed to a decoder that recovers a word $\hat{w} \in \mathcal{W}$ from $\bm{y}$ w.r.t. the context $q$ by employing a decoding function $h : \mathcal{Y}^{(n)} \times  \mathcal{Q} \to \mathcal{W}$ given that $h\in \mathcal{H}$ -- with $\mathcal{H}$ being the set of all valid decoding functions -- and $q \in \mathcal{Q}$, with $\mathcal{Q}$ being the set of all plausible contexts. For this particular setting, the average semantic error (or average semantic distortion) is defined using (\ref{semantic_distance_defn}) as \cite[eq. (7)]{SemCom_Game'18}
\begin{multline}
	\label{average_semantic_error_defn}
	D_{\theta}\big( (g, h), \mathbb{P}(Q|W , \Theta = \theta) \big) \eqdef\sum_{w\in \mathcal{W}, q\in \mathcal{Q}, \bm{y}\in \mathcal{Y}^{(n)}}   \sum_{\bm{x}\in \mathcal{X}^{(n)}} \\ 
	p(W =w, Q =q, \bm{Y}=\bm{y}, \bm{X}=\bm{x}|\Theta=\theta) d(w, h(\bm{y}, q)),
\end{multline}
where $(g, h) \in \mathcal{G} \times \mathcal{H}$ and the RV $\Theta$ characterizes a given agent's nature -- either helpful or adversarial -- via $\mathbb{P}(\Theta=\theta)$ which is defined in \cite[eq. (1)]{SemCom_Game'18}. Because average semantic error -- per (\ref{average_semantic_error_defn}) -- determines only the semantic similarity between individual words, it would be difficult to compute for large datasets \cite{Sem_Empowered_Commun'22}. This leads us to discuss a computationally easy semantic metric for the assessment of both text and speech quality named WER.

\subsection{Word Error Rate}
\label{subsec: WER_text_SemCom}
WER is defined as the edit distance normalized by the length of a sentence \cite{Luo_SemCom_Overview'22}. This text SemCom metric is therefore easy to calculate and can reflect semantic similarity to a certain extent \cite{Sem_Empowered_Commun'22}. Nevertheless, WER cannot capture the effects of synonyms or semantic similarity \cite{Sem_Empowered_Commun'22}.

We now proceed with our discussion of a text quality assessment metric that is useful for the design, analysis, and optimization of text SemCom systems -- named BLEU.

\subsection{Bilingual Evaluation Understudy}
\label{subsec: BLEU}
To evaluate the quality of a machine translated text, the BLEU score \cite{Papineni-etal-bleu'02} is a metric that is commonly used to assess the effectiveness of text SemCom systems \cite{Xie_DL-based_SemCom'21,Xie_Lite_distributed_SemCom'21,Zhou_WiCom_Letters'22,Peng_Robust_DL-Based_SemCom'22,Yao_Semantic_Coding'22,Jiang_Deep_Source-Channel_coding'22}. SemCom systems' performance can be quantified using the BLEU score -- between the transmitted sentence $\bm{s}$ and the recovered sentence $\hat{\bm{s}}$ -- which is defined as \cite{Papineni-etal-bleu'02}, \cite[eq. (14)]{Qin_Sem_Com_Principles_Apps'22}
\begin{equation}
	\label{BLEU_calc_definition}
	\log \textnormal{BLEU} \eqdef \min(1-l_{\hat{\bm{s}}}/l_{\bm{s}}, 0) + \sum_{n=1}^N u_n \log P_n,
\end{equation}
where $l_{\hat{\bm{s}}}$ and $l_{\bm{s}}$ are, respectively, the length of $\hat{\bm{s}}$ and $\bm{s}$, $u_n$ denotes the weights of the \textit{$n$-grams}, and $P_n$ is the $n$-grams score defined as \cite[eq. (15)]{Qin_Sem_Com_Principles_Apps'22}
\begin{equation}
	\label{n-grams_score_def}
	P_n \eqdef \frac{\sum_k \textnormal{min}\big( C_k(\hat{\bm{s}}), C_k(\bm{s}) \big)}{\sum_k \textnormal{min}\big( C_k(\hat{\bm{s}}) \big)},
\end{equation}
where $C_k(\cdot)$ represents the frequency count function for the $k$-th element in the $n$-th gram \cite{Qin_Sem_Com_Principles_Apps'22}. Although BLEU considers linguistic laws given that semantically consistent words often come together in a given corpus, it computes only the differences between the words in two sentences -- without providing any insight into the meaning of the sentences \cite{Yang_SemCom_meets_Edge_Intelligence'22}. More specifically, the BLUE metric cannot distinguish subtle variations in words such as \textit{polysemy}\footnote{Polysemy epitomizes the following phenomenon: when an instance of a word (or phrase) is used in different contexts to convey two or more different meanings \cite{Shi_to_Semantic_Fidelity'21}.} and synonym \cite{Xie_DL-based_SemCom'21}.    

We now continue with our discussion of another text quality assessment metric that is useful for the design, analysis, and optimization of text SemCom systems -- named CIDEr.
\subsection{Consensus-Based Image Description Evaluation}
\label{subsec: CIDEr}
The authors of \cite{CIDEr_paper'15} propose to use CIDEr as an automatic consensus metric of image description quality. CIDEr was originally used to measure the similarity {between a candidate sentence to a collection of human-generated reference sentences (i.e., ground truth sentences) describing a given image \cite{CIDEr_paper'15,Yang_SemCom_meets_Edge_Intelligence'22,SemCom_for_6G_Future_Internet'22}. As a result, CIDEr is used as semantic metric for the text SemCom system proposed by the authors of \cite{Lu_RL-powered_SemCom'21}. To define CIDEr which automatically evaluates -- for a given image $I_i$ -- how well a candidate sentence $c_i$ matches the consensus of a variety of image descriptions $S_i \eqdef \{ s_{i1}, . . . , s_{im}\}$, let all words of the candidate and reference sentences be mapped to their root forms, each sentence be represented by the set of $n$-grams present in it (where an $n$-gram $\omega_k$ is a set of one or more ordered words \cite{CIDEr_paper'15}), and $h_k(s_{ij})$ $\big( h_k(c_i) \big)$ be the number of times an $n$-gram $\omega_k$ occurs in the $j$-th reference sentence $s_{ij}$ (candidate sentence $c_i$). For this setting, the term frequency-inverse document frequency weighting $g_k(s_{ij})$ for each $n$-gram $\omega_k$ is computed as \cite[eq. (1)]{CIDEr_paper'15}
	\begin{multline}
		\label{TF-IDF_weighting_defn}
		 g_k(s_{ij}) \eqdef \frac{h_k(s_{ij})}{ \sum_{ \omega_l \in \Omega} h_l(s_{ij}) } \times \\ \log\bigg( \frac{|\mathcal{I}|}{\sum_{I_p \in \mathcal{I}} \min(1, \sum_{q} h_k(s_{pq})  )} \bigg),
	\end{multline}
	where $\Omega$ stands for the vocabulary of all $n$-grams and $\mathcal{I}$ is the set of all images in the dataset \cite{CIDEr_paper'15}. Employing (\ref{TF-IDF_weighting_defn}), the CIDEr$_n$ score for $n$-grams of length $n$ is computed using the \textit{average cosine similarity} \cite{Cosine_similarity} between the candidate sentence and the reference sentences as \cite[eq. (2)]{CIDEr_paper'15}
	\begin{equation}
		\label{CIDEr_n_def}
		\textnormal{CIDEr}_n (c_i, S_i) \eqdef \frac{1}{m} \sum_{j=1}^m \frac{\bm{g}_n (c_i) \cdot \bm{g}_n (s_{ij})}{ \| \bm{g}_n (c_i) \| \| \bm{g}_n (s_{ij}) \|},
	\end{equation}
	where $\bm{g}_n (c_i)$ denotes a vector formed by $g_k(c_i)$ that corresponds to all the $n$-grams of the candidate sentence $c_i$, and $\bm{g}_n (s_{ij})$ represents a vector formed by $g_k(s_{ij})$ that signifies all the $n$-grams of the $i$-th reference sentence $s_{ij}$ \cite{CIDEr_paper'15}. In light of (\ref{CIDEr_n_def}), longer $n$-grams are used to capture grammatical properties and richer semantics \cite{CIDEr_paper'15}. To this end, the CIDEr$_n$ scores from $n$-grams of varying lengths are combined using (\ref{CIDEr_n_def}) as follows \cite[eq. (3)]{CIDEr_paper'15}{:   
		\begin{equation}
			\label{CIDEr_combined_defn}
			\textnormal{CIDEr} (c_i, S_i) = \sum_{n=1}^ N  \textnormal{CIDEr}_n (c_i, S_i), 
		\end{equation}
		where uniform weights $w_n = 1/N$ work the best \cite{CIDEr_paper'15} and $1 \leq N \leq 4$ (as constrained by the authors of \cite{CIDEr_paper'15}). The advantage of CIDEr -- as it is defined in (\ref{CIDEr_combined_defn}) -- is that it assesses semantic similarity on the basis of a set of human-generated reference sentences having identical meaning \cite{Yang_SemCom_meets_Edge_Intelligence'22,SemCom_for_6G_Future_Internet'22} rather than a reference sentence like BLEU. On the other hand, the downside of CIDEr like BLUE is that it is based on the comparison of word groups -- CIDEr captures the semantic similarity at the word level \cite{SemCom_for_6G_Future_Internet'22,Yang_SemCom_meets_Edge_Intelligence'22}, rather than the sentence level while considering the various possible contexts of a word.
		
		To address the linguistic fact that a word can have different meanings in various contexts (e.g., \textquotedblleft mouse'' in biology and \textquotedblleft mouse'' in computer science), the authors of \cite{Xie_DL-based_SemCom'21} introduce SSM, which we discuss below. 
		
		\subsection{Semantic Similarity Metric}
		\label{subsec: SSM}
		SSM measures the semantic similarity between the transmitted sentence $\bm{s}$ and the estimated sentence $\hat{\bm{s}}$. For $\hat{\bm{s}}$ and $\bm{s}$, SSM is defined as \cite[eq. (13)]{Xie_DL-based_SemCom'21}, \cite[eq. (16)]{Qin_Sem_Com_Principles_Apps'22}
		\begin{equation}
			\label{Sen_similarity_metric_definition}
			\eta(\hat{\bm{s}}, \bm{s}) \eqdef \frac{ \bm{B}_{\bm{\Phi}}(\bm{s}) \bm{B}_{\bm{\Phi}}(\hat{\bm{s}})^T }{ \| \bm{B}_{\bm{\Phi}}(\bm{s}) \|  \| \bm{B}_{\bm{\Phi}}(\hat{\bm{s}}) \|}, 
		\end{equation}
		where $0 \leq \eta(\hat{\bm{s}}, \bm{s}) \leq 1$ and $\bm{B}_{\bm{\Phi}}(\cdot)$ denotes the output of BERT, which is an enormous pre-trained model that encompasses billions of parameters used for mining semantic information \cite{Xie_DL-based_SemCom'21}. As defined in (\ref{Sen_similarity_metric_definition}), the metric $\eta (\bm{s}, \hat{\bm{s}})$ takes values between 0 and 1 (which mirror \textit{semantic irrelevance} and \textit{semantic consistency}, respectively) \cite{Jiang_Reliable_SemCom'22}. Meanwhile, since BERT are sensitive to polysemy, semantic information is quantified by the sentence similarity metric at the sentence level \cite{Yang_SemCom_meets_Edge_Intelligence'22}. Meanwhile, the probabilistic aspect of a BERT-based SSM per (\ref{Sen_similarity_metric_definition}) can be assessed using a probabilistic metric named the \textit{upper tail probability of SSM}.
		
		\subsection{Upper Tail Probability of SSM}
		\label{subsec: upper_tail_prob_SSM}
		The upper tail probability of SSM $\eta (\bm{s}, \hat{\bm{s}})$ w.r.t. $\eta_{\textnormal{min}}\in[0,1]$ is proposed by the authors of \cite{arXiv_Getu_DeepSC_Performance_Limits'23} as a suitable metric for assessing the performance of a wireless text SemCom technique and is defined as \cite[eq. (8)]{arXiv_Getu_DeepSC_Performance_Limits'23}
		\begin{equation}
			\label{tail_probability_defn}
			p(\eta_{\textnormal{min}}) \eqdef \mathbb{P}\big( \eta (\bm{s}, \hat{\bm{s}}) \geq \eta_{\textnormal{min}} \big),  
		\end{equation}
		where $\eta (\bm{s}, \hat{\bm{s}})$ is defined in (\ref{Sen_similarity_metric_definition}) and $\eta_{\textnormal{min}}$ stands for minimum semantic similarity. The upper tail probability of SSM is useful for quantifying the probabilistic assessment of wireless/optical text SemCom techniques. To this end, the authors of \cite{arXiv_Getu_DeepSC_Performance_Limits'23} employed it to quantify the asymptotic performance of a DL-enabled semantic communication system (\textit{DeepSC} \cite{Xie_DL-based_SemCom'21}) subject to single-interferer as well as multi-interferer radio frequency interference. It is worth underscoring, however, that employing the upper tail probability of SSM to assess the performance of a text SemCom technique can lead to mathematical intractability -- especially when analyzing the non-asymptotic performance of a DL-based text SemCom technique -- due to DL models' fundamental \textit{lack of interpretability} \cite{Poggio_Theo_Issues_Dnets_2020,Toward_Science_of_Interpretable_ML'17} and the lack of a commonly agreed-upon (unified) definition of semantics / semantic information.
		
		The probabilistic metric set out in (\ref{tail_probability_defn}) is inspired by the SSM metric defined in (\ref{Sen_similarity_metric_definition}). The metric in (\ref{Sen_similarity_metric_definition}) is a cosine similarity metric using BERT. Nevertheless, the sentence embeddings that result from using a pre-trained BERT model without fine-tuning on semantic textual similarity task inadequately capture the sentences' semantic meaning due to anisotropic embedding space \cite{Li-etal-2020-sentence_embeddings,Lee_EQ2SEQ-SC'22}. We therefore discuss below another text SemCom metric termed \textit{SSM using SBERT}\footnote{SBERT: sentence-BERT.} \cite{Reimers_SBERT'19}.
		
		\subsection{SSM using SBERT}
		To begin with, \textquotedblleft child'' and \textquotedblleft children'' are semantically associated even though their lexical similarity computed using BLEU is zero \cite{Lee_EQ2SEQ-SC'22}. Despite the input and output having such a low BLEU score for lexical similarity, their semantic similarity can be high \cite{Lee_EQ2SEQ-SC'22}. To capture this notion of high semantic similarity, the authors of \cite{Lee_EQ2SEQ-SC'22} represent sentences as embeddings using an embedding model $\bm{M}$ and compute the cosine similarity between the input sentence $\bm{s}$ and the recovered sentence $\hat{\bm{s}}$ as follows \cite[eq. (4)]{Lee_EQ2SEQ-SC'22}:
		\begin{equation}
			\label{SSM_using_SBERT_defn}
			\textnormal{match}(\hat{\bm{s}}, \bm{s}) \eqdef \frac{\bm{M}(\bm{s})\bm{M}(\hat{\bm{s}})^T}{ \| \bm{M}(\bm{s})\| \| \bm{M}(\hat{\bm{s}})\|}.
		\end{equation}
		Rather than using BERT without fine-tuning on semantic textual similarity task (which will poorly capture the semantic meaning of the sentences \cite{Li-etal-2020-sentence_embeddings,Lee_EQ2SEQ-SC'22}), the authors of \cite{Lee_EQ2SEQ-SC'22} use SBERT \cite{Reimers_SBERT'19} -- fine-tuned on semantic textual similarity tasks -- as an embedding model $\bm{M}$. To this end, the definition in (\ref{SSM_using_SBERT_defn}) represents the metric SSM using SBERT provided that the SBERT model is fine-tuned on semantic textual similarity tasks to encode the sentence embedding \cite{Lee_EQ2SEQ-SC'22}.
		
		We now move on to our discussion of another text quality assessment metric that is useful for the design, analysis, and optimization of text SemCom systems -- termed METEOR.
		
		\subsection{Metric for Evaluation of Translation with Explicit Ordering}
		\label{subsec: METEOR}
		METEOR is an automatic metric for the assessment of machine translation that is based on a generalized concept of \textit{unigram matching} -- based on their surface forms, stemmed forms, and meanings -- between a translation produced by a machine and a set of reference translations produced by a human \cite{METEOR_paper'05}. It therefore expands the synonym set by introducing external knowledge sources \cite{Jiang_Reliable_SemCom'22}, such as \textit{WordNet} (see \cite{Fellbaum'00_WordNetA}). In addition, METEOR employs precision $P_m$ and recall $R_m$ to evaluate the similarity between transmitted and received texts as follows \cite[eq. (3)]{Jiang_Reliable_SemCom'22}: 
		\begin{equation}
			\label{METEOR_defn}
			\textnormal{METEOR} \eqdef (1-\textnormal{Pen})\bar{F},
		\end{equation}
		where Pen is the penalty coefficient and $\bar{F}$ is the harmonic mean that combines $P_m$ and $R_m$ as given by \cite[eq. (2)]{Jiang_Reliable_SemCom'22}
		\begin{equation}
			\label{F_bar_defn}
			\bar{F} \eqdef \frac{P_m R_m}{\alpha P_m + (1-\alpha)R_m}, 
		\end{equation}
		where $\alpha$ is the hyperparameter according to WordNet \cite{Jiang_Reliable_SemCom'22}. To summarize, the authors of \cite{METEOR_paper'05} substantiate that METEOR considerably improves correlation with human judgment. Despite this notable advantage, it is restricted to unigram matches, which makes it a strictly word-level metric \cite{Extending_METEOR'10}. This leads us to the discussion of our last text SemCom metric, called average bit consumption per sentence.
		
		\subsection{Average Bit Consumption per Sentence}
		\label{subsec: average_bit_consumption}
		The authors of \cite{Jiang_Deep_Source-Channel_coding'22} introduce \textit{average bit consumption per sentence} as a wireless text SemCom metric. This metric measures a system's performance from a communication perspective \cite{Qin_Sem_Com_Principles_Apps'22}. More specifically, the authors of \cite{Jiang_Deep_Source-Channel_coding'22} deploy this text semantic metric to evaluate the performance of their proposed text semantic transmission techniques with hybrid automatic repeat request (HARQ).
		
		The reader is referred to \cite{Chandrasekaran_Semantic_Similarity'22} for a survey on the evolution of semantic similarity and to \cite{Semantic_Textual_Similarity_Methods'16} for a survey on the methods, tools, and applications of semantic textual similarity for additional information on the possibly useful metrics applicable for text SemCom. Wrapping up, the existing semantic metrics for text quality assessment that are applicable in both wireless text SemCom and optical text SemCom are summarized along with their pros and cons in Table \ref{table: text_SemCom_metrics}.
		\begin{table*}
			\centering
			\begin{tabular}{| l | l | l | }
				\hline
				\textbf{Metrics} & \textbf{Pros} & \textbf{Cons}  \\ \hline
				(Average) semantic  & This metric uses semantic distance based on   & Since this metric only calculates the semantic similarity between  \\ 
				distance/distortion & lexical taxonomies as a distortion measure \cite{Sem_Empowered_Commun'22}. &  individual words, it would be difficult to compute for large data sets \cite{Sem_Empowered_Commun'22}.  \\ \hline
				WER & WER is easy to calculate and can reflect the   & WER can hardly capture the effects of synonyms or semantic \\ 
				& semantic similarity to a some extent \cite{Sem_Empowered_Commun'22}.   &  similarity \cite{Sem_Empowered_Commun'22}.    \\ \hline
				BLEU & BLEU observes the fundamental linguistic law  & $1)$ Rather than the semantic meaning of words in sentences, BLEU can     \\  
				& that semantically similar sentences are invariable    & only compare the differences between words in two sentences \cite{Sem_Empowered_Commun'22}.   \\  
				& in the semantic space \cite{Sem_Empowered_Commun'22}.   & $2)$ BLUE cannot distinguish more subtle variation in words such as      \\ 
				&    & polysemy and synonym \cite{Xie_DL-based_SemCom'21}.     \\   \hline    
				CIDEr   & Unlike BLEU, CIDEr does not assess semantic     & $1)$ CIDEr focuses more on the middle part of a sentence (the middle part     \\
				& similarity based on a reference sentence, but a    &  possessing more $n$-gram weight) \cite{Sem_Empowered_Commun'22}.       \\  
				& group of sentences with the same meaning \cite{Sem_Empowered_Commun'22}.      & $2)$  CIDEr captures the respective semantic similarity at the word      \\ 
				&   & level \cite{SemCom_for_6G_Future_Internet'22,Yang_SemCom_meets_Edge_Intelligence'22}, rather than at a sentence level while considering the   \\ 
				&   & several contexts of a word.    \\ \hline 
				SSM (with BERT) & Pertaining to BERT’s sensitivity to polysemy,  & $1)$ As a cause of limitation to this metric, it is not easy to generalize  \\ 
				& SSM (with BERT) can explain semantics at   &  the pre-trained BERT model on others \cite{Sem_Empowered_Commun'22}.    \\ 
				& the sentence level \cite{Sem_Empowered_Commun'22}.    & $2)$ The sentence embeddings from a pre-trained BERT model without     \\ 
				&    & fine-tuning on semantic textual similarity task inadequately capture the     \\
				&    & sentences' semantic meaning because of anisotropic embedding   \\
				&    & space \cite{Li-etal-2020-sentence_embeddings,Lee_EQ2SEQ-SC'22}.    \\    \hline
				The upper tail  & This metric captures all the probabilistic aspects    & This metric can lead to mathematical intractability due to the DL models’  \\ 
				probability of SSM &  of the SSM w.r.t. the minimum semantic   &  fundamental lack of interpretability and the lack of commonly agreed     \\ 
				& similarity $\eta_{\textnormal{min}}\in[0,1]$.   &  upon (unified) definition of semantics as well as semantic information.     \\ \hline
				SSM using SBERT & This text SemCom metric can capture a high   & The SBERT model is fine-tuned on semantic textual similarity tasks  \\  
				& semantic similarity even when the respective    & in order to encode the sentence embedding \cite{Lee_EQ2SEQ-SC'22}.    \\
				& BLEU score is low   &    \\ \hline 
				METEOR & $1)$ METEOR expands the synonym set by   & METEOR is restricted to unigram matches \cite{Extending_METEOR'10}:    \\ 
				& introducing external knowledge sources \cite{Jiang_Reliable_SemCom'22},    & $1)$ By emphasizing on only one match type per stage, the aligner misses    \\ 
				& such as WordNet (see \cite{Fellbaum'00_WordNetA}). $2)$ METEOR can    & a considerable part of the likely alignment space \cite{Extending_METEOR'10}.   \\ 
				&  considerably improve correlation with human    & $2)$ Choosing partial alignments grounded only on the least number of    \\ 
				& judgments \cite{METEOR_paper'05}.  & per-stage crossing alignment links can practically give rise to missing      \\ 
				&    &  full alignments \cite{Extending_METEOR'10}.          \\ \hline
			\end{tabular}  \\ [3mm]
			\caption{Main semantic metrics for text quality assessment along with their pros and cons -- WER: word error rate; BLEU: bilingual evaluation understudy; SSM: semantic similarity metric; BERT: bidirectional encoder representations from transformers; SBERT: sentence-BERT; METEOR: metric for evaluation of translation with the explicit ordering.}
			\label{table: text_SemCom_metrics}
		\end{table*}
		
We now continue with our discussion on state-of-the-art semantic metrics for speech quality assessment.

\section{Semantic Metrics for Speech Quality Assessment}
\label{sec: metrics_for_speech}
For speech quality assessment, the following metrics are commonly used:  \textit{signal-to-distortion ratio} (SDR), \textit{perceptual evaluation of speech quality} (PESQ), \textit{(unconditional) Fr\'echet deep speech distance} (FDSD), and \textit{(unconditional) kernel deep speech distance} (KDSD) \cite{Qin_Sem_Com_Principles_Apps'22,SemCom_for_6G_Future_Internet'22}. Recently, WER and \textit{character error rate} (CER) have been employed to assess the quality of speech recovered by the semantic receiver in a wireless audio SemCom system \cite{Weng_SemCom_Sys_Speech_Trans'21,SemCom_for_speech_signals'20,Weng_SemCom_Speech_Recognition'21,Han_Semantic-aware_Speech2Text_Transmission'22,Weng_DL-enabled_SemCom'22,Tong_FL_ASC'21}. In what follows, we discuss the following semantic metrics applicable to audio SemCom: WER, CER, SDR, PESQ, FDSD, and KDSD. We begin with a brief discussion of WER.

\subsection{Word Error Rate}
\label{subsection: WER}
Since audio data and text data are very similar, WER has also been applied to assess the accuracy of speech signal transmission \cite{Sem_Empowered_Commun'22}. To this end, it is defined in terms of the number of word substitutions ($S_W$), word deletions ($D_W$), and word insertions ($I_W$) as \cite[eq. (10)]{Weng_SemCom_Speech_Recognition'21}
\begin{equation}
	\label{WER_defn}
	\textnormal{WER} \eqdef \frac{S_W + D_W + I_W}{N_W}, 
\end{equation}
where $N_W$ stands for the number of words in the original speech transcription. As defined in (\ref{WER_defn}), WER has been applied in the design of various audio SemCom techniques including the one used in \cite{Han_Semantic-preserved_Com_System'22}.

We now continue with our discussion of a speech quality assessment metric that is useful for the design, analysis, and optimization of audio SemCom systems -- termed CER.

\subsection{Character Error Rate}
Unlike WER for the evaluation of text similarity, the CER metric operates at the character level rather than the word level to assess the accuracy of speech recognition \cite{Sem_Empowered_Commun'22,Weng_SemCom_Speech_Recognition'21}. Accordingly, similar to WER, CER is defined in terms of the number of character substitutions ($S_C$), character deletions ($D_C$), and character insertions ($I_C$) as \cite[eq. (9)]{Weng_SemCom_Speech_Recognition'21}
\begin{equation}
	\label{CER_defn}
	\textnormal{CER} \eqdef \frac{S_C + D_C + I_C}{N_C}, 
\end{equation}
where $N_C$ denotes the number of characters in the original speech transcription.

We now proceed with our discussion of another speech quality assessment metric that is useful for the design, analysis, and optimization of audio SemCom systems -- named SDR.

\subsection{Signal-to-Distortion Ratio}
SDR is a commonly used metric for speech transmission \cite{Weng_SemCom_Sys_Speech_Trans'21,SemCom_for_speech_signals'20,Vincent_Blind_Source_Separation'06}. For a given speech sample sequence $\bm{s} = [s_1, s_2, \dots , s_W ] \in \mathbb{R}^{1\times W}$ and a decoded speech sequence $\hat{\bm{s}} = [\hat{s}_1, \hat{s}_2, \dots , \hat{s}_W ] \in \mathbb{R}^{1\times W}$, SDR is defined as \cite[eq. (6)]{Weng_SemCom_Sys_Speech_Trans'21}, \cite[eq. (13)]{Vincent_Blind_Source_Separation'06}      
\begin{equation}
	\label{SDR_defn}
	\textnormal{SDR} \eqdef 10\log_{10} \bigg( \frac{ \| \bm{s}\|^2 }{ \| \bm{s}-\hat{\bm{s}} \|^2 } \bigg). 
\end{equation}
As can be inferred from (\ref{SDR_defn}), SDR and mean squared error (MSE) are related \cite{Sem_Empowered_Commun'22} such that one can be inferred from the other. Accordingly, (\ref{SDR_defn}) asserts that a lower MSE value leads to higher SDR value, and vice versa. In addition, because a difference in SDR produces a visible performance difference, it can be used to optimize DNNs \cite{Sem_Empowered_Commun'22}.

We now continue with our discussion of yet another speech quality assessment metric that is useful for the design, analysis, and optimization of audio SemCom systems -- termed PESQ.

\subsection{Perceptual Evaluation of Speech Quality}
PESQ \cite{PESQ_paper'01} is an International Telecommunication Union (ITU)-standardized\footnote{ITU standardized PESQ as the \textit{ITU-T Recommendation P.862} \cite{ITU-T_recommendation_P.862'01}.} metric for evaluating the subjective quality of speech signals under various conditions -- such as background noise, analog filtering, and variable delay -- by scoring their quality on a scale from -0.5 to 4.5 \cite{Weng_SemCom_Sys_Speech_Trans'21,PESQ_paper'01}. This metric is the result of merging the perceptual analysis measurement system (PAMS) and an enhanced version of the perceptual speech quality measure (PSQM) named PSQM99 \cite{PESQ_paper'01}. Meanwhile, the basic diagram of PESQ and its philosophy is shown in Fig. \ref{fig: PESQ_diagram_20221020}.     

\begin{figure*}[htb!]
	\centering
	\includegraphics[scale=0.55]{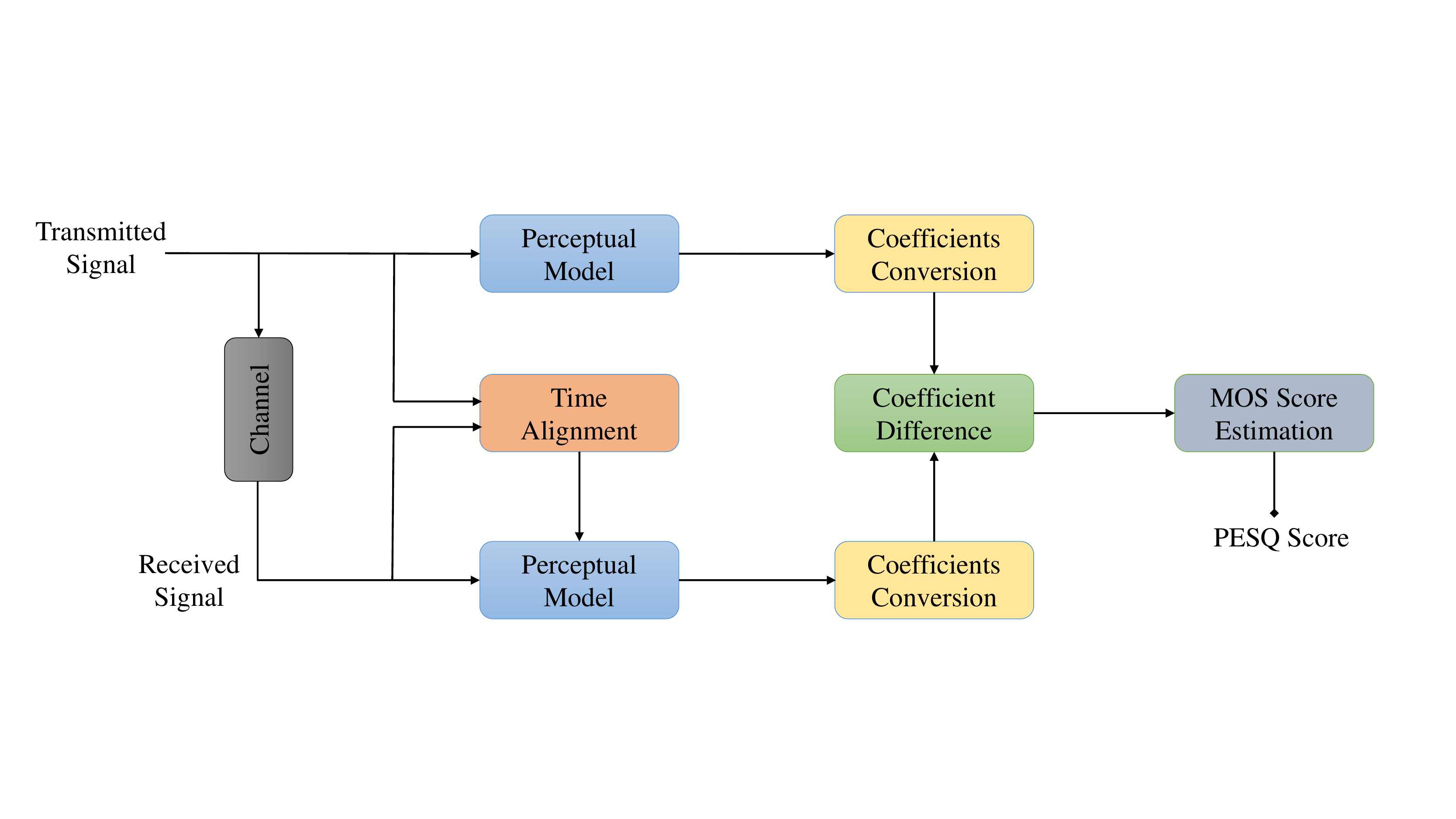}  \vspace{-2.2cm} 
	\caption{Basic diagram of PESQ \cite[Fig. 13]{SemCom_for_6G_Future_Internet'22} and its philosophy \cite[Figure 1/P.862]{ITU-T_recommendation_P.862'01} -- MOS: mean opinion score.}
	\label{fig: PESQ_diagram_20221020}
\end{figure*} 

PESQ is deployed in \cite{Weng_SemCom_Sys_Speech_Trans'21} and \cite{SemCom_for_speech_signals'20} to evaluate the performance of SemCom systems for speech transmission. PESQ presumes {that humans' perceptual memory is short, which makes it a realistic metric w.r.t. human behavior \cite{SemCom_for_6G_Future_Internet'22}. Nonetheless, PESQ quantifies the accuracy of speech transmission rather than its semantic content \cite{SemCom_for_6G_Future_Internet'22}.    
	
	We now continue with our discussion of one more speech quality assessment metric that is important for the design, analysis, and optimization of audio SemCom systems -- dubbed FDSD.
	
	\subsection{Fr\'echet Deep Speech Distance}
	\label{subsec: FDSD}
	FDSD is used to quantify the quality of synthesized speech signals \cite{Qin_Sem_Com_Principles_Apps'22,High_Fidelity_Speech_Synthesis'19}. If we let the original speech samples $\bm{D}\in\mathbb{R}^{K \times L}$ and the synthesized speech samples $\hat{\bm{D}}\in\mathbb{R}^{\hat{K} \times L}$ have \textit{means} $\bm{\mu}_D$ and $\bm{\mu}_{\hat{D}}$, respectively, FDSD can be defined mathematically as \cite[eq. (21)]{Qin_Sem_Com_Principles_Apps'22}
	\begin{equation}
		\label{FDSD_definition}
		\Gamma^2\eqdef\| \bm{\mu}_D- \bm{\mu}_{\hat{D}} \|^2+ \textnormal{tr}\big(\bm{\Sigma}_D + \bm{\Sigma}_{\hat{D}}- \sqrt{\bm{\Sigma}_D  \bm{\Sigma}_{\hat{D}}} \big),  
	\end{equation}
	where $\bm{\Sigma}_D$ and $\bm{\Sigma}_{\hat{D}}$ denote the covariance matrices of $\bm{D}$ and $\hat{\bm{D}}$, respectively. In light of (\ref{FDSD_definition}), the smaller the value of FDSD, the more similar the real and synthesized speech signals are \cite{Qin_Sem_Com_Principles_Apps'22}. FDSD is employed in the design and optimization of an audio SemCom system in \cite{Weng_DL-enabled_SemCom'22}.
	
	This leads us to the discussion of our last speech quality assessment metric that is important for the design, analysis, and optimization of audio SemCom systems -- named KDSD.
	
	\subsection{Kernel Deep Speech Distance}
	\label{subsec: KDSD}
	Like FDSD, KDSD is also utilized to assess the quality of synthesized speech signals \cite{Qin_Sem_Com_Principles_Apps'22,High_Fidelity_Speech_Synthesis'19}. Using the definitions set out in Section \ref{subsec: FDSD}, KDSD can be defined mathematically w.r.t. kernel $q(\cdot,\cdot)$ as \cite[eq. (22)]{Qin_Sem_Com_Principles_Apps'22}
	\begin{multline}
		\label{KDSD_definition}
		\Delta^2 \eqdef \frac{1}{ K(K-1)}  \sum_{1 \leq i, j \leq K: \hspace{1mm} i\neq j} q\big(\bm{D}_i, \hat{\bm{D}}_j\big)      +   \frac{1}{ \hat{K}(\hat{K}-1)}  \\  \times \sum_{1 \leq i, j \leq K: \hspace{1mm} i\neq j} q\big(\bm{D}_i, \hat{\bm{D}}_j\big)+ \sum_{i=1}^K \sum_{j=1}^{\hat{K}}  q\big(\bm{D}_i, \hat{\bm{D}}_j\big).
	\end{multline}
	When it comes to the definition in (\ref{KDSD_definition}), the smaller the KDSD values are, the more similar the real and synthesized speech signals are \cite{Qin_Sem_Com_Principles_Apps'22}. KDSD is exploited in the design and optimization of an audio SemCom system in \cite{Weng_DL-enabled_SemCom'22}.
	
	The aforementioned metrics for speech quality assessment hardly quantify performance at the level of semantic understanding \cite{SemCom_for_6G_Future_Internet'22}. Thus, the audio SemCom research field lacks semantic assessment metrics that incorporate semantic understanding, like BERT and BLEU \cite{SemCom_for_6G_Future_Internet'22}. At last, the existing metrics for speech quality assessment that are applicable to wireless audio SemCom are summarized along with their pros and cons in Table \ref{table: audio_SemCom_metrics}.
	
	\begin{table*}
		\centering
		\begin{tabular}{| l | l | l | }
			\hline
			\textbf{Metrics} & \textbf{Pros} & \textbf{Cons}  \\ \hline
			WER & WER is a computationally easy audio SemCom metric.  & WER's quantification may not be consistent with human perception.  \\ \hline
			CER & CER is also a computationally simple audio SemCom metric. & CER's evaluation may not be consistent with human perception.  \\ \hline
			SDR & $1)$ SDR is easy to calculate \cite{Sem_Empowered_Commun'22}. & The evaluation results of SDR are sensitive to the volume of audios \cite{Sem_Empowered_Commun'22}.   \\    
			& $2)$ SDR can reflect the quality of voice to a certain   &   \\ 
			& degree \cite{Sem_Empowered_Commun'22}.   &     \\ \hline 
			PESQ & $1)$ PESQ's evaluation is objective \cite{Sem_Empowered_Commun'22}.   & PESQ exhibits an intrinsically high computational complexity \cite{Sem_Empowered_Commun'22}.  \\ 
			& $2)$ PESQ's assessment is close to human perception \cite{Sem_Empowered_Commun'22}.   &   \\ \hline
			FDSD & FDSD is demonstrated experimentally that it ranks   & FDSD manifests an inherent computational complexity.   \\ 
			& models consistent with MOSes obtained through    &        \\ 
			& human evaluation \cite{High_Fidelity_Speech_Synthesis'19}.  &        \\ \hline
			KDSD & KDSD is also corroborated experimentally that it ranks  & KDSD exhibits an intrinsic computational complexity.  \\   
			& models in accordance with MOSes obtained via   &        \\ 
			& human evaluation \cite{High_Fidelity_Speech_Synthesis'19}.   &        \\ \hline
		\end{tabular}  \\ [3mm]
		
		\caption{Main semantic metrics for speech quality assessment along with their pros and cons -- WER: word error rate; CER: character error rate; SDR: signal-to-distortion ratio; PESQ: perceptual evaluation of speech quality; FDSD: unconditional Fr\'echet deep speech distance; KDSD: unconditional kernel deep speech distance; MOSes: mean opinion scores.}
		\label{table: audio_SemCom_metrics}
	\end{table*}
	
	We now continue with our discussion on the state-of-the-art semantic metrics for image quality assessment. 

\section{Semantic Metrics for Image Quality Assessment}
\label{sec: metrics_for_image}
Numerous semantic metrics have been proposed to date for image quality assessment (IQA) \cite{Liu_VQA_Survey'13}. Some of these IQA metrics have been exploited in the design, analysis, or optimization of state-of-the-art wireless image SemCom systems \cite{Eirina_JSCC'19,Kurka_Deep_JSCC-f'20,Bandwidth_Agile_Image_Transmission'21,Zhang_Wireless_Information_Transmission_of_Image'22,Xu_Wireless_Image_Transmission'22,Pan_IM-SemCom'22,Yang_WITT'22,Dai_NLT_SCC'22,Lee_Joint_Transmission_Recognition_for_IoTs'19,Hu_Robust_SemCom'22,Huang_Toward_SemCom'23} and an optical image SemCom system \cite{Yu_Optical_SemCom'22}. Applicable to these systems, image SemCom metrics such as \textit{image semantic similarity}, \textit{peak signal-to-noise ratio} (PSNR), \textit{structural similarity index measure} (SSIM), \textit{multi-scale structural similarity index measure} (MS-SSIM), \textit{learned perceptual image patch similarity} (LPIPS), \textit{mean intersection over union} (mIoU), \textit{image-to-graph semantic similarity} (ISS), and \textit{recognition accuracy} are widely used by designers of wireless as well as optical image SemCom systems. These metrics are detailed henceforward, beginning with image semantic similarity.

\subsection{Image Semantic Similarity}
\label{subsec: image_semantic_similarity}
The image semantic similarity of two images $A$ and $B$ is computed as \cite[eq. (18)]{Qin_Sem_Com_Principles_Apps'22}
\begin{equation}
	\label{image_semantic_similarity_defn}
	\Theta(f(A), f(B)) \eqdef \| f(A) - f(B)\|_2^2, 
\end{equation}
where $f(\cdot)$ denotes an image embedding function that maps an image to a point in the Euclidean space \cite{Qin_Sem_Com_Principles_Apps'22}. However, the metric defined by (\ref{image_semantic_similarity_defn}) depends on the higher-order image structure, which is often context-dependent \cite{Qin_Sem_Com_Principles_Apps'22}.   

We now move on to our discussion of a computationally simple IQA metric that is important for designing, analyzing, and optimizing image SemCom systems -- known as PSNR.

\subsection{Peak Signal-to-Noise Ratio}
\label{subsec: PSNR}
PSNR quantifies the ratio between the maximum possible power of the desired signal and the power of the noise that has contaminated the desired signal \cite{Eirina_JSCC'19}. Accordingly, PSNR is defined in a logarithmic-scale as \cite[eq. (4)]{Eirina_JSCC'19} 
\begin{equation}
	\label{PSNR_log_defn}
	\textnormal{PSNR} \eqdef 10\log_{10} \frac{\textnormal{MAX}^2}{\textnormal{MSE}} \hspace{2mm} [\textnormal{dB}],
\end{equation}
where MAX denotes the maximum possible number of image pixels and MSE represents the mean squared error between a reference image and a reconstructed image. The following conclusion can be drawn from the definition in (\ref{PSNR_log_defn}): as the MSE between the transmitted image and the reconstructed image becomes smaller, the PSNR\footnote{PSNR can also be employed to assess the quality of video transmission since a video is made of several image frames \cite{Luo_SemCom_Overview'22}.} gets larger, meaning a better-quality of reconstructed image \cite{Luo_SemCom_Overview'22}.  

We now move on to our discussion of a widely known IQA metric that is also important for designing, analyzing, and optimizing image SemCom systems -- termed SSIM.

\subsection{Structural Similarity Index Measure}
\label{subsec: SSIM}
To formally define the metric SSIM, let us first define an overall similarity measure for two non-negative image signals $\bm{x}$ and $\bm{y}$ as \cite[eq. (5)]{Wang_SSIM_Metric'04} 
\begin{equation}
	\label{Similarity_measure_defn}
	S(\bm{x},\bm{y}) \eqdef f(l(\bm{x},\bm{y}), c(\bm{x},\bm{y}), s(\bm{x},\bm{y})), 
\end{equation}
where $S(\bm{x},\bm{y})$ denotes the overall measure of similarity between $\bm{x}$ and $\bm{y}$; $l(\cdot,\cdot)$, $c(\cdot,\cdot)$, and $s(\cdot,\cdot)$ represent the luminance comparison function, the contrast comparison function, and the structure comparison function, respectively; and $f(\cdot, \cdot, \cdot)$ is the similarity measure function whose arguments are the outputs of $l(\cdot,\cdot)$, $c(\cdot,\cdot)$, and $s(\cdot,\cdot)$. In light of (\ref{Similarity_measure_defn}) and the functions $l(\cdot,\cdot)$, $c(\cdot,\cdot)$, and $s(\cdot,\cdot)$ as defined in \cite[eq. (6)]{Wang_SSIM_Metric'04}, \cite[eq. (9)]{Wang_SSIM_Metric'04}, and \cite[eq. (10)]{Wang_SSIM_Metric'04}, respectively, the SSIM between $\bm{x}$ and $\bm{y}$ is defined as \cite[eq. (12)]{Wang_SSIM_Metric'04}
\begin{equation}
	\label{SSIM_defn}
	\textnormal{SSIM}(\bm{x},\bm{y}) \eqdef [l(\bm{x},\bm{y})]^{\alpha} [c(\bm{x},\bm{y})]^{\beta} [s(\bm{x},\bm{y})]^{\gamma}, 
\end{equation} 
where $\alpha, \beta, \gamma>0$ are parameters used to adjust the relative importance of the three functions' outputs \cite{Wang_SSIM_Metric'04}. In view of (\ref{SSIM_defn}), one may require a single overall quality measure -- for the entire image in question -- which can be captured by the metric \textit{mean SSIM} (MSSIM) that is defined via (\ref{SSIM_defn}) as \cite[eq. (17)]{Wang_SSIM_Metric'04}
\begin{equation}
	\label{MSSIM_defn}
	\textnormal{MSSIM}(\bm{X},\bm{Y}) \eqdef \frac{1}{M}   \sum_{j=1}^M \textnormal{SSIM}(\bm{x}_j,\bm{y}_j),
\end{equation}
where $\bm{X}$ and $\bm{Y}$ are the reference and distorted images, respectively; $M$ is the number of local windows of the image; and $\bm{x}_j$ and $\bm{y}_j$ are the images' content at the $j$-th local window \cite{Wang_SSIM_Metric'04}.

It is worth mentioning that SSIM is less effective when assessing blurred and noisy images \cite{Sem_Empowered_Commun'22}. To overcome this limitation, SSIM variants such as \textit{three-component weighted SSIM} (3-SSIM) \cite{Li2009_3-SSIM} and \textit{feature similarity index for image quality assessment} (FSIM) \cite{Zhang_FSIM'11} are proposed. 
\begin{figure*}[t!]
	\centering
	\includegraphics[scale=0.52]{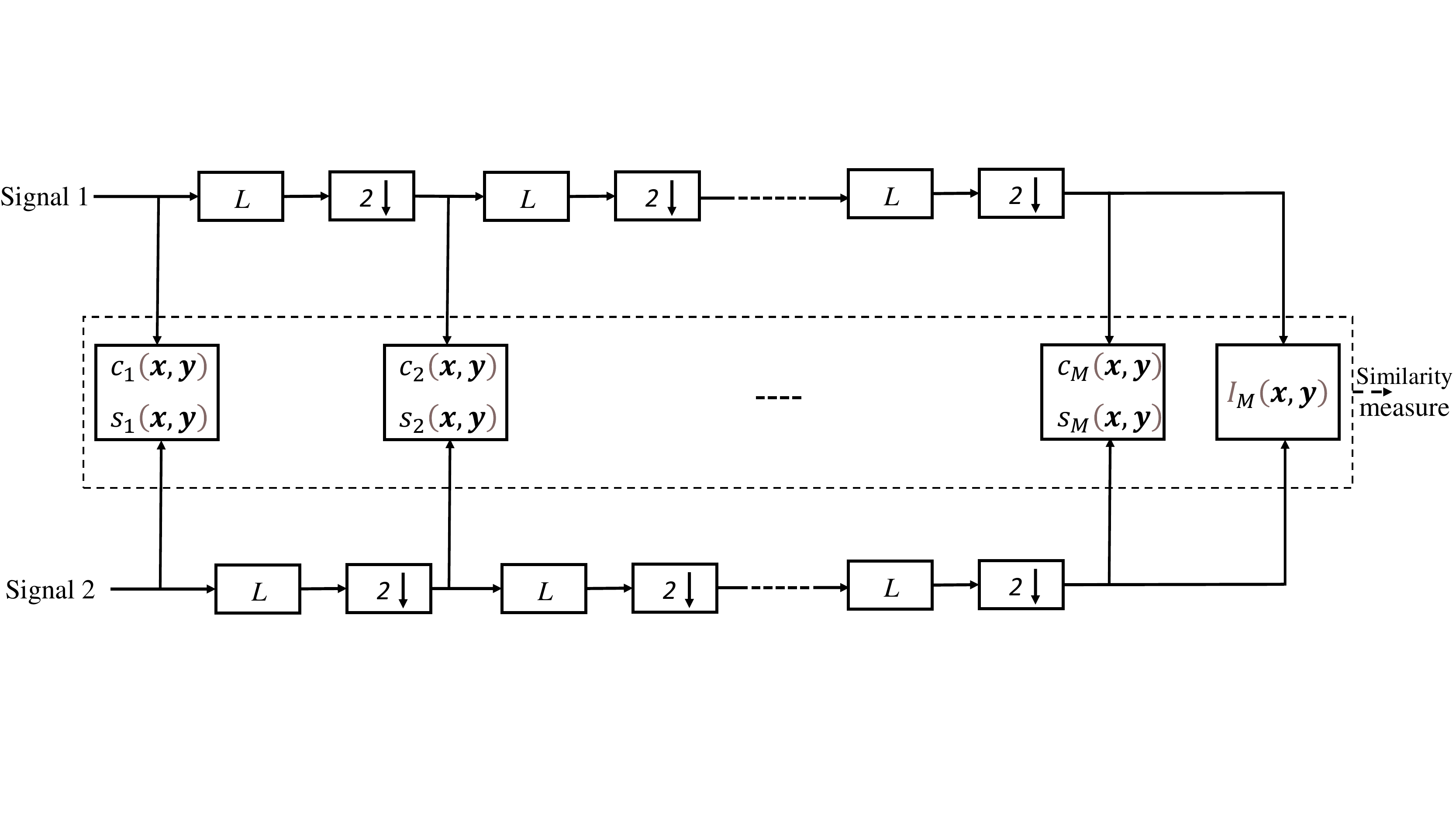}  \vspace{-2.0cm} 
	\caption{Basic diagram of the MS-SSIM system -- $L$: low-pass filtering; 2 $\downarrow$:  downsampling by 2 \cite[Fig. 1]{Wang_MS-SSIM'03}.}
	\label{fig:MS-SSIM_Fig.pdf}
\end{figure*}

We now continue with our discussion of another IQA metric that is used for designing, analyzing, and optimizing image SemCom systems -- named MS-SSIM.

\subsection{Multi-Scale Structural Similarity Index Measure}
\label{subsec: MS-SSIM}
Practically speaking, the subjective evaluation of an image varies when the following factors change: the distance from the image plane to the observer, the sampling density of the image signal, and the perceptual capability of the observer’s visual system \cite{Wang_MS-SSIM'03}. Multi-scale method is therefore convenient to incorporate the details of images captured at various resolutions \cite{Wang_MS-SSIM'03}. To this end, the authors of \cite{Wang_MS-SSIM'03} put forward the metric MS-SSIM for image quality assessment, whose system diagram is schematized in Fig. \ref{fig:MS-SSIM_Fig.pdf}. As is shown in Fig. \ref{fig:MS-SSIM_Fig.pdf}, the MS-SSIM system uses the reference and distorted image signals as the input, which are fed into the system that iteratively applies a low-pass filter and downsamples the filtered image by a factor of 2 \cite{Wang_MS-SSIM'03}. When the original image is indexed as scale 1 and the highest scale as scale $M$ (obtained after $M -1$ iterations), the MS-SSIM metric between signals $\bm{x}$ and $\bm{y}$ can be defined by combining the measurements taken at different scales as follows \cite[eq. (7)]{Wang_MS-SSIM'03}:
\begin{equation}
	\label{MS-SSIM_defn}
	\textnormal{MS-SSIM}(\bm{x},\bm{y}) \eqdef [l_M(\bm{x},\bm{y})]^{\alpha_M} \prod_{j=1}^M [c_j(\bm{x},\bm{y})]^{\beta_j} [s_j(\bm{x},\bm{y})]^{\gamma_j},
\end{equation}
where $c_j(\bm{x},\bm{y})$ and $s_j(\bm{x},\bm{y})$ are the contrast comparison and the structure comparison at the $j$-th scale, respectively; $l_M(\bm{x},\bm{y})$ denotes the luminance comparison, which is computed only at scale $M$; and the constants $\alpha_M$, $\beta_j$, and $\gamma_j$ are used to adjust the relative importance of the components mentioned \cite{Wang_MS-SSIM'03}. It is worth noting that the MS-SSIM definition in (\ref{MS-SSIM_defn}) encompasses SSIM as a special case.

In light of (MS-)SSIM, the LPIPS model \cite{Ding_Comparison'2021,Wang_Perceptual_Learned'22}, which we discuss below, is another crucial metric for image SemCom.

\subsection{Learned Perceptual Image Patch Similarity}
\label{subsec: LPIPS}
The authors of \cite{Zhang_The_Unreasonable_Effectiveness'18} introduce the metric LPIPS, whose key idea is to use \textit{deep features} to construct a loss function. This approach comprises two steps: calculating the distance from a given network -- (pre-trained) network $\mathcal{F}$ -- and then predicting perceptual judgment, to wind up with a loss function \cite[Figure 3]{Zhang_The_Unreasonable_Effectiveness'18}. The following are three possible LPIPS configurations -- namely \textit{lin}, \textit{tune}, and \textit{scratch} \cite{Zhang_The_Unreasonable_Effectiveness'18,Engman_Perceptual_Metric'20} -- depending on how the loss function was constructed:
\begin{itemize}
	\item In the lin configuration, the pre-trained network weights $\mathcal{F}$ are fixed, and the linear weights $w$ are learned on top.\footnote{In an existing feature space, this comprises the perceptual calibration of a few parameters \cite{Zhang_The_Unreasonable_Effectiveness'18}.}
	
	\item In the tune configuration, a pre-trained classification model is employed for initialization, and all the weights for network $\mathcal{F}$ are tweaked/fine-tuned.
	
	\item In the scratch configuration, a network is initialized from random normal weights and trained entirely using judgment from related studies \cite{Zhang_The_Unreasonable_Effectiveness'18}.
\end{itemize}

For the first step of LPIPS (i.e., distance calculation), the distance between a reference patch $x$ and a distorted patch $x_0$ is calculated using network $\mathcal{F}$ as follows \cite[eq. (1)]{Zhang_The_Unreasonable_Effectiveness'18}:
\begin{equation}
	\label{dist_x_x_0_defn}
	d(x,x_0) \eqdef \sum_{l} \frac{1}{H_lW_l} \sum_{h,w} \| \bm{w}_l \odot [(\hat{\bm{\mathcal{Y}}}_l)_{h,w}-(\hat{\bm{\mathcal{Y}}}_{l,0})_{h,w}]    \|_2^2, 
\end{equation}
where $H_l,W_l  \in\mathbb{N}$ are the spatial components of the $l$-th layer; $(\hat{\bm{\mathcal{Y}}}_l)_{h,w}$ and $(\hat{\bm{\mathcal{Y}}}_{l,0})_{h,w}$ are the comprising vectors of tensors $\hat{\bm{\mathcal{Y}}}_l\in\mathbb{R}^{H_l \times W_l \times C_l}$ and $\hat{\bm{\mathcal{Y}}}_{l,0}\in\mathbb{R}^{H_l \times W_l \times C_l}$, respectively, the latter of which are extracted deep feature embeddings from the $l$-th layer that have been unit-normalized in the channel dimension; and $\bm{w}_l$ is a scaling vector deployed for channel-wise \textit{activation} scaling \cite{Zhang_The_Unreasonable_Effectiveness'18}. Following the distance calculation per (\ref{dist_x_x_0_defn}), the second step of LPIPS is to predict perceptual judgment through a small network $\mathcal{G}$ that has been trained -- using cross-entropy (CE) loss -- to predict perceptual judgment $h$ from distance pair $(d_0, d_1)$ \cite{Zhang_The_Unreasonable_Effectiveness'18}. Consequently, the loss function is ultimately expressed as \cite[eq. (2.11)]{Engman_Perceptual_Metric'20}
\begin{multline}
	\label{loss_func_culmination}
	\mathcal{L}(x,x_0,x_1,h) = - h\log\mathcal{G}\big( d(x,x_0), d(x,x_1)\big)- (1-h)  \times\\
	\log \big( 1- \mathcal{G}\big( d(x,x_0), d(x,x_1)\big) \big),
\end{multline}
where $d_0$ and $d_1$ denote the distance between patches $\{x, x_0\}$ and $\{x, x_1\}$, respectively; and $h$ is the predicted perceptual judgment \cite{Zhang_The_Unreasonable_Effectiveness'18}. Furthermore, to try and cover as many properties as possible \cite{Engman_Perceptual_Metric'20}, the authors of \cite{Kaplanyan_DeepFovea'19} present a weighted version of LPIPS with two other loss functions (adversarial loss and optical flow loss for temporal dynamics).

When a system designer requires accurate semantic-level recovery, an image SemCom system can be designed/analyzed using the metric mIoU \cite{Pan_IM-SemCom'22}, which we discuss below.

\subsection{Mean Intersection over Union}
\label{subsec: mIoU}
The metric mIoU is defined as \cite[eq. (4)]{Pan_IM-SemCom'22}
\begin{equation}
	\label{mIoU_defn}
	\textnormal{mIoU} \eqdef \frac{1}{N_{cls}}  \sum_{i=1}^{N_{cls}}   \frac{P_i \bigcap G_i}{P_i\bigcup G_i},
\end{equation}
where $P_i$ represents the set of pixel regions predicted by the decoder for the $i$-th object category, $G_i$ stands for the actual set of pixel regions pertaining to the $i$-th object category, and $N_{cls}$ denotes the number of object categories (e.g., pedestrians, vehicles, and trucks) in the input image \cite{Pan_IM-SemCom'22}. For the definition in (\ref{mIoU_defn}), the higher the mIoU value, the better the image SemCom performance \cite{Pan_IM-SemCom'22}.

We now continue with our discussion of another IQA metric that is used for designing, analyzing, and optimizing image SemCom systems -- called ISS.

\subsection{Image-to-Graph Semantic Similarity}
\label{subsec: ISS}
ISS \cite{Zhang_Opt_in_Image_SemCom'23} is an important image SemCom metric for assessing the performance of cooperative image SemCom networks in which a set of servers cooperatively transmit images to a set of users using SemCom schemes (vis-à-vis the transmission of semantic information that captures the meaning of images). To formally define the metric ISS in the context of cooperative semantic communication networks, let us define the semantic information about an image $G_k$ extracted by a server $v$ and transmitted to a user $k$ as \cite[eq. (1)]{Zhang_Opt_in_Image_SemCom'23}  
\begin{equation}
	\label{Sem_Info_G_k}
	\bm{\varPsi}_{vk}   \eqdef  \big\{ \bm{\psi}_{vk}^1, \bm{\psi}_{vk}^2, \ldots, \bm{\psi}_{vk}^n, \ldots, \bm{\psi}_{vk}^{N_{vk}} \big\}, 
\end{equation}
where $N_{vk}$ is the number of semantic triples in image $G_k$; $\bm{\psi}_{vk}^n\eqdef \big( e^n_{vk,i}, l^n_{vk,ij}, e^n_{vk,j} \big)$ is a semantic triple given that $e^n_{vk,i}$ is the category of object $i$ in image $G_k$; and $l^n_{vk,ij}$ denotes the relationship between objects $e^n_{vk,i}$ and $e^n_{vk,j}$ \cite{Zhang_Opt_in_Image_SemCom'23}. Note that $l^n_{vk,ij} \neq l^n_{vk,ji}$ since $l^n_{vk,ij}$ is directional \cite{Zhang_Opt_in_Image_SemCom'23}.

Some semantic triplets in $\bm{\varPsi}_{vk}$ may contain irrelevant information. Thus, to enhance the efficiency of the SemCom model considered by the authors of \cite{Zhang_Opt_in_Image_SemCom'23}, each server $v$ transmits the semantic triples that incorporate a significant image meaning \cite{Zhang_Opt_in_Image_SemCom'23}. Thus, the partial semantic information that server $v$ transmits to a user $k$ can be equated to \cite[eq. (3)]{Zhang_Opt_in_Image_SemCom'23} 
\begin{equation}
	\label{partial_Sem_Info_G_k}
	\hat{\bm{\varPsi}}_{vk}   \eqdef  \big\{ \hat{\bm{\psi}}_{vk}^1, \hat{\bm{\psi}}_{vk}^2, \ldots, \hat{\bm{\psi}}_{vk}^n, \ldots, \hat{\bm{\psi}}_{vk}^{\hat{N}_{vk}} \big\}  \subset \bm{\varPsi}_{vk}, 
\end{equation}
where $\hat{N}_{vk}$ denotes the number of selected semantic triples in $\hat{\bm{\varPsi}}_{vk}$.

The authors of \cite{Zhang_Opt_in_Image_SemCom'23} employ ISS to evaluate the performance of cooperative image SemCom networks per the aforementioned scenario. Furthermore, whereas SSIM measures the differences in a set of pixels, ISS captures the correlation between the meaning of the image and that of its corresponding semantic information \cite{Zhang_Opt_in_Image_SemCom'23}. Meanwhile, the authors of \cite{Zhang_Opt_in_Image_SemCom'23} deploy a DNN-based encoder to \textit{vectorize} the original image $G_k$ and the semantic information $\hat{\bm{\varPsi}}_{vk}$ which are, respectively, defined as \cite{Zhang_Opt_in_Image_SemCom'23}
\begin{subequations}
	\begin{align}
		\label{C_G_k_defn}
		\bm{C}\big( G_k \big) & \eqdef  \Big\{  \bm{C}\big(\bm{\psi}_{vk}^1 \big),   \ldots,  \bm{C}\big(\bm{\psi}_{vk}^n  \big) , \ldots,  \bm{C}\big(  \bm{\psi}_{vk}^{N_{vk}} \big)  \Big\}   \\  
		\label{O_vk_defn}
		\bm{O}_{vk}   & \eqdef  \Big\{ \bm{C}\big( \hat{\bm{\psi}}_{vk}^1\big), \ldots, \bm{C}\big( \hat{\bm{\psi}}_{vk}^n \big), \ldots, \bm{C}\big( \hat{\bm{\psi}}_{vk}^{\hat{N}_{vk}} \big)  \Big\}, 
	\end{align}
\end{subequations}
where $\bm{C}\big( \cdot\big)$ represents a \textit{vectorization} function that forms the relationship between the image and the input semantic information by matching text-image pairs with similar meanings \cite{Zhang_Opt_in_Image_SemCom'23}.

ISS is defined as the cosine angle between an image vector and its corresponding normalized semantic triple vectors \cite{Zhang_Opt_in_Image_SemCom'23}. Accordingly, for the formulations in (\ref{Sem_Info_G_k})-(\ref{O_vk_defn}), the ISS of $\hat{\bm{\varPsi}}_{vk}$ that is transmitted from server $v$ to user $k$ is defined as \cite[eq. (6)]{Zhang_Opt_in_Image_SemCom'23}
\begin{multline}
	\label{ISS_metric_defn}
	E\big(\hat{\bm{\varPsi}}_{vk}, \bm{a}_{vk} \big)  \eqdef \sum_{q=1}^Q a_{vk}^q \times \\  \frac{ \big\| \sum_{n=1}^{\hat{N}_{vk}} \big| \overline{\bm{C}\big( \hat{\bm{\psi}}_{vk}^n \big)} \cdot \bm{C}\big( G_k \big)^T  \big| \overline{\bm{C}\big( \hat{\bm{\psi}}_{vk}^n \big)} \big\| }{ \| \bm{C}\big( G_k \big) \|},
\end{multline}
where $Q$ denotes the number of downlink orthogonal resource blocks (RBs), $\bm{a}_{vk} \eqdef  \big[ a_{vk}^1, \ldots, a_{vk}^Q \big]$ represents an RB allocation vector for user $k$ of server $v$ given that $a_{vk}^q \in \{0,1\}$ is the user-server connection index, and $\overline{\bm{O}_{vk}} \eqdef  \Big\{ \overline{\bm{C}\big( \hat{\bm{\psi}}_{vk}^1\big)}, \ldots, \overline{\bm{C}\big( \hat{\bm{\psi}}_{vk}^n \big)}, \ldots, \overline{\bm{C}\big( \hat{\bm{\psi}}_{vk}^{\hat{N}_{vk}} \big)}  \Big\}$ is the Gram-Schmidt orthogonalized version of $\bm{O}_{vk}$ per (\ref{O_vk_defn}). It is evident from (\ref{ISS_metric_defn}) that the value of ISS increases with the number of transmitted semantic triples, in line with the objectives of human cognition \cite{Zhang_Opt_in_Image_SemCom'23}.

We now proceed with a discussion on our last IQA metric that is used for designing, analyzing, and optimizing image SemCom systems -- called recognition accuracy.

\subsection{Recognition Accuracy}
\label{subsec: recognition_accuracy}
Recognition accuracy is a metric for assessing the quality of reconstructed images that is proposed by the authors of \cite{Lee_Joint_Transmission_Recognition_for_IoTs'19} for a joint transmission-recognition scheme for an image SemCom system also proposed by them.  

Other major IQA metrics are \textit{complex-wavelet SSIM} (CW-SSIM) \cite{Zhou_CW-SSIM'05}, \textit{fast SSIM} and \textit{fast MS-SSIM} \cite{Chen_Fast_SSIM'11}, \textit{information content weighted SSIM} (IW-SSIM) \cite{Wang_IW-SSIM'11}, \textit{information fidelity criterion} (IFC) \cite{Sheikh_IFC'05}, \textit{visual information fidelity} (VIF) \cite{Sheikh_VIF'06}, \textit{multi-scale geometric analysis-based IQA} (MGA-based IQA) \cite{Gao_MGA-based_IQA'09}, the \textit{detail loss metric} (DLM) \cite{Li_DLM'11}, \textit{multi-metric fusion} (MMF) \cite{Liu_MMF'13}, \textit{most apparent distortion} (MAD) \cite{Larson_MAD'10}, \textit{peak signal-to-noise ratio-human vision system modified} (PSNR-HVS-M) \cite{Ponomarenko_Ponomarenko'07}, the \textit{noise quality measure} (NQM) \cite{Damera-Venkata_NQM'00}, and \textit{visual signal-to-noise ratio} (VSNR) \cite{Chandler_VSNR'07}. These metrics are also crucial for the design, analysis, and optimization of image SemCom systems. The image SemCom metrics defined in Sections \ref{subsec: image_semantic_similarity} through \ref{subsec: recognition_accuracy} are often employed to evaluate the semantic similarity between the natural images transmitted and those received. Generative adversarial networks (GANs) \cite{GANs_NIPS2014,Creswell_GANs_18,Wang2019GenerativeAN}, on the other hand, are being exploited to produce natural-looking synthetic images whose similarity is also assessed in comparison with natural images. To this end, metrics such as \textit{adversarial loss} \cite{GANs_NIPS2014}, \textit{inception score} (IS) \cite{Tim_Improved_Training_for_GANs'16}, \textit{Fr\'echet inception distance} (FID) \cite{Heusel_FID_metric'17}, and \textit{kernel inception distance} (KID) \cite{MMD_GANs'18}\footnote{Because FID and KID aim to compare the distribution of generated images with the distribution of real images, they cannot fully utilize the spatial relationship between features \cite{Sem_Empowered_Commun'22}. On the other hand, it is worth underscoring the following assessment-related concepts: $1)$ a lower FID value is demonstrated to correlate well with higher-quality images; $2)$ a lower KID score indicates better sampling quality, as KID quantifies the \textit{maximum mean discrepancy} in a classifier's feature space \cite{Sem_Empowered_Commun'22}.} have been proposed to measure the similarity between natural and GAN-generated images. At last, the existing semantic metrics that are used for image quality assessment and applicable to both wireless image SemCom and optical image SemCom are summarized along with their pros and cons in Table \ref{table: image_SemCom_metrics}.
\begin{table*}
	\centering
	\begin{tabular}{| l | l | l | }
		\hline
		\textbf{Metrics} & \textbf{Pros} & \textbf{Cons}  \\ \hline
		PSNR & $1)$ PSNR is a simple and computationally inexpensive IQA    & $1)$ PSNR is a shallow function that fails to count the many     \\ 
		&  metric \cite{Liu_VQA_Survey'13}. $2)$ PSNR can roughly reflect the image   & nuances of human perception \cite{Qin_Sem_Com_Principles_Apps'22}. $2)$ PSNR usually correlates     \\ 
		& similarity \cite{Sem_Empowered_Commun'22}.   & poorly with subjective visual quality \cite{Liu_VQA_Survey'13,Wang_MSE_SPM'09}. $3)$ PSNR is not   \\ 
		&    & continually consistent with human perception \cite{Sem_Empowered_Commun'22}.     \\ \hline
		SSIM & $1)$ SSIM is an easy metric to implement \cite{Liu_VQA_Survey'13}.  & $1)$ SSIM is also a shallow function that fails to count the many \\       
		& $2)$ SSIM exhibits good correlation with subjective scores \cite{Liu_VQA_Survey'13}.   &  nuances of human perception \cite{Qin_Sem_Com_Principles_Apps'22}. $2)$ SSIM is sensitive to relative      \\  
		&$3)$ SSIM is more consistent -- compared with PSNR -- with    & translations, rotations, and scalings of image \cite{Liu_VQA_Survey'13,Wang_MSE_SPM'09}. $3)$ SSIM       \\  
		& human perception in IQA \cite{Sem_Empowered_Commun'22}.  & is less effective when it is employed to assess among blurred and      \\ 
		&     & noisy images \cite{Sem_Empowered_Commun'22}. $4)$ SSIM reflects a higher evaluation than the       \\ 
		&     & actual scale \cite{Sem_Empowered_Commun'22}.  \\ \hline
		MS-SSIM & $1)$ MS-SSIM is a convenient approach to incorporate image & MS-SSIM exhibits considerable computational complexity  \\ 
		& details at various resolutions \cite{Wang_MS-SSIM'03}. $2)$ MS-SSIM manifests & as $M$ gets large.       \\ 
		& better correlation with subjective scores than SSIM \cite{Liu_VQA_Survey'13}.     &   \\ \hline
		LPIPS & LPIPS is based on the feature maps of different DNN  & LPIPS has an inherent computational complexity which can also   \\ 
		& architectures that have sound effectiveness in accounting    & be aggravated by a significant training cost of a deep network.  \\ 
		& for human perception of image quality \cite{Wang_Perceptual_Learned'22}.    &   \\ \hline 
		Image   & This metric is a computationally easy semantic metric.  & This metric depends on the higher-order image structure, which  \\  
		semantic &   & is often context-dependent \cite{Qin_Sem_Com_Principles_Apps'22}.     \\  
		similarity&   &   \\ \hline 
		ISS  & In comparison with SSIM, the ISS metric can capture   & The ISS metric is computationally complex.  \\ 
		& the correlation of the meaning between the image and   &   \\  
		& its corresponding semantic information \cite{Zhang_Opt_in_Image_SemCom'23}.     &    \\ \hline 
		FID & FID manifests distinctive robustness to noise \cite{Sem_Empowered_Commun'22}. & Since FID aims to compare the distribution of generated images  \\ 
		&  & with the distribution of real images, it can not entirely utilize  \\ 
		&  & the spatial relationship between features \cite{Sem_Empowered_Commun'22}.   \\ \hline
		KID & KID exhibits peculiar robustness to noise \cite{Sem_Empowered_Commun'22}. & Because KID revolves around comparing the distribution of      \\ 
		&  & generated images with the distribution of real images, it can     \\ 
		&  & not completely utilize the spatial relationship between     \\ 
		& & features \cite{Sem_Empowered_Commun'22}. \\ \hline
	\end{tabular}  \\ [3mm]
	
	\caption{Main semantic metrics for IQA along with their pros and cons -- PSNR: peak signal-to-noise ratio; SSIM: structural similarity index measure; MS-SSIM: multi-scale structural similarity index measure; LPIPS: learned perceptual image patch similarity; ISS: image-to-graph semantic similarity; FID: Fr\'echet inception distance; KID: kernel inception distance.}
	\label{table: image_SemCom_metrics}
\end{table*}

We now proceed with our discussion on the state-of-the-art semantic metrics for video quality and 3D human sensing assessment. 

\section{Semantic Metrics for Video Quality and 3D Human Sensing Assessment}
\label{sec: metrics_for_video_and_3D_human_sensing}
Video quality assessment (VQA) metrics can be classified based on the availability of reference \cite{Liu_VQA_Survey'13}. When no reference signal is available to compare the distorted/test signal with, the VQA metric is termed a \textit{no-reference} (NR) metric \cite{Liu_VQA_Survey'13}. On the other hand, if information is available for part of the reference medium (for instance, a group of extracted features), the VQA metric is called a \textit{reduced-reference} (RR) metric \cite{Liu_VQA_Survey'13}. Contrary to RR metric, the \textit{full-reference} (FR) VQA metric requires the entire reference medium to assess the distorted/test medium \cite{Liu_VQA_Survey'13}. The FR VQA metric is expected to have the best video quality prediction performance since it has complete information about the original medium \cite{Liu_VQA_Survey'13}.

VQA metrics can also be categorized into five types based on their assessment methodology \cite{Liu_VQA_Survey'13}:
\begin{enumerate}
	\item \textit{Image/video fidelity metrics}: these metrics operate based only on the direct accumulation of errors and thus are often FR \cite{Liu_VQA_Survey'13}. Even though these VQA metrics are the simplest ones that are still widely used, they are usually not a good reflection of perceived visual quality when the distortion is not additive \cite{Liu_VQA_Survey'13}.
	
	\item \textit{Human visual system (HVS) model-based metrics}: these VQA metrics typically deploy frequency-based decomposition and take into consideration various aspects of the HVS, such as contrast and orientation sensitivity, spatial and temporal masking effects, frequency selectivity, and color perception \cite{Liu_VQA_Survey'13}. They can therefore can become very complex and computationally expensive \cite{Liu_VQA_Survey'13}.
	
	\item \textit{Signal structure (information or other feature)-based metrics}: some of these metrics quantify visual fidelity on the basis of the assumption that a high-quality image/video is one whose structural content -- such as object boundaries or regions of high entropy -- parallels that of the original image/video \cite{Liu_VQA_Survey'13}. The other metrics of this type are contingent on the assumption that the HVS understands an image mainly through its low-level features \cite{Liu_VQA_Survey'13}. Consequently, image deterioration can be perceived by comparing the low-level features of the reference and distorted images \cite{Liu_VQA_Survey'13}. 
	
	\item \textit{Packet analysis-based metrics}: these metrics center upon evaluating the impact of network impairments on visual quality \cite{Liu_VQA_Survey'13}. To do so and measure quality loss, they often exploit the parameters extracted from the transport stream \cite{Liu_VQA_Survey'13}. Meanwhile, these metrics have the advantage of being able to measure the quality of several image/video streams in parallel and are becoming more prevalent because of the increasing popularity of network-based video delivery service, such as Internet streaming \cite{Liu_VQA_Survey'13}.
	
	\item \textit{Learning-oriented metrics}: these metrics extract particular features from the image/video and then employ ML techniques to obtain a trained model \cite{Liu_VQA_Survey'13}. The perceived quality of images/videos is then predicted using the trained model \cite{Liu_VQA_Survey'13}.
\end{enumerate}

In light of the aforementioned classifications of VQA metrics, we discuss below \textit{traditional VQA metrics}, the \textit{motion-based video integrity evaluation} (MOVIE) index \cite{Seshadrinathan_MOVIE'10}, \textit{fusion-based video quality assessment} (FVQA) \cite{Lin_FVQA'14}, the \textit{video quality metric} (VQM) \cite{Pinson_VQM'04}, the \textit{video quality model for variable frame delay} (VQM$\_$VFD) \cite{Wolf_VQM_VFD_NTIA'11},  and \textit{video multi-method assessment fusion} (VMAF) \cite{Rassool_VMAF'17,NetFlix_VMAF_Blog'16}, which are applicable for designing wireless video SemCom systems \cite{Jiang_Wireless_SemCom'22,Wang_Wireless_Deep_Video_Transmission,Tung_DeepWiVe'21,Huang_IS-SemCom'22}. We begin with traditional VQA metrics.

\subsection{Traditional VQA Metrics}
\label{subsec: traditional_VQA_Metrics}
Two techniques have traditionally been employed in \textit{video codec} research and development to assess video quality: 1) subjective visual testing, and 2) calculating simple (and computationally inexpensive) objective metrics such as PSNR, or more recently, SSIM \cite{NetFlix_VMAF_Blog'16}. Subjective VQA metrics such as the \textit{mean opinion score} (MOS) can be useful for obtaining human judgment on video quality \cite{Rassool_VMAF'17}. However, many researchers' findings echo the fact that MOS is a time-consuming and (financially) expensive metric that requires specialized expertise \cite{Rassool_VMAF'17}. Besides, MOS cannot produce real-time quality ratings across a distribution network \cite{Rassool_VMAF'17}. Objective IQA metrics such as MSE, PSNR, and SSIM are usually utilized within codecs to optimize coding decisions and report the final quality of an encoded video \cite{NetFlix_VMAF_Blog'16}. In this respect, PSNR remains the \textit{de facto} standard for codec comparison and standardization, though researchers and practitioners are cognizant that it does not consistently reflect human perception \cite{NetFlix_VMAF_Blog'16}.

IQA metrics such as fast MS-SSIM \cite{Chen_Fast_SSIM'11} have been adapted for streamed video and commercialized as \textit{SSIMWAVE} \cite{Rassool_VMAF'17}. \textit{Video Quality Monitor}, on the other hand, is a commercial tool that measures video quality in relation to format and protocol specifications while requiring no reference \cite{Rassool_VMAF'17}. This NR VQA metric can objectively evaluate blockiness, blurriness, and frame rate \cite{Rassool_VMAF'17}. 

We now proceed with our brief discussion of the VQA metric dubbed the MOVIE index \cite{Pinson_VQM'04}.

\begin{figure*}[t!]
	\centering
	\includegraphics[scale=0.52]{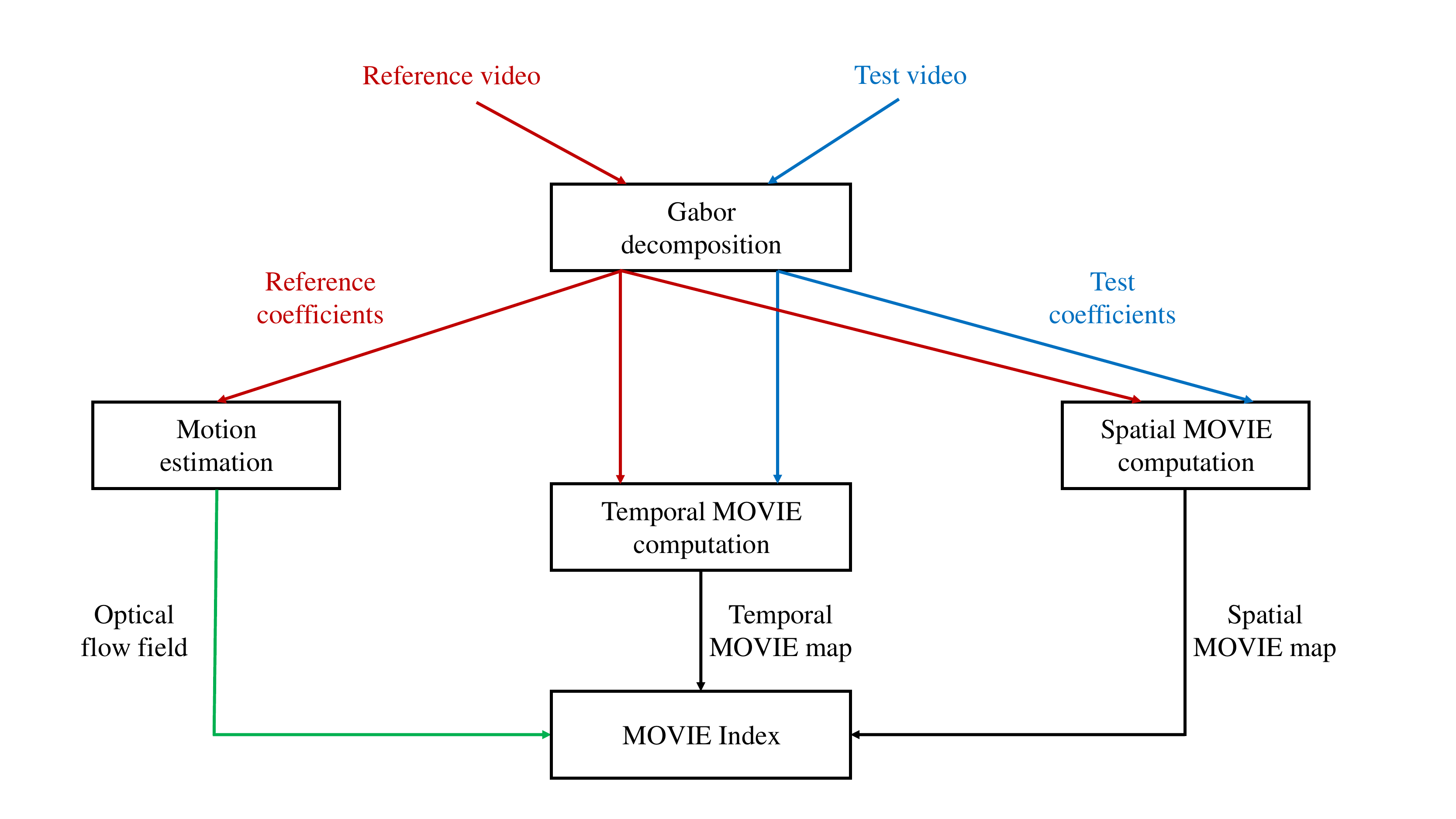}  \vspace{3mm} 
	\caption{Block diagram of the MOVIE index \cite[Fig. 1]{Seshadrinathan_MOVIE'10}.}
	\label{fig: MOVIE_Index_20230128.pdf}
\end{figure*}

\subsection{The Motion-Based Video Integrity Evaluation Index}
\label{subsec: MOVIE}
The MOVIE index is an FR, HVS model-based VQA metric that is proposed by the authors of \cite{Seshadrinathan_MOVIE'10} and uses optical flow estimation to adaptively guide spatial–temporal filtering by exploiting three-dimensional Gabor filter banks \cite{Seshadrinathan_MOVIE'10,Liu_VQA_Survey'13}. A subset of Gabor filters is selected adaptively at each location based on the direction and speed of motion \cite{Seshadrinathan_MOVIE'10,Liu_VQA_Survey'13}. To this end, the principal axis of the filter set is oriented in the frequency domain along the direction of motion \cite{Seshadrinathan_MOVIE'10,Liu_VQA_Survey'13}.

As can be seen in Fig. \ref{fig: MOVIE_Index_20230128.pdf}, the MOVIE index computation begins with the Gabor decomposition of the reference and test videos. These videos undergo linear decomposition using a Gabor filter family \cite{Seshadrinathan_MOVIE'10}. Following decomposition, three major computations are carried out -- as schematized in Fig. \ref{fig: MOVIE_Index_20230128.pdf} -- to compute the MOVIE index:
\begin{enumerate}
	\item \textit{Motion estimation} uses the output of the 3D Gabor decomposition of the reference video to determine its \textit{optical flow field}; see Fig. \ref{fig: MOVIE_Index_20230128.pdf}.  
	
	\item \textit{Temporal MOVIE computation} captures temporal degradation in the video following 3D Gabor decomposition by using motion information from the reference video and assesses the quality of the test video along the reference video's motion trajectories \cite{Seshadrinathan_MOVIE'10,Liu_VQA_Survey'13}. This produces the \textit{temporal MOVIE map}; see Fig. \ref{fig: MOVIE_Index_20230128.pdf}.
	
	\item \textit{Spatial MOVIE computation} employs the output of the multi-scale Gabor decomposition of the reference and test videos to gauge spatial distortions in the video and produce the \textit{spatial MOVIE map} \cite{Seshadrinathan_MOVIE'10,Liu_VQA_Survey'13}.
\end{enumerate}
The above-computed spatial MOVIE map, temporal MOVIE map, and optical flow field must be combined per Fig. \ref{fig: MOVIE_Index_20230128.pdf} to obtain the VQA metric MOVIE index \cite{Seshadrinathan_MOVIE'10,Liu_VQA_Survey'13}. 

We now detail another crucial VQA metric -- FVQA \cite{Lin_FVQA'14}.

\subsection{Fusion-Based Video Quality Assessment}
\label{subsec: FVQA}
FVQA is an FR, fusion-based, and learning-oriented VQA metric that is used to predict the visual quality of a streaming video \cite{Lin_FVQA'14}. To this end, FVQA is composed of two main steps \cite{Lin_FVQA'14} as depicted in Fig. \ref{fig: FVQA_method_20230128.pdf}:
\begin{itemize}
	\item First, video sequences are grouped based on their content complexity while minimizing content diversity within each group according to their content complexity \cite{Lin_FVQA'14}.
	
	\item Second, various existing VQA techniques -- particularly, FR VQA methods -- are applied to the reference and distorted videos, and their scores are fused to produce the final video quality score \cite{Lin_FVQA'14}. The corresponding fusion coefficients are learned from training video samples that belong to the same group as the videos assessed \cite{Lin_FVQA'14}.
\end{itemize}
Once the two steps have been completed, the FVQA index's performance is assessed by cross validation \cite{Lin_FVQA'14}.

Video grouping -- by classifying videos of similar content into a group -- makes it possible to build a more accurate quality prediction model within each group \cite{Lin_FVQA'14}. To this end and for the purpose of VQA, the authors of \cite{Lin_FVQA'14} opt for considering the spatial and the temporal information defined in the \textit{ITU-T Recommendation P.910} \cite{ITU-T_recommendation_P.910'99} to portray the spatio-temporal characteristics of source videos and apply them to grouped video content. This leads to there being two groups, named \textit{Group I} and \textit{Group II} \cite[Fig. 3]{Lin_FVQA'14}. In addition to this grouping, compression and resizing are two types of distortion that exist in the \textit{MCL-V} \cite{Lin_MCL-V'15} video quality database \cite{Lin_FVQA'14}. Consequently, there ends up being four groups, as shown in Fig. \ref{fig: FVQA_method_20230128.pdf} below. For each group, the authors of \cite{Lin_FVQA'14} consider the FVQA technique that fuses the scores of five VQA indices by employing a support vector machine (SVM) as a supervised learning algorithm to determine their weight coefficients. After that, a fused decision is obtained for each video group -- as can be seen in Fig. \ref{fig: FVQA_method_20230128.pdf}. 

Finally, the authors of \cite{Lin_FVQA'14} corroborate FVQA's superior performance compared to other VQA methods using the MCL-V \cite{Lin_MCL-V'15} (video) database. 

We now move on to our discussion of the nationally and internationally standardized VQA metric known as VQM.

\begin{figure*}[t!]
	\centering
	\includegraphics[scale=0.52]{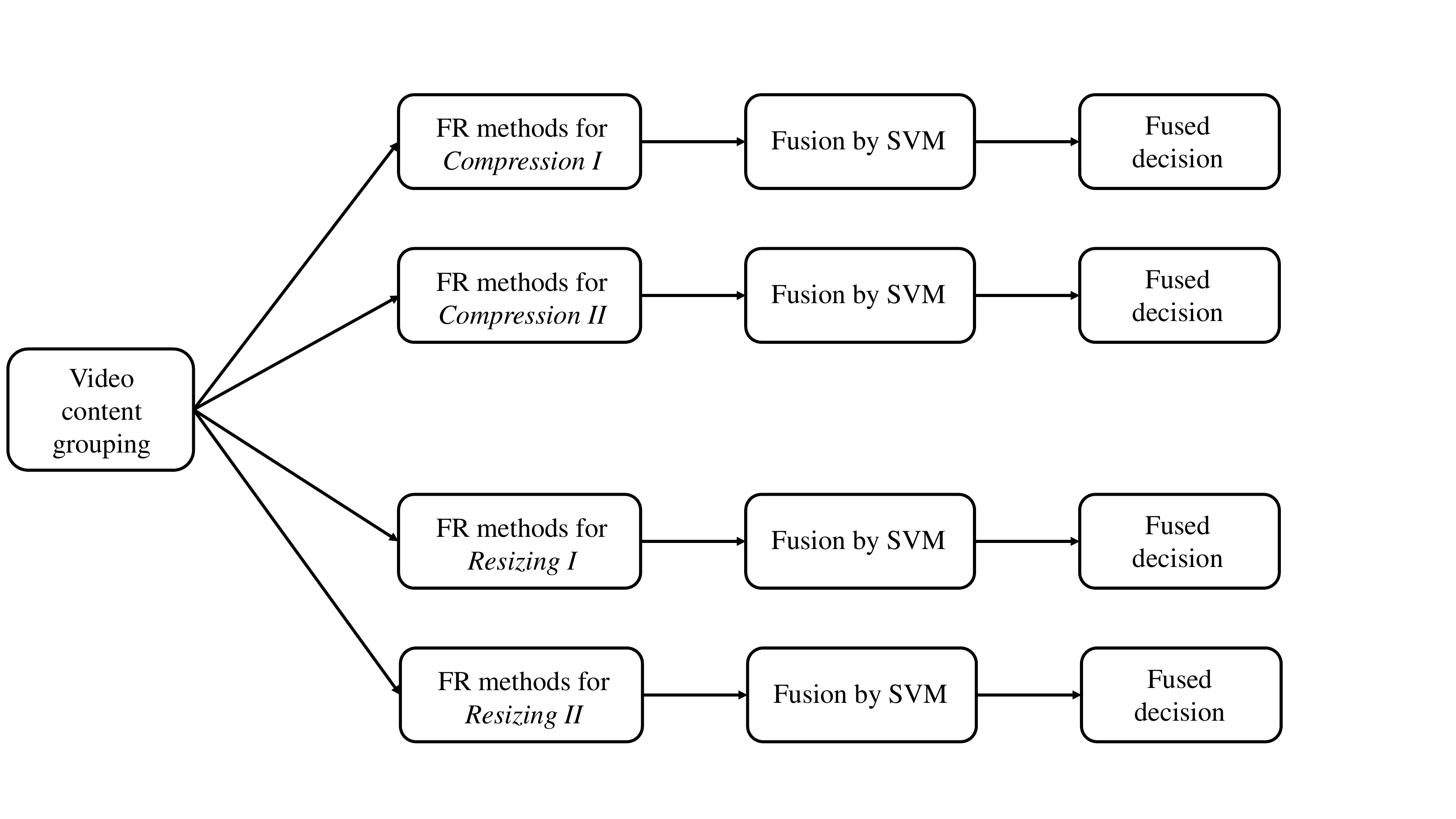}  \vspace{2mm} 
	\caption{Block diagram of the FVQA metric -- FR: full-reference; SVM: support vector machine \cite[Fig. 2]{Lin_FVQA'14}.}
	\label{fig: FVQA_method_20230128.pdf}
\end{figure*}

\subsection{Video Quality Metric}
\label{subsec: VQM}
VQM is an RR, HVS model-based VQA metric that was developed by the National Telecommunications and Information Administration (NTIA) to deliver an objective measurement of perceived video quality \cite{Pinson_VQM'04,Liu_VQA_Survey'13}. VQM supports the following quality models depending on the video sequence considered, with various calibration options available prior to feature extraction: $1)$ the television model, $2)$ the video conferencing model \cite{Liu_VQA_Survey'13}, and $3)$ the general model. As for the general model, it comprises seven independent parameters \cite{Pinson_VQM'04,Liu_VQA_Survey'13}: 
\begin{itemize}
	\item four of which (\texttt{si$\_$loss}, \texttt{hv$\_$loss}, \texttt{hv$\_$gain}, and \texttt{si$\_$gain}) reflect the features extracted from the spatial gradients of the $Y$ luminance component \cite{Pinson_VQM'04,Liu_VQA_Survey'13}; 
	
	\item two of which (\texttt{chroma$\_$spread}, \texttt{chroma$\_$extreme}) are based on the features extracted from the vector formed by the two chrominance components ($Cb$, $Cr$) \cite{Pinson_VQM'04,Liu_VQA_Survey'13}; and
	
	\item the last of which (\texttt{ct$\_$ati$\_$gain}) is contingent on the product of the features that evaluate contrast and motion, both of which are extracted from the $Y$ luminance component \cite{Pinson_VQM'04,Liu_VQA_Survey'13}.
\end{itemize}

VQM exploits the linear combination of the seven parameters and uses the original video and the processed video as inputs for computation \cite{Pinson_VQM'04,Liu_VQA_Survey'13}. VQM and its associated calibration techniques were adopted by the American National Standards Institute (ANSI) as a North American Standard in 2003 because it produced the best results for the Video Quality Experts Group (VQEG) Phase II full reference television (FR-TV) test \cite{Pinson_VQM'04}. VQM has also been considered a normative method in two draft recommendations made by the ITU \cite{Pinson_VQM'04}.

We now move on to discuss a VQA metric -- called VQM$\_$VFD -- that was developed by NTIA, inspired by VQM, and appropriately accounts for the perceptual impact of variable frame delay (VFD) \cite{Wolf_VQM_VFD_NTIA'11}. 

\subsection{Video Quality Model for Variable Frame Delay}
\label{subsec: VQM_VFD}
VQM$\_$VFD is an {RR, HVS model-based, and learning-oriented VQA metric \cite{Wolf_VQM_VFD_NTIA'11}. It also employs perceptual features distilled from the spatio-temporal (ST) blocks of a fixed angular extent \cite{Wolf_VQM_VFD_NTIA'11}. These features enable it to track subjective quality over a wide range of image sizes and viewing distances \cite{Wolf_VQM_VFD_NTIA'11}. To this end, VQM$\_$VFD relies on eight objective video quality parameters \cite{Wolf_VQM_VFD_NTIA'11}:
	\begin{itemize}
		\item The \texttt{HV$\_$Loss} parameter, which is derived from the one used in VQM, but incorporates four differences \cite{Wolf_VQM_VFD_NTIA'11}.
		
		\item The \texttt{HV$\_$Gain}, \texttt{SI$\_$Loss}, and \texttt{SI$\_$Gain} parameters, which are similar to the ones used in VQM \cite{Wolf_VQM_VFD_NTIA'11}.
		
		\item The \texttt{TI$\_$Gain} parameter, which is used to compute an ST block's root mean square (rms) motion energy or temporal information \cite{Wolf_VQM_VFD_NTIA'11}.
		
		\item The \texttt{RMSE$\_$Gain} parameter, which is calculated as the rms error between the VFD-matched original clip and the ST blocks in the processed clip \cite{Wolf_VQM_VFD_NTIA'11}.
		
		\item The \texttt{VFD$\_$Par1} parameter, which captures the perceptual impact of repeated or dropped frames and variable video delays \cite{Wolf_VQM_VFD_NTIA'11}. 
		
		\item The \texttt{VFD$\_$Par1$\cdot$PSNR$\_$VFD} parameter captures the perceptual attributes of both PSNR and VFD (computed as the product of \texttt{VFD$\_$Par1} and \texttt{PSNR$\_$VFD}\footnote{\texttt{PSNR$\_$VFD} is the PSNR calculated after the original clip has been VFD-matched to the processed clip \cite{Wolf_VQM_VFD_NTIA'11}.}) \cite{Wolf_VQM_VFD_NTIA'11}. 
		
	\end{itemize}
	
	The eight parameters are mapped to subjective quality estimates by training a two-layer neural network \cite{Wolf_VQM_VFD_NTIA'11}. The authors of \cite{Wolf_VQM_VFD_NTIA'11} corroborated that VQM$\_$VFD can attain 0.9 correlation to subjective quality using subjective datasets at image sizes ranging from Quarter Common Intermediate Format (QCIF) to High Definition TV (HDTV) by testing it on a trained two-layer neural network \cite{Wolf_VQM_VFD_NTIA'11}. VQM$\_$VFD is therefore recognized by many as the state-of-the-art in the field of VQA. 
	
	We note that the traditional as well as one or more other VQA metrics discussed above do not work well with Netflix content \cite{NetFlix_VMAF_Blog'16}. To address this limitation, Netflix researchers adopted an ML-based model to design the metric VMAF \cite{NetFlix_VMAF_Blog'16,Rassool_VMAF'17}, which seeks to reflect the HVS in terms of video quality \cite{NetFlix_VMAF_Blog'16}. VMAF is similar\footnote{The main difference between VQM$\_$VFD and VMAF is that the former extracts features at lower levels than the latter does, such as spatial and temporal gradients \cite{NetFlix_VMAF_Blog'16}.} to VQM$\_$VFD in spirit \cite{NetFlix_VMAF_Blog'16} and therefore discussed next.   
	
	\subsection{Video Multi-Method Assessment Fusion}
	\label{subsec: VMAF}
	VMAF is an FR, HVS model-based, and learning-oriented VQA metric that predicts subjective quality by merging multiple elementary quality metrics \cite{Rassool_VMAF'17,NetFlix_VMAF_Blog'16}. This metrics merging is inspired by the basic rationale that each constituent elementary metric can have its own strengths and weaknesses w.r.t. the source content's characteristics, type of artifacts, and degree of distortion \cite{NetFlix_VMAF_Blog'16}. VMAF can preserve all the strengths of the individual metrics and produce a more accurate final score by combining the elementary metrics using an ML algorithm \cite{NetFlix_VMAF_Blog'16}.  
	
	A huge sample of MOS scores are used as the ground truth to train a quality estimation model using an ML algorithm known as an SVM regressor in view of the fact that VMAF was formulated by Netflix to correlate strongly with subjective MOS scores \cite{Rassool_VMAF'17}. To this end, the current version of VMAF employs three image fidelity metrics -- DLM \cite{Li_DLM'11}, VIF \cite{Sheikh_VIF'06}, and anti-noise SNR (AN-SNR) -- and one temporal signal -- mean co-located pixel difference (MCPD) -- as elementary metrics that have been fused together by an SVM regression \cite{NetFlix_VMAF_Blog'16,Rassool_VMAF'17}. As for MCPD, the MCPD (the temporal component) of a frame w.r.t. the previous frame is a crucial parameter that is usually lacking metrics that compare only the reference image and the decoded image \cite{Rassool_VMAF'17}. 
	
	The authors of \cite{NetFlix_VMAF_Blog'16} thoroughly assess VMAF's performance and demonstrate -- using a Netflix dataset and three popular public datasets -- that VMAF outperforms a number of VQA metrics, including VQM$\_$VFD \cite{Wolf_VQM_VFD_NTIA'11}. In addition, the authors of \cite{Rassool_VMAF'17} corroborate the following results:
	\begin{itemize}
		\item There exists a strong correlation (0.948) between subjective MOS and the VMAF score computed \cite{Rassool_VMAF'17}.
		
		\item While VMAF is a robust predictor of a collective subjective opinion of video quality, the results reported in \cite{Rassool_VMAF'17} confirm that the value computed often (85\% of the time) overestimates the subjective quality.
		
		\item If a video service provider were to encode a video to attain a VMAF score of approximately 93, the service provider could then be confident that it is optimally serving the vast majority of its subscribers \cite{Rassool_VMAF'17}.\footnote{Content is indistinguishable from the original or includes noticeable (but not annoying) distortion \cite{Rassool_VMAF'17}.}
		
	\end{itemize}
	
	Apart from the afore-discussed VQA metrics, there exist a number of other VQA metrics: \textit{Speed-SSIM} (an FR, signal structure-based metric) \cite{Wang_Speed-SSIM'07}, \textit{digital video quality} (DVQ) (an FR, HVS model-based metric) \cite{Watson_DVQ'01}, \textit{continuous video quality} (CVQ) (an NR, learning-oriented metric) \cite{Yoshikazu_CVQ'08}, \textit{TetraVQM} (an FR, HVS model-based metric) \cite{Barkowsky_TetraVQM'09}, \textit{V-Factor} (an NR, packet-analysis-based metric) \cite{Winkler_V-Factor'08}, \textit{spatial–temporal assessment of quality} (STAQ) (an RR, HVS model-based metric) \cite{Amirshahi_STAQ'11}, and \textit{spatiotemporal MAD} (ST-MAD) (an FR, HVS model-based metric) \cite{Vu_ST-MAD'11}. These VQA metrics are also useful for the design, analysis, and optimization of 6G systems based on video SemCom. 
	
	We now move on to our brief discussion of a 3D human sensing metric dubbed mean per joint position error (MPJPE).
	
	\subsection{Mean Per Joint Position Error}
	\label{subsec: MPJPE}
	Since MPJPE is a semantic metric that is applicable to 3D human pose estimation, it is defined for a frame $f$ and a skeleton $\mathcal{S}$ as \cite[eq. (8)]{Ionescu_Human3.6M'14}
	\begin{equation}
		\label{MPJPE_defn}
		E_{MPJPE}(f, \mathcal{S})  \eqdef \frac{1}{N_{\mathcal{S}}} \sum_{i=1}^{N_{\mathcal{S}}} \| m_{\textnormal{\textbf{f}}, \mathcal{S}}^{(f)}(i) - m_{\textnormal{\textbf{gt}}, \mathcal{S}}^{(f)}(i)   \|_2,
	\end{equation}
	where $N_{\mathcal{S}}$ denotes the number of joints in skeleton $\mathcal{S}$; $\textnormal{\textbf{f}}$ is the pose estimator; $\textnormal{\textbf{gt}}$ denotes the ground truth; $m_{\textnormal{\textbf{f}}, \mathcal{S}}^{(f)}(i)$ is a function that returns the coordinates of the $i$-th joint of skeleton $\mathcal{S}$ at frame $f$ from  $\textnormal{\textbf{f}}$; and $m_{\textnormal{\textbf{gt}}, \mathcal{S}}^{(f)}(i)$ is the $i$-th joint of the ground truth frame $f$. MPJPE is employed as an evaluation metric in goal-oriented SemCom and a sensing technique proposed by the authors of \cite{Zhang_Semantic_Sensing_and_Commun'22} for a 3D human mesh construction task. The authors of \cite{Ionescu_Human3.6M'14} also propose the metrics \textit{mean per joint angle error} (MPJAE) \cite[eq. (9)]{Ionescu_Human3.6M'14} and \textit{mean per joint localization error} (MPJLE) \cite[eq. (10)]{Ionescu_Human3.6M'14} to assess the quality of 3D human sensing.
	
	The design, analysis, and optimization of video SemCom systems can be guided by not only the video SemCom metrics that are presented and discussed in Sections \ref{subsec: traditional_VQA_Metrics} through \ref{subsec: MPJPE} above, but also age of information- and value of information-based semantic metrics, which we discuss next.

\section{Age of Information- and Value of Information-Based Semantic Metrics}
\label{sec: AoI_and_VoI_based_semantic_metrics}
Semantic metrics that are based on the age of information (AoI) and the value of information (VoI) -- also known as \textit{effectiveness-level metrics} \cite{Sem_Empowered_Commun'22} -- can be used in the design, analysis, and optimization of many classical SemCom systems (in both the wireless and optical domains) \cite{SemCom_for_6G_Future_Internet'22}. Classical SemCom systems can involve text SemCom \cite{SemCom_Game'18,Farsad_DL_JSCC'18,Xie_DL-based_SemCom'21,Xie_Lite_distributed_SemCom'21,Zhou_WiCom_Letters'22,Peng_Robust_DL-Based_SemCom'22,Yao_Semantic_Coding'22,Lu_RL-powered_SemCom'21,Rethinking_modern_com_Lu_2022,Luo_SemCom_with_relay'21,Jiang_Deep_Source-Channel_coding'22,Liu_Context-Based_SemCom'22}; audio SemCom \cite{Weng_SemCom_Sys_Speech_Trans'21,SemCom_for_speech_signals'20,Weng_SemCom_Speech_Recognition'21,Han_Semantic-aware_Speech2Text_Transmission'22,Weng_DL-enabled_SemCom'22,Tong_FL_ASC'21}; image SemCom \cite{Eirina_JSCC'19,Kurka_Deep_JSCC-f'20,Bandwidth_Agile_Image_Transmission'21,Zhang_Wireless_Information_Transmission_of_Image'22,Xu_Wireless_Image_Transmission'22,Pan_IM-SemCom'22,Yang_WITT'22,Dai_NLT_SCC'22,Lee_Joint_Transmission_Recognition_for_IoTs'19,Hu_Robust_SemCom'22,Huang_Toward_SemCom'23}; video SemCom \cite{Jiang_Wireless_SemCom'22,Wang_Wireless_Deep_Video_Transmission,Tung_DeepWiVe'21,Huang_IS-SemCom'22}; multimodal SemCom \cite{WANG_Multimodal_SemCom'23}; or cross-modal SemCom \cite{Li_Cross-Modal_SemCom'22}. AoI- and VoI-based semantic metrics also have several applications in goal-oriented SemCom \cite{Sem_Empowered_Commun'22}. Consequently, we discuss below \textit{AoI-based semantic metrics}, \textit{VoI-based semantic metrics}, and \textit{combined semantic metrics}. We begin with AoI-based semantic metrics.

\subsection{AoI-Based Semantic Metrics}
\label{subsec: AoI-Based_semantic_metrics}
Let us start by formally defining AoI\footnote{In the literature, the phrases \textit{age of information} (AoI), \textit{status age}, or \textit{plain age} are used interchangeably \cite{AoI_Foundations_and_Trends'17}.} \cite{AoI_Survey_Introduction'21,Uysal_AoI_in_Practice'21,Maatouk_AoII'20,AoI_Foundations_and_Trends'17,Maatouk_Opt_Wireless_Networks'20,Optimizing_AoI'20,Kaul_Rreal_time_status'12,Minimizing_AoI_TMC'21,Perspectives_on_time'22,Sun_AoI_Book'19}.
\begin{definition}[{AoI \cite[Definition 2.1.1]{AoI_Foundations_and_Trends'17}}]
	\label{AoI_Definition}
	Consider a system involving a communication duo of a source and a destination. Regarding this duo of source and destination, suppose $t'_k$ be the times at which the status updates are received at the destination. The index of the most recently received update at time $\xi$ is given by \cite[eq. (2.1)]{AoI_Foundations_and_Trends'17}
	\begin{equation}
		\label{index-defn}
		N(\xi) \eqdef \textnormal{max}\big\{ k|t'_k \leq \xi \big\}. 
	\end{equation}
	Using (\ref{index-defn}), the timestamp of the most recently received update is defined as \cite[eq. (2.2)]{AoI_Foundations_and_Trends'17}
	\begin{equation}
		\label{time_stamp_recent_update_defn}
		u(\xi) \eqdef t_{N(\xi)}.
	\end{equation}
	Employing (\ref{time_stamp_recent_update_defn}) and (\ref{index-defn}), the AoI of the source $s$ at the destination $d$ is a random process defined as \cite[eq. (2.3)]{AoI_Foundations_and_Trends'17}
	\begin{equation}
		\label{AoI_eqn_defn}
		\Delta(t) \eqdef t-u(t).
	\end{equation}
\end{definition}
In light of (\ref{AoI_eqn_defn}), the age of the most recently received packet is defined as the difference between the current time and the timestamp of the packet \cite{Kaul_Rreal_time_status'12}. Using (\ref{AoI_eqn_defn}), meanwhile, the \textit{time average AoI} (\textit{time average age of a status update}) \cite{Kaul_Rreal_time_status'12,AoI_Foundations_and_Trends'17} is defined below.
\begin{definition}[{Time average AoI \cite[Definition 2.1.2]{AoI_Foundations_and_Trends'17}}]
	\label{Average_AoI_Definition}
	For an interval of observation $(0, T)$ and $\Delta(t)$ being the AoI per Definition \ref{AoI_Definition}, the time average age of a status update system is expressed as \cite[eq. (2.7)]{AoI_Foundations_and_Trends'17}, \cite[eq. (1)]{Kaul_Rreal_time_status'12}
	\begin{equation}
		\label{Average_AoI_Defn}
		\Delta_T \eqdef \frac{1}{T} \int_{0}^T \Delta(t) dt. 
	\end{equation}
\end{definition}
Note that the integral in (\ref{Average_AoI_Defn}) amounts to the area under $\Delta(t)$. Inspired by AoI and time average AoI, the authors of \cite{Costa_PAoI_paper'14} and \cite{Costa_PAoI_journal'16} propose the metric \textit{peak age of information} (PAoI), which is defined below.
\begin{definition}[{PAoI \cite[Definition 3]{Costa_PAoI_journal'16}}]
	\label{PAoI_defn}
	Let the RVs $T_{k-1}$ and $Y_k$ be the time in the system for the previously transmitted packet and the interdeparture time (or the time elapsed between service completion of the $(k-1)$-th packet and service completion of the $k$-th packet), respectively. The value of age attained immediately before receiving the $k$-th update is termed peak AoI (\textit{peak age}) and defined as \cite[eq. (10)]{Costa_PAoI_journal'16}
	\begin{equation}
		\label{PAoI_defn_eqn}
		A_k \eqdef T_{k-1} + Y_k.
	\end{equation}
\end{definition}
Regarding its advantage of a simpler formulation, PAoI can be used instead of AoI \cite{AoI_Foundations_and_Trends'17}. To this end, PAoI can be employed in applications in which there is interest in knowing/inferring the worst-case age or a need to apply a threshold restriction on age \cite{AoI_Foundations_and_Trends'17}.

AoI and the aforementioned AoI-related\footnote{Another AoI-related metric that also has many applications in status update systems is the metric \textit{relative age of information} (rAoI) \cite{Zou_rAoI'19}. The rAoI metric is defined as the AoI observed at the receiver \textit{relative to} the AoI at the transmitter \cite{Zou_rAoI'19}. Thus, rAoI is also an important semantic metric.} metrics have numerous applications in monitoring systems -- where only the most recent state generated by the source is of interest to the destination -- such as vehicular monitoring systems, industrial sensor networks, unmanned aerial vehicle path planning, and surveillance videos \cite{AoI_Survey_Introduction'21,SemCom_for_6G_Future_Internet'22}. AoI also has applications in caching and data analytics \cite{AoI_Foundations_and_Trends'17}, remote estimation, multi-server scenarios, and multi-hop networks \cite{Maatouk_Opt_Wireless_Networks'20}. Despite such broad applicability, AoI has inherent limitations due to the fact that its definition does not consider the current VoI process and its estimate at the monitor \cite{Maatouk_AoII'20,Maatouk_Opt_Wireless_Networks'20,AoII_metric_journal'20}. Consequently, age-optimal sampling policies have been found to be sub-optimal\footnote{Should the service times follow a heavy-tail distribution, age-optimal sampling, periodic sampling, and zero-wait sampling policies are hardly optimal \cite{Sun_Sampling_Wiener_Process'20}.} in several remote estimation applications \cite{Maatouk_AoII'20,Sun_Sampling_Wiener_Process'20}. Meanwhile, AoI and AoI-related metrics ignore the validity of the recovered data, and in some cases the monitor is concerned with only abnormal and abrupt states at a source \cite{Sun_Sampling_Wiener_Process'20,SemCom_for_6G_Future_Internet'22}. Furthermore, because AoI does not consider the value of current states, some pointless updates are transmitted to the monitor, which results in resources being wasted \cite{SemCom_for_6G_Future_Internet'22}. This justifies the need for VoI-based semantic metrics. 

\subsection{VoI-Based Semantic Metrics}
\label{subsec: VoI-Based_semantic_metrics}
Before VoI was introduced to communication systems -- especially networked control systems \cite{SemCom_Net_Systems'21,Soleymani_dissertation_19} -- as a new metric \cite{SemCom_Net_Systems'21}, the concept of VoI was well-known in the information analysis community, which defined it as the price a decision maker is willing to pay to take the information into account \cite{Howard_Info_Value_Theory'1966,SemCom_for_6G_Future_Internet'22}. In the context of conventional communications, on the other hand, VoI can be viewed as a measure of uncertainty reduction from the source's information set with successful transmission \cite{SemCom_for_6G_Future_Internet'22,Ayan_AoI_vs_VoI'19}. However, when it comes to communications with specific tasks, VoI needs to be redefined yet again and is employed to assess the relevance of a piece of information to a given communication task -- whereas AoI and AoI-related metrics focus on freshness and ignore content \cite{SemCom_for_6G_Future_Internet'22}. To underscore this VoI-guided design strategy, in the context of a remote temperature control system as discussed in \cite{Maatouk_AoII'20}, the overarching design goal is to guarantee that the controller reacts promptly to any abnormal increase in temperature compared to the real-time temperature variation of the sources \cite{SemCom_for_6G_Future_Internet'22}. Accordingly, the data concerning any abnormal temperature increase should be assigned high VoI \cite{SemCom_for_6G_Future_Internet'22}. In this respect, the image classification task studied in \cite{Yang_SemCom_with_AI_Tasks'21} is assessed by the (metric) VoI pertaining to the importance of the extracted features for the accurate classification of the images. Furthermore, VoI can be employed as part of a prioritizing scheduler \cite{Molin_Estimators_Based_on_VoI'19,Sem_Empowered_Commun'22}. 

Despite being a crucial semantic metric for both SemCom and goal-oriented SemCom, the definition of VoI is largely task-dependent, which makes the derivation of an explicit function for VoI challenging \cite{SemCom_for_6G_Future_Internet'22}. In this vein, deriving a definite function of VoI is a cumbersome task for complex systems and the research that has been diving into VoI-based metrics is relatively insufficient, to date \cite{SemCom_for_6G_Future_Internet'22}. Moreover, even though the value of data is usually decided by not only the content but also the communication context, the state-of-the-art VoI calculations do not take into consideration the factors mentioned \cite{SemCom_for_6G_Future_Internet'22}. On the other hand, factors such as context and content have inspired the following combined semantic metrics.

\subsection{Combined Semantic Metrics}
\label{subsec: combined_semantic_metrics}
The discussed error-based, AoI-based, and VoI-based semantics metrics focus merely on one attribute of the information conveyed by the recovered data \cite{SemCom_for_6G_Future_Internet'22}. To address this limitation, combined metrics -- applicable as combined SemCom metrics -- have been proposed. The authors of \cite{Optimizing_AoI'20} integrate VoI into AoI-based metrics and propose the \textit{age of information at query} (QAoI); and the authors of \cite{Maatouk_AoII'20} and \cite{AoII_metric_journal'20} integrate AoI into error-based metrics and put forward \textit{age of incorrect information} (AoII).

The definition of AoI and the afore-discussed AoI-related metrics implicitly assume that new information is used at any time \cite{Optimizing_AoI'20}. Nonetheless, the instants -- at which information is collected and used -- are not always contingent on a certain query process \cite{Optimizing_AoI'20}. To address this issue w.r.t. the fact that discrete-time systems involve queries wherein the monitoring process samples available information \cite{Optimizing_AoI'20}, the authors of \cite{Optimizing_AoI'20} put forward a model that accounts for the discrete-time nature of many monitoring processes and formally define QAoI as follows.    
\begin{definition}[{QAoI \cite{Optimizing_AoI'20}}]
	\label{QAoI_definition}
	Consider a time-slotted system indexed by $t = 1, 2, \ldots,$ and let $t_{q,1}, t_{q,2}, \ldots,$ be the query arrival times at the edge node. For this setting, the long-term expected QAoI is defined as \cite[eq. (3)]{Optimizing_AoI'20}
	\begin{equation}
		\label{long_term_QAoI_eqn}
		\tau_{\infty} \eqdef \lim_{t \to \infty} \mathbb{E} \Big\{ \sum_{i: t_{q,i} \leq t}  \Delta(t_{q,i}) \Big\}, 
	\end{equation}
	where $\Delta(t)$ is the AoI per Definition \ref{AoI_Definition}. 
\end{definition}
QAoI generalizes AoI by sampling $\Delta(t)$ per an arbitrary querying process while considering only the instants at which a query arrives \cite{Optimizing_AoI'20}. Accordingly, the QAoI-based scheme is likely to produce fresh updates when a query arrives, though its average AoI can be worse than that of an AoI-based scheme \cite{Optimizing_AoI'20}. On the other hand, age-optimal sampling policies have been found to be sub-optimal -- as noted above -- in various remote estimation applications \cite{Maatouk_AoII'20,Sun_Sampling_Wiener_Process'20}. For remote estimation applications in the context of SemCom and goal-oriented SemCom, the authors of \cite{Maatouk_AoII'20} and \cite{AoII_metric_journal'20} put forward the AoII metric, which is defined below.  
\begin{definition}[{AoII \cite{Maatouk_AoII'20,AoII_metric_journal'20}}]
	\label{AoII_metric_defn}
	Consider a basic transmitter-receiver system subjected to a process that can possibly change at any time instant $t$. Let a process $X_t$ be observed by the transmitter at time $t$ and $\hat{X}_t$ be the estimate created by the monitor (receiver). For this setup, the AoII metric is defined as \cite{Maatouk_AoII'20}
	\begin{equation}
		\label{AoII_metric_defn_equation}
		\Delta_{\textnormal{AoII}}(X_t, \hat{X}_t, t) \eqdef f(t) \times g(X_t, \hat{X}_t),  
	\end{equation}
	where $f : [0, \infty) \mapsto [0, \infty)$ is a non-decreasing function whose role is penalizing the system increasingly the more prolonged the mismatch between $X_t$ and $\hat{X}_t$ is and $g: \mathcal{D} \times \mathcal{D} \mapsto [0, \infty)$ -- for $\mathcal{D}$ being the state space of $X_t$ -- is a function that mirrors the gap between $X_t$ and $\hat{X}_t$.
\end{definition}
As for the function $g$ on the right-hand side (RHS) of (\ref{AoII_metric_defn_equation}), one can adopt the standard error-based metrics such as the indicator error (ind) function, the squared error (sq) function, and the threshold error (threshold) function, which are defined in (\ref{ind_function}), (\ref{sq_function}), and (\ref{threshold_function}), respectively, as follows \cite[eqs. (5)-(7)]{Maatouk_AoII'20}:
\begin{subequations}
	\begin{align}
		\label{ind_function}
		g_{\textnormal{ind}}(X_t, \hat{X}_t) & \eqdef \mathbb{I}\{X_t \neq \hat{X}_t\}      \\
		\label{sq_function}
		g_{\textnormal{sq}}(X_t, \hat{X}_t) & \eqdef (X_t - \hat{X}_t)^2, \hspace{2mm} \textnormal{and}    \\
		\label{threshold_function}
		g_{\textnormal{threshold}}(X_t, \hat{X}_t)  & \eqdef \mathbb{I}\{ |X_t - \hat{X}_t|\geq c\},
	\end{align}
\end{subequations}
where $c\in\mathbb{R}^+$ stands for a predefined threshold. Per (\ref{ind_function})-(\ref{threshold_function}), the functions $g_{\textnormal{ind}}$, $g_{\textnormal{sq}}$, and $g_{\textnormal{threshold}}$ are chosen when any mismatch between $X_t$ and $\hat{X}_t$ (regardless of its value) harms the system’s performance, when the system’s performance is impacted more significantly the larger the gap between $X_t$ and $\hat{X}_t$ is, and when the system’s performance is unsusceptible to small mismatches between $X_t$ and $\hat{X}_t$, respectively \cite{Maatouk_AoII'20}. Meanwhile, the function $f$ on the RHS of (\ref{AoII_metric_defn_equation}) can take the form of the linear time-dissatisfaction (linear) function, the  degree $m$ monomial (monomial) function, and the  time-threshold dissatisfaction (threshold) function. These functions are defined in (\ref{f_linear_defn}), (\ref{f_monomial_defn}), and (\ref{f_threshold_defn}), respectively, w.r.t. $V_t$ -- the last time instant whereupon $g(X_t, \hat{X}_t)$ was equal to 0 (or the last time instant wherein the monitor had sufficiently accurate information regarding $X_t$) \cite{Maatouk_AoII'20} -- as follows \cite[eqs. (8)-(10)]{Maatouk_AoII'20}:
\begin{subequations}
	\begin{align}
		\label{f_linear_defn}
		f_{\textnormal{linear}}(t)  & \eqdef t-V_t     \\
		\label{f_monomial_defn}
		f_{\textnormal{monomial}} (t) & \eqdef (t-V_t)^m , \hspace{2mm} \textnormal{and}  \\
		\label{f_threshold_defn}
		f_{\textnormal{threshold}} (t)  & \eqdef \mathbb{I}\{ t-V_t \geq c \}, 
	\end{align} 
\end{subequations}
where $m > 1$ is a positive integer and $c\in\mathbb{R}^+$ is a fixed threshold. According to (\ref{f_linear_defn})-(\ref{f_threshold_defn}), the functions $f_{\textnormal{linear}}$, $f_{\textnormal{monomial}}$, and $f_{\textnormal{threshold}}$ are chosen when the system's impact due to the mismatch between $X_t$ and $\hat{X}_t$ grows steadily with time, when the system’s performance deteriorates quickly as a result of the mismatch between $X_t$ and $\hat{X}_t$, and when the system’s performance is resistant to the mismatch between $X_t$ and $\hat{X}_t$ for a certain time duration $c$, respectively \cite{Maatouk_AoII'20}.  

At last, the existing AoI- and VoI-based metrics that are applicable to both SemCom and goal-oriented SemCom are summarized along with their pros and cons in Table \ref{table: AoI-VoI-Based_SemCom_metrics}. 

\begin{table*}
	\centering
	\begin{tabular}{| l | l | l | }
		\hline
		\textbf{Metrics} & \textbf{Pros} & \textbf{Cons}  \\ \hline
		AoI-based   & $1)$ AoI can reflect the freshness of information \cite{Sem_Empowered_Commun'22}. & $1)$ AoI may misjudge the value of information \cite{Sem_Empowered_Commun'22}.  \\ 
		semantic metrics & $2)$ PAoI can reflect the freshness of information \cite{Sem_Empowered_Commun'22}.    & $2)$ PAoI may misconstrue the value of information \cite{Sem_Empowered_Commun'22}.    \\   
		& $3)$ With the advantage of a simpler formulation, PAoI can be  & $3)$ The definition of AoI does not consider the current    \\ 
		& used instead of AoI \cite{AoI_Foundations_and_Trends'17}.   & value of the information process and its estimate at   \\ 
		& $4)$ PAoI can be employed in applications where there is interest    & the monitor \cite{Maatouk_AoII'20,Maatouk_Opt_Wireless_Networks'20,AoII_metric_journal'20}.   \\ 
		& in the worst case age or the need to apply a threshold restriction    & $4)$ AoI and AoI-related metrics ignore the validity of    \\   
		& on age \cite{AoI_Foundations_and_Trends'17}.   & the recovered data and, in some cases, the monitor is     \\   
		&    & only concerned with the abnormal and abrupt states  \\ 
		&   &   at the source \cite{Sun_Sampling_Wiener_Process'20,SemCom_for_6G_Future_Internet'22}.  \\  
		&   & $5)$ Since AoI does not consider the value of current     \\
		&   & states, some pointless updates are transmitted to the   \\
		&   & monitor -- yielding resource wastage \cite{SemCom_for_6G_Future_Internet'22}.    \\
		&   &    \\   \hline
		VoI-based & $1)$ These semantic metrics capture the value of information \cite{Sem_Empowered_Commun'22}. & $1)$ For some complicated systems, it is not definitely   \\ 
		semantic metrics & $2)$ VoI is a crucial semantic metric for design, analysis, and & easy to design the VoI function \cite{Sem_Empowered_Commun'22}.   \\ 
		& optimization of both SemCom and goal-oriented SemCom systems.   & $2)$ As the definition of VoI is largely task-dependent,   \\
		&    & the derivation of an explicit function for VoI is certainly    \\
		&    & challenging \cite{SemCom_for_6G_Future_Internet'22}.   \\     
		&    & $3)$Although the value of data is usually decided not only     \\
		&    & by the content but also by the communication context,   \\
		&    & the state-of-the-art VoI calculations have not taken   \\
		&    & into consideration these relevant factors \cite{SemCom_for_6G_Future_Internet'22}.  \\     \hline
		Combined  & $1)$ Combined semantic metrics rightfully consider crucial   & $1)$ Concerning AoII, the optimal estimation of the       \\ 
		semantic metrics & factors such as context and content.  & penalty function ought to be further investigated \cite{Sem_Empowered_Commun'22}.   \\   
		& $2)$ AoII blends the age and value of information to reflect the   & $2)$ The discussed combined semantic metrics (i.e.,     \\ 
		& significance of updates \cite{Sem_Empowered_Commun'22}.   & QAoI and AoII) can be computationally demanding.  \\ \hline 
	\end{tabular}  \\ [3mm]
	
	\caption{AoI- and VoI-Based semantic metrics for image quality assessment along with their pros and cons -- AoI: age of information; QAoI: the age of information at query; PAoI: peak age of information; AoII: age of incorrect information; VoI: value of information.}
	\label{table: AoI-VoI-Based_SemCom_metrics}
\end{table*} 

Combined semantic metrics such as QAoI and AoII are useful in the design, analysis, and optimization of communications systems based on (goal-oriented) SemCom. Apart from inspiring the design, analysis, and optimization of several systems based on (goal-oriented) SemCom, AoI- and VoI-based semantic metrics have also inspired resource allocation-oriented optimization across multiple classical SemCom networks and semantic-aware networks. To optimize SemCom for semantic-aware networking, a system designer needs resource allocation semantic metrics, which are discussed below.

\section{Resource Allocation Semantic Metrics}
\label{sec: resource_allocation_semantic_metrics}
The optimization of usually scarce resources -- for optimality and efficiency across one or more networks -- across wireless or optical SemCom networks is one of the key problems facing classical SemCom systems. Resource allocation semantic metrics are therefore crucial to optimize several types of classical SemCom systems -- in either the wireless or optical domain -- such as text SemCom \cite{SemCom_Game'18,Farsad_DL_JSCC'18,Xie_DL-based_SemCom'21,Xie_Lite_distributed_SemCom'21,Zhou_WiCom_Letters'22,Peng_Robust_DL-Based_SemCom'22,Yao_Semantic_Coding'22,Lu_RL-powered_SemCom'21,Rethinking_modern_com_Lu_2022,Luo_SemCom_with_relay'21,Jiang_Deep_Source-Channel_coding'22,Liu_Context-Based_SemCom'22}; audio SemCom \cite{Weng_SemCom_Sys_Speech_Trans'21,SemCom_for_speech_signals'20,Weng_SemCom_Speech_Recognition'21,Han_Semantic-aware_Speech2Text_Transmission'22,Weng_DL-enabled_SemCom'22,Tong_FL_ASC'21}; image SemCom \cite{Eirina_JSCC'19,Kurka_Deep_JSCC-f'20,Bandwidth_Agile_Image_Transmission'21,Zhang_Wireless_Information_Transmission_of_Image'22,Xu_Wireless_Image_Transmission'22,Pan_IM-SemCom'22,Yang_WITT'22,Dai_NLT_SCC'22,Lee_Joint_Transmission_Recognition_for_IoTs'19,Hu_Robust_SemCom'22,Huang_Toward_SemCom'23}; video SemCom \cite{Jiang_Wireless_SemCom'22,Wang_Wireless_Deep_Video_Transmission,Tung_DeepWiVe'21,Huang_IS-SemCom'22}; multimodal SemCom \cite{WANG_Multimodal_SemCom'23}; and cross-modal SemCom \cite{Li_Cross-Modal_SemCom'22}. The following resource allocation semantic metrics are largely relevant to optimize the mentioned SemCom systems in either a wireless or an optical network setting: the \textit{metric of semantic similarity} (MSS), \textit{semantic transmission rate} (S-R), \textit{semantic spectral efficiency} (S-SE), \textit{quality-of-experience} (QoE), and \textit{system throughput in message} (STM). We present these semantic metrics below, beginning with MSS.

\subsection{Metric of Semantic Similarity}
\label{subsec: MSS}
The authors of \cite{Wang_SemCom_Per_Optimization'21} and \cite{Wang_JSAC_Performance_Optimi_SemCom'22} define MSS for a semantic-driven network transmitting semantic information on its downlink using orthogonal frequency division multiple access (OFDMA) technology over $Q$ downlink orthogonal RBs. W.r.t. the $Q$ RBs, let $\bm{\alpha}_i \eqdef [\alpha_{i,1}, \ldots, \alpha_{i,q}, \ldots, \alpha_{i,Q}]$ be the resource allocation vector of user $i$, $\mathcal{G}_i'$ be the partial semantic information (modeled by a KG) that the base station (BS) transmits to user $i$, and $L_i'(\bm{\alpha}_i, \mathcal{G}_i')$ be the recovered text. For this setting, the MSS of $L_i'(\bm{\alpha}_i, \mathcal{G}_i')$ is defined as follows \cite[eq. (11)]{Wang_JSAC_Performance_Optimi_SemCom'22}: 
\begin{equation}
	\label{MSS_defn}
	E_i(\bm{\alpha}_i, \mathcal{G}_i') \eqdef 	\xi_i \frac{A_i(\bm{\alpha}_i, \mathcal{G}_i')R_i(\bm{\alpha}_i, \mathcal{G}_i')}{\varphi A_i(\bm{\alpha}_i, \mathcal{G}_i')+(1-\varphi) R_i(\bm{\alpha}_i, \mathcal{G}_i')}, 
\end{equation}
where $\xi_i$ is a penalty (regarding a short text) that is defined in \cite[eq. (12)]{Wang_JSAC_Performance_Optimi_SemCom'22}, $\varphi \in (0,1)$ is the weight parameter employed to adjust the semantic accuracy $A_i(\bm{\alpha}_i, \mathcal{G}_i')$ and the semantic completeness $R_i(\bm{\alpha}_i, \mathcal{G}_i')$ of $L_i'(\bm{\alpha}_i, \mathcal{G}_i')$, which are defined in \cite[eq. (9)]{Wang_JSAC_Performance_Optimi_SemCom'22} and \cite[eq. (10)]{Wang_JSAC_Performance_Optimi_SemCom'22}, respectively.

We now proceed to discuss another resource allocation semantic metric -- named S-R.

\subsection{Semantic Transmission Rate}
\label{subsec: S-R}
S-R is defined as the amount of semantic information effectively transmitted per second and measured in \textit{suts/s} \cite{Resource_allocation_text_SemCom'22}. For a text SemCom system, the S-R $\Gamma_{n,m}$ of the $n$-th user over the $m$-th channel is defined as \cite[eq. (4)]{Resource_allocation_text_SemCom'22}
\begin{equation}
	\label{S-R_defn}
	\Gamma_{n,m} \eqdef \frac{WI}{k_nL} \xi_{n,m},
\end{equation}
where $W$, $I$, and $k_nL$ are the channel bandwidth, the amount of semantic information, and the average number of semantic symbols at the $n$-th user, respectively, and $\xi_{n,m}$ is the semantic similarity -- per \cite[eq. (13)]{Xie_DL-based_SemCom'21} -- of the $n$-th user over the $m$-th channel \cite{Resource_allocation_text_SemCom'22}.

We now continue with our discussion of another resource allocation semantic metric -- termed S-SE.

\subsection{Semantic Spectral Efficiency}
\label{subsec: S-SE}
S-SE is defined as the rate at which semantic information can be successfully transmitted over a unit of bandwidth and is measured in \textit{suts/s/Hz} \cite{Resource_allocation_text_SemCom'22}. For a text SemCom system, the S-SE of the $n$-th user over the $m$-th channel is defined via (\ref{S-R_defn}) as \cite[eq. (5)]{Resource_allocation_text_SemCom'22} 
\begin{equation}
	\label{Phi_defn}
	\Phi \eqdef  \frac{\Gamma_{n,m}}{W} =\frac{I}{k_nL} \xi_{n,m}.
\end{equation}

We now move on to discuss another resource allocation semantic metric -- dubbed QoE.

\subsection{Quality-of-Experience}
\label{subsec: QoE}
In the context of semantic-aware resource allocation in a multi-cell multi-task network, the QoE of the $q$-th user group in the $b$-th cell -- which is denoted by $QoE_q^b$ -- is defined as \cite[eq. (8)]{Yan_QoE_Aware_Resource_Allocation_SemCom'22}
\begin{subequations}
	\begin{align}
		\label{QoE_defn_1}
		QoE_q^b &\eqdef \sum_{u \in \mathcal{G}_q^b}  w_u G_u^R + (1-w_u) G_u^A  \\
		\label{QoE_defn_2}
		&= \sum_{u \in \mathcal{G}_q^b}  \frac{w_u}{1+e^{\beta_u (\varphi_u^{\textnormal{req}}-\varphi_u)}} +  \frac{(1-w_u)}{1+e^{\lambda_u (\xi_u^{\textnormal{req}}-\xi_q^b)}}  , 
	\end{align}
\end{subequations}
where $u$ denotes the user index, $\mathcal{G}_q^b$ is the user group of the $q$-th user in the $b$-th cell, $w_u$ and $1-w_u$ are the weights of the semantic rate and the semantic accuracy of the $u$-th user, respectively, $G_u^R$ and $G_u^A$ are the scores of the semantic rate and the semantic accuracy of the $u$-th user, respectively, $\beta_u$ and $\lambda_u$ denote the growth rates of $G_u^R$ and $G_u^A$, respectively, and $\varphi_u^{\textnormal{req}}$ and $\xi_u^{\textnormal{req}}$ symbolize the minimum semantic rate and semantic accuracy required to attain 50\% of the scores, respectively \cite{Yan_QoE_Aware_Resource_Allocation_SemCom'22}.

We now continue to our discussion of another resource allocation semantic metric -- termed STM.

\subsection{System Throughput in Message}
\label{subsec: STM}
STM represents network performance from a semantic perspective and is proposed by the authors of \cite{Xia_Resource_Management_SemCom'22} in the broader context of intelligent SemCom (iSemCom) and an iSemCom-enabled heterogeneous network (iSemCom-HetNet). For an iSemCom-HetNet, let $\mathcal{B} \eqdef \{BS_1, BS_2, . . . , BS_L\}$ be a set of BSs for $BS_j$ -- the $j$-th BS in a network served by $L$ BSs, $\mathcal{U} \eqdef \{MU_1,MU_2, . . . ,MU_M\}$ be the set of all mobile users (MUs) for $MU_i$ -- the $i$-th MU, and $x_{ij} \in \{0,1\}$ be an association indicator, where $x_{ij} = 1$ if $MU_i$ is associated with $BS_j$, and $x_{ij} = 0$ otherwise. For this setting, the STM -- denoted by $T_M$ -- is defined as \cite[eq. (7)]{Xia_Resource_Management_SemCom'22} 
\begin{equation}
	\label{STM_defn}
	T_M \eqdef \sum_{i \in \mathcal{U}}  \sum_{j \in \mathcal{B}}  x_{ij} S_i(b_{ij}),  
\end{equation}
where $S_i(\cdot)$ represents a universal bit-to-message transformation function pertaining to $MU_i$ under a given channel condition \cite{Xia_Resource_Management_SemCom'22}, and $b_{ij}$ stands for the downlink bit rate of $MU_i$ (served by $BS_j$ with $n_{ij}$ bandwidth) defined as follows \cite[eq. (3)]{Xia_Resource_Management_SemCom'22}:
\begin{equation}
	\label{downlink_bit-rate_defn}
	b_{ij} \eqdef n_{ij} \log_2 (1 + \gamma_{ij}), 
\end{equation}
where $\gamma_{ij}$ is the signal-to-interference-plus-noise ratio (SINR) experienced by $MU_i$ from $BS_j$ \cite{Xia_Resource_Management_SemCom'22}.

While the semantic metrics discussed above in Sections \ref{subsec: MSS} through \ref{subsec: STM} are chiefly applicable for resource allocation optimization in many types of SemCom systems, the design, analysis, and optimization of several types of SemCom systems have been inspired by the generic semantic metrics of SemCom, which we discuss below.

\section{Generic Semantic Metrics of SemCom}
\label{sec: generic_semantic_metrics_of_SemCom}
The generic semantic metrics of SemCom that we discuss in this section are metrics that are applicable to the design, analysis, and optimization of a wide variety of classical SemCom systems -- in either the wireless or optical domain -- including text SemCom \cite{SemCom_Game'18,Farsad_DL_JSCC'18,Xie_DL-based_SemCom'21,Xie_Lite_distributed_SemCom'21,Zhou_WiCom_Letters'22,Peng_Robust_DL-Based_SemCom'22,Yao_Semantic_Coding'22,Lu_RL-powered_SemCom'21,Rethinking_modern_com_Lu_2022,Luo_SemCom_with_relay'21,Jiang_Deep_Source-Channel_coding'22,Liu_Context-Based_SemCom'22}; audio SemCom \cite{Weng_SemCom_Sys_Speech_Trans'21,SemCom_for_speech_signals'20,Weng_SemCom_Speech_Recognition'21,Han_Semantic-aware_Speech2Text_Transmission'22,Weng_DL-enabled_SemCom'22,Tong_FL_ASC'21}; image SemCom \cite{Eirina_JSCC'19,Kurka_Deep_JSCC-f'20,Bandwidth_Agile_Image_Transmission'21,Zhang_Wireless_Information_Transmission_of_Image'22,Xu_Wireless_Image_Transmission'22,Pan_IM-SemCom'22,Yang_WITT'22,Dai_NLT_SCC'22,Lee_Joint_Transmission_Recognition_for_IoTs'19,Hu_Robust_SemCom'22,Huang_Toward_SemCom'23}; video SemCom \cite{Jiang_Wireless_SemCom'22,Wang_Wireless_Deep_Video_Transmission,Tung_DeepWiVe'21,Huang_IS-SemCom'22}; multimodal SemCom \cite{WANG_Multimodal_SemCom'23}; and cross-modal SemCom \cite{Li_Cross-Modal_SemCom'22}. We present below the following generic semantic metrics of SemCom that have inspired the materialization of the mentioned SemCom systems: the \textit{general quality index of semantic service}, \textit{triplet drop probability} (TDP), \textit{semantic mutual information} (SMI), the \textit{semantic impact}, the \textit{communication symmetry index}, and \textit{reasoning capacity}. We start with the general quality index of semantic service.

\subsection{General Quality Index of Semantic Service}
\label{subsec: GQI_semantic_service}
The authors of \cite{Dong_Innovative_SemCom'22} propose the general quality index of semantic service which is defined as \cite[eq. (1)]{Dong_Innovative_SemCom'22}
\begin{equation}
	\label{SS_defn}
	SS \eqdef \frac{ST(\hat{S})}{ST(S)},
\end{equation}
where the function $ST(\cdot)$ captures how well the source performs a given task, and $S$ and $\hat{S}$ denote the unprocessed information at the transmitter and the information recovered through semantics at the receiver, respectively. For the definition in (\ref{SS_defn}), the authors of \cite{Dong_Innovative_SemCom'22} suggest to convert the output of $ST(\cdot)$ to a range $[0,1]$ using \textit{sigmoid} and other similar functions.

We now move on to our discussion of another generic semantic metric of SemCom -- dubbed TDP.

\subsection{Triplet Drop Probability}
\label{subsec: TDP}
The TDP $P_k$ is an important semantic metric for both SemCom and goal-oriented SemCom \cite{Kang_Personalized_Saliency'23} and is defined as \cite[eq. (17)]{Kang_Personalized_Saliency'23}
\begin{equation}
	\label{TDP_defn}
	P_k \eqdef \sum_{j=D_E+1}^{D_T}  E_k^j (1-E_k)^{D_T-j}, 
\end{equation}
where $D_T$ denotes the bit length, $D_E$ represents the maximum number of bits in error, and $E_k$ is the $k$-th user's average bit error rate (BER) and defined as \cite[eq. (15)]{Kang_Personalized_Saliency'23}, \cite[eq. (13)]{Zhang_New_Results'18}
\begin{equation}
	\label{average_BER_defn}
	E_k \eqdef  \int_{0}^{\infty}  \displaystyle \frac{\Gamma(\lambda_2, \lambda_1\gamma)}{2\Gamma(\lambda_2)}    f_{\gamma_k}(\gamma)   d\gamma,
\end{equation}
where (\ref{average_BER_defn}) is valid under a variety of modulation formats, $\frac{\Gamma(\lambda_2, \lambda_1\gamma)}{2\Gamma(\lambda_2)}$ equates to the conditional bit error probability, $\lambda_1$ and $\lambda_2$ are modulation-specific parameters that take different values under different modulation schemes, $\gamma_k$ is the SINR of the $k$-th user, and $f_{\gamma_k}(\cdot)$ symbolizes the probability distribution function (PDF) of $\gamma_k$.

We now discuss another generic semantic metric of SemCom -- named SMI.

\subsection{Semantic Mutual Information}
\label{subsec: SMI}
SMI is proposed by the authors of \cite{Sun_semantic-assisted'22} and aims to quantify the semantic-level distortion present during the compression process for specific downstream AI task. When a downstream AI task processes the pixel-level information of the input images, the feature-level information and then semantic-level information can be acquired \cite{Sun_semantic-assisted'22}. Semantic-level information is the meaning that is eventually understood by the downstream AI task and contained in perceptual results \cite{Sun_semantic-assisted'22}.  

SMI quantifies the mutual information (MI) of all the perceptual results, which comprise all the semantic-level information of the downstream AI task \cite{Sun_semantic-assisted'22}. To define this metric formally (using the notation of \cite{Sun_semantic-assisted'22}), let $\bm{y}_b$ be the perceptual results of the original image $\bm{x}_b$ and $\bm{y}'_b$ be the perceptual results of the compressed image $\bm{x}'_b$. Estimating the SMI of the perceptual results $\bm{y}_b$ and $\bm{y}'_b$ is challenging because the \textit{entropy} of the original image dataset is mathematically intractable \cite{Sun_semantic-assisted'22}. To overcome this intractability and estimate their respective SMI, the authors of \cite{Sun_semantic-assisted'22} exploit contrastive log-ratio upper bound (CLUB) \cite{Cheng_CLUB'20}\footnote{When the conditional PDF $p(\bm{y}|\bm{x})$ is known, the MI CLUB is defined for two random multivariate RVs $\bm{x}$ and $\bm{y}$ as \cite[eq. (10)]{Cheng_CLUB'20}:   
	\begin{equation}  
		\label{MI_CLUB_defn}
		I_{\textnormal{CLUB}} (\bm{x}; \bm{y}) \eqdef \mathbb{E}_{p(\bm{x},\bm{y})} \{\log p(\bm{y} | \bm{x})\} - \mathbb{E}_{p(\bm{x})} \mathbb{E}_{p(\bm{y})} \{\log p(\bm{y} | \bm{x}) \},
	\end{equation}where $p(\bm{x} , \bm{y})$ is the joint PDF. The simplification of (\ref{MI_CLUB_defn}) leads to a theorem \cite[Theorem 3.1]{Cheng_CLUB'20} on an important inequality that is given by \cite[eq. (12)]{Cheng_CLUB'20}
	\begin{equation}
		\label{MI_upper_bound}
		I (\bm{x}; \bm{y})    \leq I_{\textnormal{CLUB}} (\bm{x}; \bm{y}), 
	\end{equation}
	where $I (\bm{x}; \bm{y})$ is the MI and equality is attained if and only if (iff) $ \bm{x}$ and $\bm{y}$ are independent RVs \cite{Cheng_CLUB'20}. Hence, $I_{\textnormal{CLUB}} (\bm{x}; \bm{y})$ is an upper bound of $I (\bm{x}; \bm{y})$.} -- an MI estimator -- due to its ability to produce reliable estimates. To this end, the authors of \cite{Sun_semantic-assisted'22} employ $\bm{y}_b$ and $\bm{y}'_b$ as inputs to train an SMI estimation network from which they can obtain the mean and variance of $\bm{y}'_b$ \cite{Sun_semantic-assisted'22}. Using this computed mean and variance values, the authors of \cite{Sun_semantic-assisted'22} compute the conditional PDF $p(\bm{y}'_b | \bm{y}_b)$. With the $p(\bm{y}'_b | \bm{y}_b)$ computed, the SMI is defined -- through the MI CLUB \cite[eq. (10)]{Cheng_CLUB'20}; see also (\ref{MI_CLUB_defn}) -- as \cite[eq. (11)]{Sun_semantic-assisted'22}    
\begin{multline}
	\label{SMI_defn}
	I_{\textnormal{CLUB}} (\bm{y}_b; \bm{y}'_b) \eqdef \mathbb{E}_{p(\bm{y}_b,\bm{y}'_b)} \{\log p(\bm{y}'_b | \bm{y}_b) \} \\ - \mathbb{E}_{p(\bm{y}_b)} \mathbb{E}_{p(\bm{y}'_b)} \{\log p(\bm{y}'_b | \bm{y}_b)\},
\end{multline}
where $p(\bm{y}_b,\bm{y}'_b)$ is a joint PDF. Underscoring its advantage as defined in (\ref{SMI_defn}), SMI can reflect/capture the semantic-level distortion \cite{Sem_Empowered_Commun'22}. However, SMI needs to be estimated by an additional module \cite{Sem_Empowered_Commun'22}.

The previously discussed generic SemCom metrics do not necessarily take into account the fact that the apprentice can leverage reasoning and causality to generate the originally transmitted message \cite{Chaccour_Building_NG_SemCom_Networks'22}. In such a scenario, which is at the heart of reasoning-driven SemCom systems \cite{Chaccour_Building_NG_SemCom_Networks'22}, the overall situation can change drastically, and there is a need for a suite of novel and generic SemCom metrics that can qualify the level of symmetry between a teacher and an apprentice \cite{Chaccour_Building_NG_SemCom_Networks'22}. Such metrics are proposed by the authors of \cite{Chaccour_Building_NG_SemCom_Networks'22} and named as the \textit{semantic impact}, the \textit{communication symmetry index}, and \textit{reasoning capacity}. These metrics are presented below, beginning with semantic impact.

\subsection{Semantic Impact}  
\label{sec: Semantic_Impact}
If we consider a particular semantic representation\footnote{According to the authors of \cite{Chaccour_Building_NG_SemCom_Networks'22}, semantic content embodies the \textquotedblleft meaningful'' part of the data and the semantic representation is the \textquotedblleft minimal way to represent this meaning.''} $Z_i$ and its semantic content element $Y_i$, the significance of $Z_i$ is equivalent to the number of data packets one would have needed to convey the exact same message \cite{Chaccour_Building_NG_SemCom_Networks'22}. To this end, semantic impact is formally defined as follows.
\begin{definition}[{Semantic impact \cite[Definition 12]{Chaccour_Building_NG_SemCom_Networks'22}}]
	\label{Semantic_impact_defn}
	Let $Z_i$ be a particular semantic representation and its semantic content element be $Y_i$. The semantic impact $\iota_{\tau}$ generated by $Z_i$ during a time duration $\tau$ is defined as the number of packets that would have been needed to be transmitted to regenerate $Y_i$.
\end{definition}
Per Definition \ref{Semantic_impact_defn}, semantic impact is part of another generic metric of SemCom dubbed the communication symmetry index \cite{Chaccour_Building_NG_SemCom_Networks'22}, which is presented below.

\subsection{Communication Symmetry Index}   
\label{sec: communication_symmetry_index}
The communication symmetry index is proposed by the authors of \cite{Chaccour_Building_NG_SemCom_Networks'22} and formally defined below.
\begin{proposition}[{Communication symmetry index \cite[Proposition 2]{Chaccour_Building_NG_SemCom_Networks'22}}]
	\label{Communication_symmetry_index_prop}
	For a transmission session $\tau$, the communication symmetry index $\eta_{b,d,\tau}$ between a teacher $b$ and an apprentice $d$ is given by \cite[eq. (13)]{Chaccour_Building_NG_SemCom_Networks'22}    
	\begin{equation}
		\label{Communication_symmetry_index_defn}
		\eta_{b,d,\tau} \eqdef \frac{\zeta_{d, \tau}}{\nu_{b, \tau}} \times \iota_{\tau, Y_i},
	\end{equation}
	where $\zeta_{d, \tau}$ is the number of query packets demanded by the apprentice to reason over the transmitted message, $\nu_{b, \tau}$ is the number of raw data packets transmitted by the teacher to accompany the transmitted semantic representation, and $\iota_{\tau, Y_i}$ is the semantic impact w.r.t. the generation of the semantic content element $Y_i$. Note that one can find $\zeta_{d, \tau}$ by applying the concept of semantic impact on the employed representation provided that the queries are communicated via a semantic representation to the teacher.
\end{proposition}

As defined in Proposition \ref{Communication_symmetry_index_prop}, the communication symmetry index makes it possible to characterize the reasoning state of the teacher and apprentice as well as the equilibrium they attain \cite{Chaccour_Building_NG_SemCom_Networks'22}. Accordingly, the following five settings are in order.
\begin{itemize}
	\item \textit{If $\eta_{b,d,\tau} \leq 1$ and $\iota_{\tau} >1$}: in this setting, the apprentice has little to no knowledge base \cite{Chaccour_Building_NG_SemCom_Networks'22}. Thus, this SemCom setting asymptotically mirrors the classical communication scenario, wherein most of the data is sent in its raw form to complement the semantic representation \cite{Chaccour_Building_NG_SemCom_Networks'22}.

	\item \textit{If $\eta_{b,d,\tau} \rightarrow \iota_{\tau}$ and $\iota_{\tau} >1$}: in this setting, the apprentice has considerable knowledge/reasoning faculties, and the teacher complements their transmitted information with raw data \cite{Chaccour_Building_NG_SemCom_Networks'22}. Nonetheless, the apprentice intervenes regularly to understand the data's causal structure (and progressively counts on semantic representations) \cite{Chaccour_Building_NG_SemCom_Networks'22}.

	\item \textit{If $\eta_{b,d,\tau} = \iota_{\tau}$ and $\iota_{\tau} >1$}: in this setting, the apprentice intervenes in the same way as current receivers transmit an acknowledgment, and the teacher depends on the transmission of raw data only to describe the \textit{unlearnable} part of the data \cite{Chaccour_Building_NG_SemCom_Networks'22}.

	\item \textit{If $\eta_{b,d,\tau} > \iota_{\tau}$ and $\iota_{\tau} >1$}: in this setting, the datastream is mostly learnable (i.e., not \textit{memorizable}) \cite{Chaccour_Building_NG_SemCom_Networks'22}. Consequently, the teacher depends mainly on semantic representations, and the apprentice actively intervenes to produce the transmitted message from the set of received semantic representations \cite{Chaccour_Building_NG_SemCom_Networks'22}.

	\item \textit{If $\eta_{b,d,\tau} > \iota_{\tau}$ and $\iota_{\tau} \leq 1$}: in this challenging setting, the apprentice demands a greater number of queries than the number of raw data transmissions the teacher sends \cite{Chaccour_Building_NG_SemCom_Networks'22}. As a result, the teacher is unable to extract a fitting semantic representation to be communicated to the apprentice \cite{Chaccour_Building_NG_SemCom_Networks'22}.
\end{itemize}
It is worth mentioning that unless a defect in reasoning is observed, $\eta_{b,d,\tau}$ does not go considerably below 1 \cite{Chaccour_Building_NG_SemCom_Networks'22}.

Should the receiver become an apprentice that counts on learning the data content rather than simply recovering it in a bit-by-bit fashion, the apprentice's understanding and impact on the reconstruction process are fittingly KPIs of a reliable communication link (between teacher and apprentice) \cite{Chaccour_Building_NG_SemCom_Networks'22}. This leads us to the generic metric of SemCom dubbed reasoning capacity, which we explain below.

\subsection{Reasoning Capacity}   
\label{sec: reasoning_capacity}
The following factors are essential to characterize the apprentice's understanding: $1)$ reasoning as evaluated by the number of queries made by the apprentice; $2)$ efficiency and minimalism (as assessed by the number of raw messages sent to supplement the semantic representation), as well as the semantic representation's impact \cite{Chaccour_Building_NG_SemCom_Networks'22}. To this end, the formal definition of reasoning capacity is provided below.
\begin{proposition}[{Reasoning capacity \cite[Proposition 3]{Chaccour_Building_NG_SemCom_Networks'22}}]
	\label{Prop_reasoning_capacity}
	The reasoning capacity between a teacher $b$ and an apprentice $d$ is expressed as \cite[eq. (15)]{Chaccour_Building_NG_SemCom_Networks'22}
	\begin{equation}
		\label{Reasoning_capacity_defn}
		C_R \eqdef \Omega \log_2 (1+ \eta_{b,d}), 
	\end{equation}
	where $\Omega$ is the maximum computing capability of the server deployed to represent/generate the semantic representation and $\eta_{b,d}$ is the communication symmetry index per second.
\end{proposition}
Reasoning capacity is universal in the sense that it is independent of the type of semantic representation employed \cite{Chaccour_Building_NG_SemCom_Networks'22}. When the datastream includes both a learnable component and a memorizable component, the total achievable capacity $C_T$ can be expressed as \cite[eq. (16)]{Chaccour_Building_NG_SemCom_Networks'22} 
\begin{equation}
	\label{Total_achievable_capacity_defn}
	C_T \eqdef C_C + C_R = W \log_2 (1+\gamma) + \Omega \log_2 (1+\eta_{b,d}), 
\end{equation}
where $C_C$ and $ C_R$ are the Shannon capacity and reasoning capacity, respectively, $W$ is the bandwidth, and $\gamma$ is the SINR \cite{Chaccour_Building_NG_SemCom_Networks'22}.

The generic semantic metrics presented above in Sections \ref{subsec: GQI_semantic_service} through \ref{sec: reasoning_capacity} are important semantic metrics for the design, analysis, and optimization of classical (wireless and optical) SemCom systems. In addition to classical SemCom systems, there also exist quantum SemCom systems whose design, analysis, and optimization are informed (or guided) by the following semantic metrics of quantum SemCom.

\section{Semantic Metrics of Quantum SemCom}   
\label{sec: semantic_metrics_of_QSemCom}
The authors of \cite{Chehimi_Quantum_SemCom'22} propose to assess the performance of their quantum SemCom system (named QSC) using the metric \textit{fidelity} \cite{Wilde_QIT'2017,Watrous_QIT'2018,Wilde_QIT_arXiv'19}, which is widely known in the quantum research community. Fidelity is a measure of the closeness of two quantum states \cite{Wilde_QIT_arXiv'19}. Hence, it is a useful semantic metric for the design, analysis, and optimization of quantum SemCom systems. We therefore discuss below three of its well-known variations, namely \textit{pure-state fidelity}, \textit{expected fidelity}, and \textit{Uhlmann fidelity} \cite[Ch. 9]{Wilde_QIT_arXiv'19}, beginning with pure-state fidelity.

\subsection{Pure-State Fidelity}
\label{subsec: PS_fidelity}
Pure-state fidelity is formally defined as follows.
\begin{definition}[{Pure-state fidelity \cite[Definition 9.2.1]{Wilde_QIT_arXiv'19}}]
	\label{PS_fidelity_formal_defn}
	Let $\mathcal{H}$ be a Hilbert space and $\ket{\psi}, \ket{\phi} \in \mathcal{H}$ be pure states. The pure state fidelity is the squared overlap of $\ket{\psi}$ and $\ket{\phi}$ defined as \cite[eq. (9.85)]{Wilde_QIT_arXiv'19}
	\begin{equation}
		\label{PS_fidelity_defn}
		F(\psi, \phi)  \eqdef | \bra{\psi}\ket{\phi}|^2.
	\end{equation}
\end{definition}
As defined in (\ref{PS_fidelity_defn}), $F(\psi, \phi)$ (i.e., pure-state fidelity) can be operationally interpreted as the probability that the output state $\ket{\phi}$ would pass a test -- carried out by someone who knows the input state -- for being the same as the input state $\ket{\psi}$ \cite{Wilde_QIT_arXiv'19}. As for the commutativity of the inner product, it follows directly from (\ref{PS_fidelity_defn}) that $F(\psi, \phi)=F(\phi, \psi)$. This metric fulfills the following bounds \cite[eq. (9.86)]{Wilde_QIT_arXiv'19}:
\begin{equation}
	\label{PS_fidelity_bounds} 
	0 \leq F(\psi, \phi) \leq 1, 
\end{equation}
where $F(\psi, \phi)=0$ iff the two corresponding states are orthogonal to each other \cite{Wilde_QIT_arXiv'19} and $F(\psi, \phi)=1$ iff the two respective states are the same. Regarding the latter case being contradictory to a distance measure that should be equal to zero when the two states are equal, the fidelity measure is not a distance measure in the strict mathematical sense \cite{Wilde_QIT_arXiv'19}. This brings us to our discussion on a quantum SemCom metric that measures the closeness between a pure state and a mixed state -- named expected fidelity.

\subsection{Expected Fidelity}
\label{subsec: expected_fidelity}
Generally, a quantum information-processing protocol is noisy and can map the pure input state $\ket{\psi}$ to a mixed state $\rho$ \cite{Wilde_QIT_arXiv'19}. These two states' closeness can be quantified by the metric expected fidelity \cite{Wilde_QIT_arXiv'19}, which is defined as follows.
\begin{definition}[{Expected fidelity \cite[Definition 9.2.2]{Wilde_QIT_arXiv'19}}]
	\label{Expected_fidelity_defn}
	Let $\ket{\psi} \in \mathcal{H}$ be a pure state and $\rho \in \mathcal{D(H)}$ be a mixed state. The expected fidelity $F(\psi, \rho)$ between these two states is given by \cite[eq. (9.89)]{Wilde_QIT_arXiv'19}   
	\begin{equation}
		\label{Expected_fidelity_defn_eqn}
		F(\psi, \rho) \eqdef \bra{\psi} \rho \ket{\psi}.
	\end{equation}
\end{definition}

The definition in (\ref{Expected_fidelity_defn_eqn}) follows directly from decomposing $\rho$ per the spectral decomposition\footnote{In this case, $p_X(x)=\mathbb{P}(X=x)$ is the \textit{probability mass function} (PMF) of a discrete RV $X$.} $\rho = \sum_x p_X(x)  \ket{\phi_x} \bra{\phi_x}$ and applying expectation w.r.t. $X$ to (\ref{PS_fidelity_defn}). In contrast to (\ref{PS_fidelity_defn}), (\ref{Expected_fidelity_defn_eqn}) characterizes fidelity when the input state is pure and the output state is mixed \cite{Wilde_QIT_arXiv'19}. Note that $F(\psi, \rho)$ per (\ref{Expected_fidelity_defn_eqn}) is a generalization of the pure-state fidelity definition given in (\ref{PS_fidelity_defn}) and obeys the same bounds \cite[eq. (9.95)]{Wilde_QIT_arXiv'19}:
\begin{equation}
	\label{expected_fidelity_bounds}
	0 \leq F(\psi, \rho)   \leq 1, 
\end{equation}
where $F(\psi, \rho) = 1$ iff the mixed state $\rho$ is equal to $\ket{\psi}$$\bra{\psi}$ and $F(\psi, \rho) = 0$ iff the support of $\rho$ is orthogonal to $\ket{\psi}$$\bra{\psi}$ \cite{Wilde_QIT_arXiv'19}. This measure, however, cannot be applied when both states are mixed. The closeness between two mixed states can be quantified using the metric Uhlmann fidelity, which we discuss below.

\subsection{Uhlmann Fidelity}
\label{subsec: Uhlmann_fidelity}
To formalize Uhlmann fidelity, we borrow an idea from pure-state fidelity (per Definition \ref{PS_fidelity_formal_defn}) to determine the fidelity between two mixed states $\rho_A$ and $\sigma_A$ that represent different states of a quantum system A \cite{Wilde_QIT_arXiv'19}. To do so, let $\ket{\phi^{\rho}}_{RA}$ and $\ket{\phi^{\sigma}}_{RA}$ stand for certain \textit{purifications} of the mixed states $\rho_A$ and $\sigma_A$, respectively, to some reference system $R$ \cite{Wilde_QIT_arXiv'19}.\footnote{For this specific scenario, it is assumed that the reference system has the same dimensions as system $A$ \cite{Wilde_QIT_arXiv'19}.} The Uhlmann fidelity $F(\rho_A,\sigma_A)$ between $\rho_A$ and $\sigma_A$ (mixed states) can now be defined as the maximum overlap between their respective purifications and given by \cite[eq. (9.97)]{Wilde_QIT_arXiv'19}
\begin{equation}
	\label{Uhlmann_fidelity_1}
	F(\rho_A,\sigma_A) \eqdef \max_{\ket{\phi^{\rho}}_{RA}, \ket{\phi^{\sigma}}_{RA}}  | \bra{\phi^{\rho}}\ket{\phi^{\sigma}}_{RA}|^2 ,
\end{equation}
where the maximization is w.r.t. all purifications $\ket{\phi^{\rho}}_{RA}$ and $\ket{\phi^{\sigma}}_{RA}$ of the corresponding mixed states $\rho_A$ and $\sigma_A$ \cite{Wilde_QIT_arXiv'19}. The RHS of (\ref{Uhlmann_fidelity_1}) can instead be maximized over \textit{unitaries} pursuant to the theorem that all purifications are equivalent up to unitaries on the reference system \cite{Wilde_QIT_arXiv'19}. This leads us to the following formal definition of Uhlmann fidelity.
\begin{definition}[{Uhlmann fidelity \cite[Definition 9.2.3]{Wilde_QIT_arXiv'19}}]
	\label{Uhlmann_fidelity_2}
	For two mixed states $\rho_A$ and $\sigma_A$, the Uhlmann fidelity $F(\rho_A,\sigma_A)$ is the maximum overlap between their respective purifications and is given by \cite[eq. (9.100)]{Wilde_QIT_arXiv'19}
	\begin{equation}
		\label{Uhlmann_fidelity_3}
		F(\rho_A,\sigma_A) \eqdef \max_{\bm{U}}  | \bra{\phi^{\rho}}_{RA} \bm{U}_R \otimes \bm{I}_A  \ket{\phi^{\sigma}}_{RA}|^2, 
	\end{equation}
	where the maximization is w.r.t. all unitaries $\bm{U}$ acting on the purification system $R$ \cite{Wilde_QIT_arXiv'19}.
\end{definition}

The Uhlmann fidelity definition in (\ref{Uhlmann_fidelity_3}) then leads us to the following important theorem.
\begin{theorem}[{Uhlmann's Theorem \cite[Theorem 9.2.1]{Wilde_QIT_arXiv'19}}]
	\label{Uhlmann_thm}
	The underneath two expressions for fidelity are equal \cite[eq. (9.102)]{Wilde_QIT_arXiv'19}:
	\begin{multline}
		\label{Uhlmann_fidelity_4}
		F(\rho_A,\sigma_A) = \max_{\bm{U}}  | \bra{\phi^{\rho}}_{RA} \bm{U}_R \otimes \bm{I}_A  \ket{\phi^{\sigma}}_{RA}|^2   \\ = \| \sqrt{\rho_A} \sqrt{\sigma_A} \|_1^2. 
	\end{multline}
\end{theorem}
For Theorem \ref{Uhlmann_thm} and (\ref{Uhlmann_fidelity_4}), it is worth remarking that Uhlmann fidelity generalizes both the pure-state fidelity defined in (\ref{PS_fidelity_defn}) and the expected fidelity defined in (\ref{Expected_fidelity_defn_eqn}) \cite{Wilde_QIT_arXiv'19}. The reader is referred to \cite[Ch. 9]{Wilde_QIT_arXiv'19} for many more important properties of fidelity. The reader is also referred to \cite{Fonseca_HD_quantum_teleportation'19} for definitions and computations of fidelity pertaining to high-dimensional quantum states such as qudits.

The aforementioned semantic metrics have inspired the design, analysis, and optimization of quantum SemCom networks as well as various quantum systems that are based on quantum SemCom. Quantum SemCom like any other type of communication system -- such as wireless SemCom and optical SemCom -- is not an end but a means to achieve specific goals \cite{Kountouris_Semantics_EmpoweredCF'21}. This goal-centric standpoint rationalizes the need for goal-oriented wireless SemCom techniques and hence the following semantic metrics of goal-oriented wireless SemCom.

\section{Semantic Metrics of Goal-Oriented Wireless SemCom}
\label{sec: Goal_oriented_SemCom_semantic_metrics}
To capture the role of data in achieving the goal of communication, a number of semantic metrics have been developed to date for goal-oriented wireless SemCom. These goal-oriented semantic metrics are chiefly crucial for the design, analysis, and optimization of goal-oriented wireless SemCom systems. Accordingly, we discuss below the following semantic metrics of goal-oriented wireless SemCom\footnote{The authors of \cite{Sem_Empowered_Commun'22} present some metrics of goal-oriented SemCom under the heading \textquotedblleft effectiveness-level metrics.''}: the \textit{$\tau$ metric}, the \textit{real-time reconstruction error}, the \textit{cost of actuation error}, \textit{multiple object detection accuracy} (MODA), \textit{value of information} (VoI), \textit{mean per joint position error} (MPJPE), \textit{triplet drop probability} (TDP), \textit{age of incorrect information} (AoII), \textit{semantic impact}, \textit{communication symmetry index}, and \textit{reasoning capacity}. We commence our discussion with the $\tau$ metric.

\subsection{The $\tau$ Metric}
\label{subsec: tau_metric}
The authors of \cite{Sana_Learning_Semantics'21} introduce $\tau$ as a generic goal-oriented wireless SemCom metric that can quantify the effectiveness of multiple transmission tasks and equates to \cite[eq. (14)]{Sana_Learning_Semantics'21}, \cite[eq. (17)]{Qin_Sem_Com_Principles_Apps'22}
\begin{equation}
	\label{generic_eta_metric_defn}
	\tau \eqdef \frac{1-\psi(\bm{s}, \hat{\bm{s}})}{\mathbb{E}\{n\}},
\end{equation}
where $\mathbb{E}\{n\}$ designates the average number of symbols per transmitted message and $\psi(\bm{s}, \hat{\bm{s}})$ quantifies the semantic error between $\bm{s}$ and $\hat{\bm{s}}$, which can take different context-dependent forms (e.g., BLEU score, MSE, or CE) \cite{Sana_Learning_Semantics'21}.   

We now proceed to discuss another goal-oriented wireless SemCom metric -- named the real-time reconstruction error. 

\subsection{Real-Time Reconstruction Error}
\label{subsec: Real-time_reconstruction_error}
Real-time reconstruction error is proposed by the authors of \cite{Kountouris_Semantics_EmpoweredCF'21} and evaluates the divergence -- in real-time as time evolves -- of values between the original source and the reconstructed source  \cite{Kountouris_Semantics_EmpoweredCF'21}. This error specifically reflects the discrepancy in real-time data exchange \cite{Sem_Empowered_Commun'22}.     

To formally define real-time reconstruction error and time-averaged real-time reconstruction error, let the original source and the reconstructed source -- at time-slot $t$ -- be denoted by $X_t$ and $\hat{X}_t$, respectively. Using these parameters, real-time reconstruction error is given by \cite{Pappas_Goal-oriented_Commun'21}
\begin{equation}
	\label{RT_reconstruction_error}
	E_t \eqdef \mathbb{I}\{X_t \neq \hat{X}_t \}, 
\end{equation}
where $E_t$ has a value of 0 or 1 for a two-state discrete-time Markov chain (DTMC). Accordingly, the system can be in either an erroneous state ($E_t = 1$) or a synced state ($E_t = 0$), and the time-averaged real-time reconstruction error is given by \cite[eq. (1)]{Pappas_Goal-oriented_Commun'21}
\begin{equation}
	\label{average_RT_reconstruction_error}
	\bar{E} \eqdef \lim_{T \to \infty} \frac{\sum_{t=1}^T E_t}{T}. 
\end{equation}
In many analytical and numerical studies, the evolution of the state of the system (i.e., $E_t$) is described by a Markov Chain; see \cite[Fig. 2]{Pappas_Goal-oriented_Commun'21}. This brings us to another relevant goal-oriented wireless SemCom metric known as the cost of actuation error.

\subsection{Cost of Actuation Error}
\label{subsec: Cost_actuation_error}
The cost of actuation error was put forward by the authors of \cite{Kountouris_Semantics_EmpoweredCF'21} and captures the significance of the error at the actuation point considering the fact that some errors ay be non-commutative and have a higher impact than others \cite{Kountouris_Semantics_EmpoweredCF'21}. There are three possible cases in which this type of error occurs in a time-slotted system \cite{Kountouris_Semantics_EmpoweredCF'21}:
\begin{itemize}
	\item \textit{The original source is in the first state, but the reconstructed source believes that it is in the second state}: in this case, the cost of actuation error is low \cite{Kountouris_Semantics_EmpoweredCF'21}.
	
	\item \textit{The original source is in the second state, but the reconstructed source believes that it is in the first state}: this pertains to a scenario in which the penalty/loss from taking a wrong action upon a misconceived system’s state is high \cite{Kountouris_Semantics_EmpoweredCF'21}. Accordingly, in this case, the cost of actuation error is presumed to be relatively high \cite{Kountouris_Semantics_EmpoweredCF'21}.
	
	\item \textit{Both the original source and the reconstructed source are in the same (first/second) state}: in this case, the states match and there is no cost of actuation error \cite{Kountouris_Semantics_EmpoweredCF'21}.
	
\end{itemize}

In the itemized cases, some errors can have larger impact than others \cite{Pappas_Goal-oriented_Commun'21}. To quantify the average impact, let $C_{i,j}$ be the cost -- at time-slot $t$ -- of being in state $i$ at the original source and in state $j \neq i$ at the reconstructed source (i.e., $E_t = 1$) \cite{Pappas_Goal-oriented_Commun'21}. It is assumed that $C_{i,j}$ doesn't change over time and that $C_{0,1} \neq C_{1,0}$ \cite{Pappas_Goal-oriented_Commun'21}. The authors of \cite{Pappas_Goal-oriented_Commun'21}, on the other hand, calculate the average cost of actuation error using a two-dimensional Markov chain that can characterize the joint status of the system for the current state at the original source whether or not the reconstructed source is synced. The average cost of actuation error is therefore given by \cite[eq. (6)]{Pappas_Goal-oriented_Commun'21}
\begin{equation}
	\label{Ave_cost_of_actuation_error_defn}
	\bar{C}_A \eqdef \pi_{(0,1)} C_{0,1} + \pi_{(1,0)} C_{1,0}, 
\end{equation}
where $\pi_{(0,1)}$ and $\pi_{(1,0)}$ are obtained from the stationary distribution of the two-dimensional DTMC \cite{Pappas_Goal-oriented_Commun'21}. This formulation offers a general view of the system, which can be deployed to derive optimal online policies using Markov decision processes or deep reinforcement learning \cite{Pappas_Goal-oriented_Commun'21}.

We now move on to discuss another goal-oriented wireless SemCom metric -- termed MODA.

\subsection{Multiple Object Detection Accuracy}
\label{subsec: GO-SemCom_MODA}
The authors of \cite{Shao_TO_Commun'22} assess the performance of their proposed goal-oriented wireless SemCom scheme by employing the metric MODA \cite{asturi_Performance_Evaluation_Framework'09}. It is defined w.r.t. each frame $t$ as \cite[eq. (7)]{asturi_Performance_Evaluation_Framework'09}
\begin{equation}
	\label{MODA_defn}
	MODA(t) \eqdef 1- \frac{c_m(m_t)+ c_f(fp_t)}{N_G^{(t)}}, 
\end{equation}
where $m_t$ and $fp_t$ are the number of misses and the number of false positives, respectively, for every frame $t$, $c_m(\cdot)$ and $c_f(\cdot)$ are the cost functions of the missed detects and false positives, respectively, and $N_G^{(t)}$ denotes the number of ground truth objects in the $t$-th frame \cite{asturi_Performance_Evaluation_Framework'09}. Normalized MODA (N-MODA) is {another important goal-oriented {wireless SemCom metric and is defined in \cite[eq. (8)]{asturi_Performance_Evaluation_Framework'09}.
		
		We now move on to discuss another goal-oriented wireless SemCom metric -- called VoI.
		
		\subsection{Value of Information}
		\label{subsec: VoI_GO_SemCom}
		VoI, as it is defined in Section \ref{subsec: VoI-Based_semantic_metrics}, specifically gauges the advantage of transmitting data packets for a communication goal, and considers not only the {packets' content, but also their respective cost of transmission \cite{SemCom_Net_Systems'21,SemCom_for_6G_Future_Internet'22}. Accordingly, the VoI metric can be quantified as the difference between the benefit a given sample affords and how much it costs to transmit it \cite{SemCom_Net_Systems'21}. VoI-based transmission policies, thus, have the potential to considerably reduce data traffic to achieve a given level of control performance, especially in networked control systems \cite{SemCom_Net_Systems'21,Soleymani_dissertation_19}. Apart from in those systems, VoI is of more interest than accuracy in resource-constrained communications, where the relevance of data packets awaiting transmission is evaluated w.r.t. the system objective \cite{SemCom_for_6G_Future_Internet'22}. VoI-based metrics are a better fit for goal-oriented wireless SemCom systems than error-based metrics \cite{SemCom_for_6G_Future_Internet'22}.
			
			Apart from VoI, MODA, the cost of actuation error, real-time reconstruction error, and the $\tau$ metric, which are highlighted above in Sections \ref{subsec: tau_metric} through \ref{subsec: VoI_GO_SemCom}, there are also generic SemCom metrics such as MPJPE, TDP, AoII, semantic impact, communication symmetry index, and reasoning capacity that are applicable to the design, analysis, and optimization of goal-oriented wireless SemCom systems. These metrics and their applications are highlighted below as miscellaneous metrics of goal-oriented SemCom.
			
			\subsection{Miscellaneous Metrics of Goal-Oriented SemCom}  
			\subsubsection{Mean Per Joint Position Error}
			\label{subsubsec: GO-SemCom_MPJPE}
			MPJPE is an important metric for evaluating the performance of goal-oriented wireless SemCom schemes such as goal-oriented SemCom for 3D human mesh construction tasks \cite{Zhang_Semantic_Sensing_and_Commun'22}. It is defined in (\ref{MPJPE_defn}).
			
			\subsubsection{Triplet Drop Probability}
			\label{subsec: GO-SemCom_}
			TDP is a generic semantic metric that is also a crucial performance analysis/optimization metric for wireless systems that are based on goal-oriented wireless SemCom \cite{Kang_Personalized_Saliency'23}. It is defined in (\ref{TDP_defn}).
			
			\subsubsection{Age of Incorrect Information}
			\label{subsec: AoII_GO_SemCom}
			AoII \cite{Maatouk_AoII'20,Maatouk_Opt_Wireless_Networks'20,AoII_metric_journal'20} is an age-based metric that facilitates goal-oriented wireless SemCom. When it comes to goal-oriented SemCom, the authors of \cite{Maatouk_AoII'20} demonstrate that AoII is able to capture the data’s role in achieving the communication goal. It is defined in Definition \ref{AoII_metric_defn}.
			
			\subsubsection{Semantic Impact}  
			\label{sec: GSemCom_Semantic_Impact}
			Semantic impact is a generic semantic metric that is applicable to the design, analysis, and optimization of systems that are based on goal-oriented wireless SemCom. It is defined in Definition \ref{Semantic_impact_defn}.
			
			\subsubsection{Communication Symmetry Index}   
			\label{sec: GSemCom_communication_symmetry_index}
			Communication symmetry index is a generic semantic metric that is applicable to the design, analysis, and optimization of goal-oriented wireless SemCom systems. It is defined in Proposition \ref{Communication_symmetry_index_prop}.
			
			\subsubsection{Reasoning Capacity}   
			\label{sec: GSemCom_reasoning_capacity} 
			Reasoning capacity is a generic semantic metric that is also applicable to the design, analysis, and optimization of systems that are based on goal-oriented wireless SemCom. It is defined in Proposition \ref{Prop_reasoning_capacity}.
			
			We now conclude this work below with our concluding summary and research outlook.
			
			\section{Concluding Summary and Research Outlook}
			\label{sec: conc_summary_and_research_outlook}
			The semantic-centric design in SemCom and goal-oriented SemCom helps to minimize power usage, bandwidth consumption, and transmission delay. These crucial advantages of SemCom and goal-oriented SemCom can mitigate some of the fundamental challenges of 6G. Consequently, SemCom and goal-oriented SemCom have been widely advocated as promising enablers of 6G and developing rapidly. Despite the upsurge in their rapid development, the design, analysis, optimization, and realization of robust and intelligent SemCom as well as goal-oriented SemCom face many fundamental challenges. Amongst these challenges, the important one is the lack of unified/universal performance assessment metrics for SemCom and goal-oriented SemCom. To put this specific challenge in perspective and stimulate fundamental research, this survey paper offered a detailed discussion on the existing metrics for SemCom and goal-oriented SemCom. More specifically, it presented semantic metrics used for text, speech, and image quality assessment; semantic metrics used for video quality and 3D human sensing assessment; AoI- and VoI-based semantic metrics; resource allocation semantic metrics; generic semantic metrics of SemCom; semantic metrics of quantum SemCom; and semantic metrics of goal-oriented wireless SemCom. By presenting all these metrics used for designing SemCom and goal-oriented SemCom systems, this paper intends to inspire the design, analysis, and optimization of many types of SemCom and goal-oriented SemCom systems. This article also invigorates the development of unified/universal performance assessment metrics of SemCom and goal-oriented SemCom, as the existing metrics are purely statistical and hardly applicable to reasoning-type tasks that constitute the heart of 6G and beyond.

\section*{Disclaimer} 
The identification of any commercial product or trade name does not imply endorsement or recommendation by the National Institute of Standards and Technology, nor is it intended to imply that the materials or equipment identified are necessarily the best available for the purpose.


\balance

\begin{IEEEbiography}[{\includegraphics[width=1.1in,height=1.25in,clip,keepaspectratio]{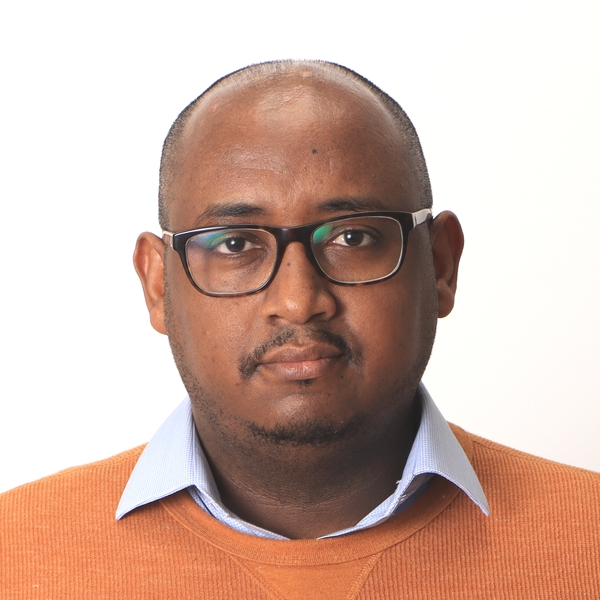}}]{\textbf{Tilahun M. Getu}} (M'19) earned the Ph.D. degree (with highest honor) in electrical engineering from the \'Ecole de Technologie Sup\'erieure (\'ETS), Montreal, QC, Canada in 2019. He is currently a Guest Researcher with the National Institute of Standards and Technology (NIST), Gaithersburg, MD, USA and a Post-doctoral Fellow with the \'ETS, Montreal, QC, Canada. His specialist and generalist fundamental research interests span the numerous fields of classical and quantum \textbf{STEM} (\textbf{S}cience, \textbf{T}echnology, \textbf{E}ngineering, and \textbf{M}athematics) at the nexus of communications, signal processing, and networking (all types); intelligence (both artificial and natural); robotics; computing; security; optimization; high-dimensional statistics; and high-dimensional causal inference.   
	
Dr. Getu has received several awards, including the 2019 ÉTS Board of Director’s Doctoral Excellence Award in recognition of his Ph.D. dissertation selected as the 2019 \'ETS all-university best Ph.D. dissertation.
\end{IEEEbiography}

\begin{IEEEbiography}[{\includegraphics[width=1.1in,height=1.25in,clip,keepaspectratio]{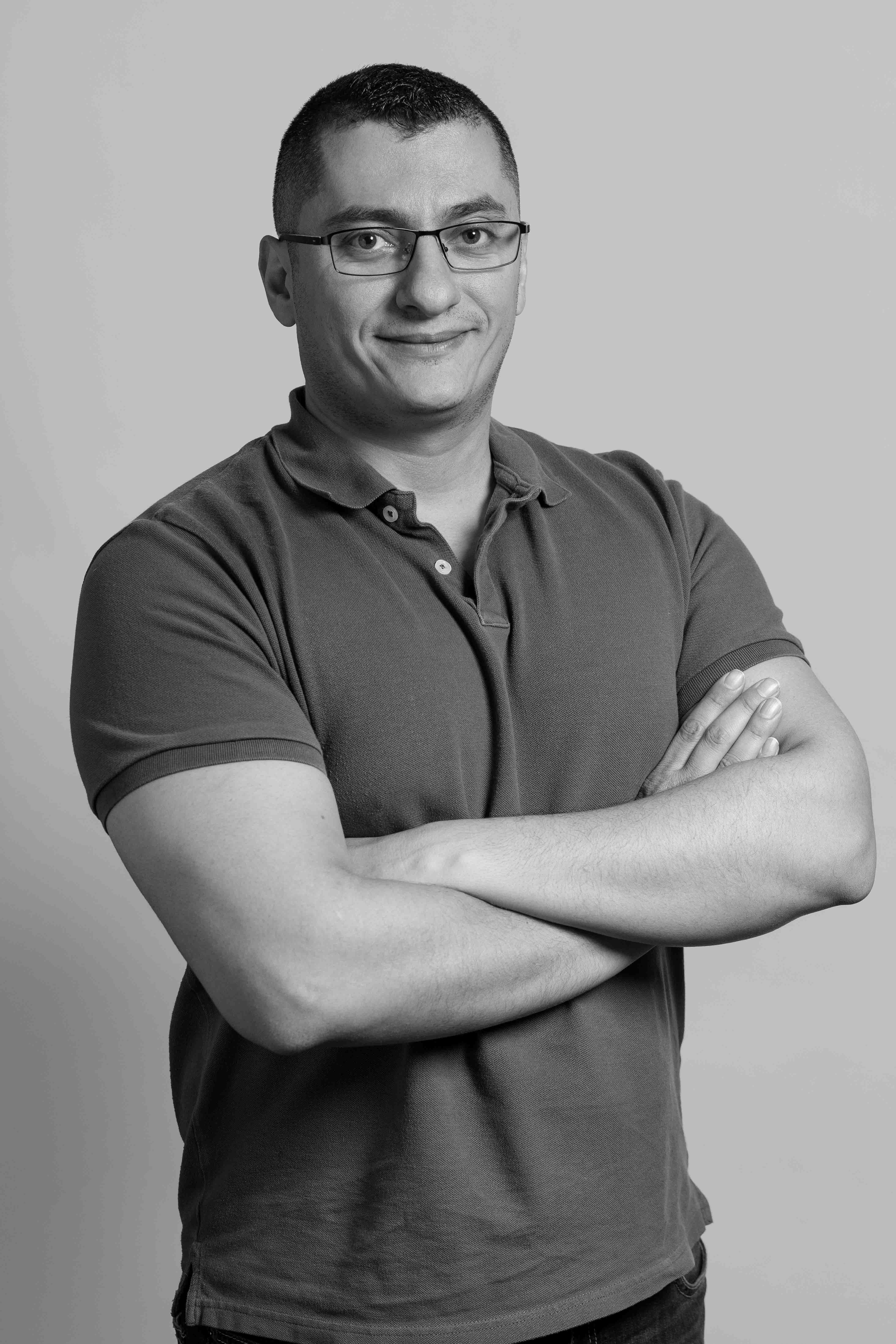}}]{\textbf{Georges Kaddoum}} (M'11--SM'20) is a professor and Tier 2 Canada Research Chair with the École de Technologie Supérieure (ÉTS), Université du Québec, Montréal, Canada. He is also a Faculty Fellow in the Cyber Security Systems and Applied AI Research Center at Lebanese American University. His recent research activities cover 5G/6G networks, tactical communications, resource allocations, and security.  Dr. Kaddoum has received many prestigious national and international awards in recognition of his outstanding research outcomes. Currently, Prof. Kaddoum serves as an Area Editor for the IEEE Transactions on Machine Learning in Communications and Networking and an Associate Editor for IEEE Transactions on Information Forensics and Security, and IEEE Transactions on Communications. 
\end{IEEEbiography}



\begin{IEEEbiography}[{\includegraphics[width=1.1in,height=1.25in,clip,keepaspectratio]{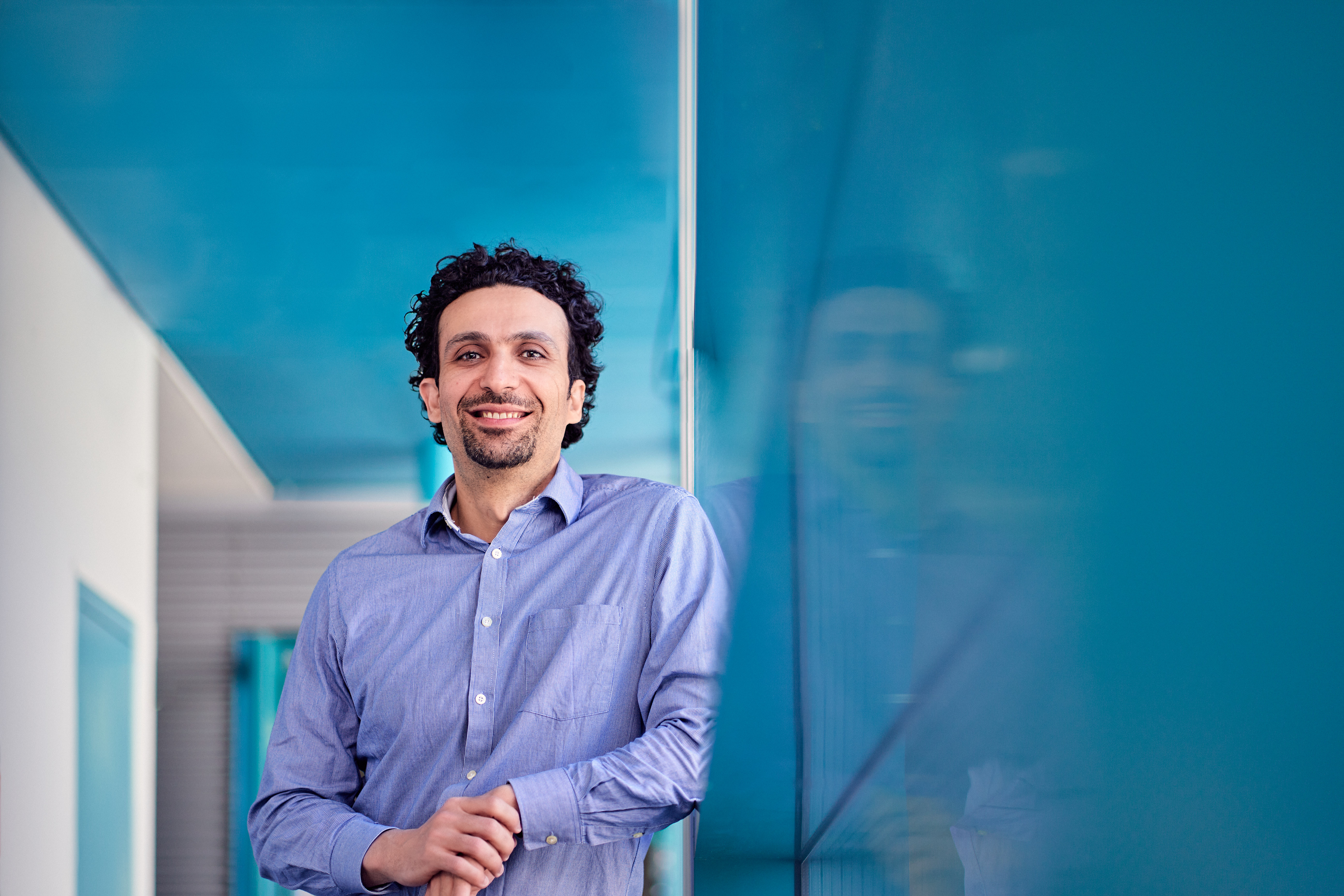}}]{\textbf{Mehdi Bennis}} (F’20) is a full tenured Professor at the Centre for Wireless Communications, University of Oulu, Finland and head of the \textbf{I}ntelligent \textbf{CO}nnectivity and \textbf{N}etworks/Systems Group (\textbf{ICON}). His main research interests are in radio resource management, game theory and distributed AI in 5G/6G networks. He has published more than 200 research papers in international conferences, journals and book chapters. He has been the recipient of several prestigious awards. Dr. Bennis is an editor of IEEE TCOM and Specialty Chief Editor for Data Science for Communications in the Frontiers in Communications and Networks journal.
\end{IEEEbiography}

\EOD


\begin{thebibliography}{100}
	\providecommand{\url}[1]{#1}
	\csname url@samestyle\endcsname
	\providecommand{\newblock}{\relax}
	\providecommand{\bibinfo}[2]{#2}
	\providecommand{\BIBentrySTDinterwordspacing}{\spaceskip=0pt\relax}
	\providecommand{\BIBentryALTinterwordstretchfactor}{4}
	\providecommand{\BIBentryALTinterwordspacing}{\spaceskip=\fontdimen2\font plus
		\BIBentryALTinterwordstretchfactor\fontdimen3\font minus
		\fontdimen4\font\relax}
	\providecommand{\BIBforeignlanguage}[2]{{%
			\expandafter\ifx\csname l@#1\endcsname\relax
			\typeout{** WARNING: IEEEtran.bst: No hyphenation pattern has been}%
			\typeout{** loaded for the language `#1'. Using the pattern for}%
			\typeout{** the default language instead.}%
			\else
			\language=\csname l@#1\endcsname
			\fi
			#2}}
	\providecommand{\BIBdecl}{\relax}
	\BIBdecl
	
	\bibitem{Saad_6G_Vision_20}
	W.~{Saad}, M.~{Bennis}, and M.~{Chen}, ``A vision of {6G} wireless systems:
	Applications, trends, technologies, and open research problems,'' \emph{IEEE
		Netw.}, vol.~34, no.~3, pp. 134--142, 2020.
	
	\bibitem{Letaief_Edge_AI_Vision'22}
	K.~B. Letaief, Y.~Shi, J.~Lu, and J.~Lu, ``Edge artificial intelligence for
	{6G}: Vision, enabling technologies, and applications,'' \emph{IEEE J. Sel.
		Areas Commun.}, vol.~40, no.~1, pp. 5--36, 2022.
	
	\bibitem{Alwis_Survey_GG_Networks'21}
	C.~D. Alwis, A.~Kalla, Q.-V. Pham, P.~Kumar, K.~Dev, W.-J. Hwang, and
	M.~Liyanage, ``Survey on {6G} frontiers: Trends, applications, requirements,
	technologies and future research,'' \emph{IEEE Open J. Commun. Soc.}, vol.~2,
	pp. 836--886, 2021.
	
	\bibitem{Akyildiz_6G_and_Beyond'20}
	I.~F. Akyildiz, A.~Kak, and S.~Nie, ``{6G} and beyond: The future of wireless
	communications systems,'' \emph{IEEE Access}, vol.~8, pp. 133\,995--134\,030,
	2020.
	
	\bibitem{Alsabah_6G_Wireless_Commun_Network'21}
	M.~Alsabah, M.~A. Naser, B.~M. Mahmmod, S.~H. Abdulhussain, M.~R. Eissa,
	A.~Al-Baidhani, N.~K. Noordin, S.~M. Sait, K.~A. Al-Utaibi, and F.~Hashim,
	``{6G} wireless communications networks: A comprehensive survey,'' \emph{IEEE
		Access}, vol.~9, pp. 148\,191--148\,243, 2021.
	
	\bibitem{Dang_Alouini_6G_20}
	{S. Dang \textit{et al.}}, ``What should {6G} be?'' \emph{Nat. Electron.},
	vol.~3, pp. 20--29, 2020.
	
	\bibitem{You_Towards_6G'21}
	{X. You \textit{et al.}}, ``Towards {6G} wireless communication networks:
	vision, enabling technologies, and new paradigm shifts,'' \emph{Sci. China
		Inf. Sci.}, vol.~64, 2021.
	
	\bibitem{Road_towards_6G'21}
	W.~Jiang, B.~Han, M.~A. Habibi, and H.~D. Schotten, ``The road towards {6G}: A
	comprehensive survey,'' \emph{IEEE Open J. Commun. Soc.}, vol.~2, pp.
	334--366, 2021.
	
	\bibitem{Roadmap_6G_Privacy_and_Security'21}
	P.~Porambage, G.~Gür, D.~P.~M. Osorio, M.~Liyanage, A.~Gurtov, and
	M.~Ylianttila, ``The roadmap to {6G} security and privacy,'' \emph{IEEE Open
		J. Commun. Soc.}, vol.~2, pp. 1094--1122, 2021.
	
	\bibitem{6G_Ecosystem'21}
	J.~R. Bhat and S.~A. Alqahtani, ``{6G} ecosystem: Current status and future
	perspective,'' \emph{IEEE Access}, vol.~9, pp. 43\,134--43\,167, 2021.
	
	\bibitem{Survey_6G_Networks'21}
	\BIBentryALTinterwordspacing
	A.~Shahraki, M.~Abbasi, M.~J. Piran, and A.~Taherkordi, ``A comprehensive
	survey on {6G} networks: Applications, core services, enabling technologies,
	and future challenges,'' 2021. [Online]. Available:
	\url{https://arxiv.org/pdf/2101.12475.pdf}
	\BIBentrySTDinterwordspacing
	
	\bibitem{Lu_6G_survey_20}
	Y.~Lu and X.~Zheng, ``{6G}: A survey on technologies, scenarios, challenges,
	and the related issues,'' \emph{J. Ind. Inf. Integr.}, vol.~19, p. 100158,
	2020.
	
	\bibitem{Yaacoub_PIEEE_20}
	E.~{Yaacoub} and M.~{Alouini}, ``A key {6G} challenge and
	opportunity---connecting the base of the pyramid: A survey on rural
	connectivity,'' \emph{Proc. IEEE}, vol. 108, no.~4, pp. 533--582, 2020.
	
	\bibitem{zhao2019survey_IRS_19}
	\BIBentryALTinterwordspacing
	J.~Zhao and Y.~Liu, ``A survey of intelligent reflecting surfaces {(IRSs)}:
	Towards {6G} wireless communication networks,'' 2019. [Online]. Available:
	\url{https://arxiv.org/pdf/1907.04789.pdf}
	\BIBentrySTDinterwordspacing
	
	\bibitem{Chowdhury_6G_2020}
	M.~Z. {Chowdhury}, M.~{Shahjalal}, S.~{Ahmed}, and Y.~M. {Jang}, ``{6G}
	wireless communication systems: Applications, requirements, technologies,
	challenges, and research directions,'' \emph{IEEE Open J. Commun. Soc.},
	vol.~1, pp. 957--975, 2020.
	
	\bibitem{Viswanathan_6G_2020}
	H.~{Viswanathan} and P.~E. {Mogensen}, ``Communications in the {6G} era,''
	\emph{IEEE Access}, vol.~8, pp. 57\,063--57\,074, 2020.
	
	\bibitem{Bariah_6G_2020}
	L.~{Bariah}, L.~{Mohjazi}, S.~{Muhaidat}, P.~C. {Sofotasios}, G.~K. {Kurt},
	H.~{Yanikomeroglu}, and O.~A. {Dobre}, ``A prospective look: Key enabling
	technologies, applications and open research topics in {6G} networks,''
	\emph{IEEE Access}, 2020.
	
	\bibitem{Tataria_6G_Wire_Systems'21}
	H.~Tataria, M.~Shafi, A.~F. Molisch, M.~Dohler, H.~Sj\"{o}land, and
	F.~Tufvesson, ``{6G} wireless systems: Vision, requirements, challenges,
	insights, and opportunities,'' \emph{Proc. IEEE}, vol. 109, no.~7, pp.
	1166--1199, 2021.
	
	\bibitem{Fettweis_6G_Per_Tactile_Internet'21}
	G.~P. Fettweis and H.~Boche, ``{6G}: The personal tactile internet - and open
	questions for information theory,'' \emph{IEEE BITS the Information Theory
		Magazine}, pp. 1--1, 2021.
	
	\bibitem{Uusitalo_6G_Hexa-X'21}
	{M. A. Uusitalo \textit{et al.}}, ``{6G} vision, value, use cases and
	technologies from european {6G} flagship project {Hexa-X},'' \emph{IEEE
		Access}, vol.~9, pp. 160\,004--160\,020, 2021.
	
	\bibitem{De_Lima_6G'21}
	{C. De Lima \textit{et al.}}, ``Convergent communication, sensing and
	localization in {6G} systems: An overview of technologies, opportunities and
	challenges,'' \emph{IEEE Access}, vol.~9, pp. 26\,902--26\,925, 2021.
	
	\bibitem{Khan_6G_20}
	L.~U. {Khan}, I.~{Yaqoob}, M.~{Imran}, Z.~{Han}, and C.~S. {Hong}, ``{6G}
	wireless systems: A vision, architectural elements, and future directions,''
	\emph{IEEE Access}, vol.~8, pp. 147\,029--147\,044, 2020.
	
	\bibitem{Rappaport_wires_commun_above_100GHz}
	T.~S. {Rappaport}, Y.~{Xing}, O.~{Kanhere}, S.~{Ju}, A.~{Madanayake},
	S.~{Mandal}, A.~{Alkhateeb}, and G.~C. {Trichopoulos}, ``Wireless
	communications and applications above {100 GHz}: Opportunities and challenges
	for {6G} and beyond,'' \emph{IEEE Access}, vol.~7, pp. 78\,729--78\,757,
	2019.
	
	\bibitem{6G_mailbox_theory'21}
	Y.~Hao, Y.~Miao, M.~Chen, H.~Gharavi, and V.~C.~M. Leung, ``{6G} cognitive
	information theory: A mailbox perspective,'' \emph{Big Data Cogn. Comput.},
	vol.~5, no.~4, 2021.
	
	\bibitem{Chen_6G_2020}
	S.~{Chen}, Y.~{Liang}, S.~{Sun}, S.~{Kang}, W.~{Cheng}, and M.~{Peng},
	``Vision, requirements, and technology trend of {6G}: How to tackle the
	challenges of system coverage, capacity, user data-rate and movement speed,''
	\emph{IEEE Wireless Commun.}, vol.~27, no.~2, pp. 218--228, Apr. 2020.
	
	\bibitem{Shaping_Future_6G_Networks'22}
	{E. Bertin, N, Crespi, and T. Magedanz (Eds.)}, \emph{Shaping Future {6G}
		Networks: Needs, Impacts, and Technologies}.\hskip 1em plus 0.5em minus
	0.4em\relax Hoboken, NJ, USA: Wiley, 2022.
	
	\bibitem{IoT_Connectivity_in_6G'21}
	\BIBentryALTinterwordspacing
	P.~Popovski, F.~Chiariotti, V.~Croisfelt, A.~E. Kalør, I.~Leyva-Mayorga,
	L.~Marchegiani, S.~R. Pandey, and B.~Soret, ``Internet of things ({IoT})
	connectivity in {6G}: An interplay of time, space, intelligence, and value,''
	2021. [Online]. Available: \url{https://arxiv.org/pdf/2111.05811.pdf}
	\BIBentrySTDinterwordspacing
	
	\bibitem{Spec_stu_6G_19}
	\BIBentryALTinterwordspacing
	F.~Tariq, M.~R.~A. Khandaker, K.~Wong, M.~A. Imran, M.~Bennis, and M.~Debbah,
	``A speculative study on {6G},'' 18 Feb. 2019. [Online]. Available:
	\url{https://arxiv.org/pdf/1902.06700.pdf}
	\BIBentrySTDinterwordspacing
	
	\bibitem{InoT_Maier'20}
	M.~Maier, A.~Ebrahimzadeh, S.~Rostami, and A.~Beniiche, ``The internet of no
	things: Making the internet disappear and \textquotedblleft see the
	invisible'','' \emph{IEEE Commun. Mag.}, vol.~58, no.~11, pp. 76--82, 2020.
	
	\bibitem{Edge_Enabled_Metaverse'22}
	\BIBentryALTinterwordspacing
	M.~Xu, W.~C. Ng, W.~Y.~B. Lim, J.~Kang, Z.~Xiong, D.~Niyato, Q.~Yang, X.~S.
	Shen, and C.~Miao, ``A full dive into realizing the edge-enabled metaverse:
	Visions, enabling technologies, and challenges,'' 2022. [Online]. Available:
	\url{https://arxiv.org/pdf/2203.05471.pdf}
	\BIBentrySTDinterwordspacing
	
	\bibitem{Gui_6G_20}
	G.~{Gui}, M.~{Liu}, F.~{Tang}, N.~{Kato}, and F.~{Adachi}, ``{6G}: Opening new
	horizons for integration of comfort, security and intelligence,'' \emph{IEEE
		Wireless Commun.}, pp. 1--7, 2020.
	
	\bibitem{Wireless_DCN_18}
	\BIBentryALTinterwordspacing
	{A. Celik, B. Shihadah, and M.-S. Alouini}, ``Wireless data center networks:
	Adavnces, challenges, and opportunities,'' 28 Nov. 2018. [Online]. Available:
	\url{https://arxiv.org/pdf/1811.11717.pdf}
	\BIBentrySTDinterwordspacing
	
	\bibitem{RWH_6G_VTM_19}
	{R. W. Heath}, ``Going toward {6G},'' \emph{IEEE Signal Process. Mag.}, pp.
	3--4, May 2019.
	
	\bibitem{Scoring_Terabit_per_second_goal'20}
	\BIBentryALTinterwordspacing
	{{N. Rajatheva \textit{et al.}}}, ``Scoring the terabit/s goal:broadband
	connectivity in {6G},'' 2020. [Online]. Available:
	\url{https://arxiv.org/pdf/2008.07220.pdf}
	\BIBentrySTDinterwordspacing
	
	\bibitem{6G_BBC_20_White_Paper}
	\BIBentryALTinterwordspacing
	{N. Rajatheva \textit{et al.}}, ``White paper on broadband connectivity in
	{6G},'' 2020. [Online]. Available: \url{https://arxiv.org/pdf/2004.14247.pdf}
	\BIBentrySTDinterwordspacing
	
	\bibitem{KDHB_18}
	K.~{David} and H.~{Berndt}, ``{6G} vision and requirements: Is there any need
	for beyond {5G}?'' \emph{IEEE Veh. Technol. Mag.}, vol.~13, no.~3, pp.
	72--80, Sep. 2018.
	
	\bibitem{Latvaaho2019KeyDAC}
	\BIBentryALTinterwordspacing
	M.~Latva-aho and K.~L. \hspace{2mm} (eds.), ``Key drivers and research
	challenges for {6G} ubiquitous wireless intelligence,'' Sep. 2019. [Online].
	Available: \url{http://jultika.oulu.fi/Record/isbn978-952-62-2354-4}
	\BIBentrySTDinterwordspacing
	
	\bibitem{Shannon_Weaver_Math_Theory_Commun'49}
	C.~E. Shannon and W.~Weaver, \emph{The Mathematical Theory of
		Communication}.\hskip 1em plus 0.5em minus 0.4em\relax Urbana, IL, USA: Univ.
	Illinois Press, 1949.
	
	\bibitem{Gunduz_Beyond_Transmitting_Bits'22}
	\BIBentryALTinterwordspacing
	D.~Gunduz, Z.~Qin, I.~E. Aguerri, H.~S. Dhillon, Z.~Yang, A.~Yener, K.~K. Wong,
	and C.-B. Chae, ``Beyond transmitting bits: Context, semantics, and
	task-oriented communications,'' 2022. [Online]. Available:
	\url{https://arxiv.org/pdf/2207.09353.pdf}
	\BIBentrySTDinterwordspacing
	
	\bibitem{hong_Theory_Semantic_Info'17}
	Y.~Zhong, ``A theory of semantic information,'' \emph{China Commun.}, vol.~14,
	no.~1, pp. 1--17, 2017.
	
	\bibitem{Zhong_Theory_Sem_Information_Book_Chapter}
	Y.~Zhong and G.~Dodig-Crnković, \emph{A Theory of Semantic Information in the
		Context of its Ecology}, 2020, ch. Chapter 5, pp. 81--112.
	
	\bibitem{Knowledge_Graphs_Survey'22}
	S.~Ji, S.~Pan, E.~Cambria, P.~Marttinen, and P.~S. Yu, ``A survey on knowledge
	graphs: Representation, acquisition, and applications,'' \emph{IEEE Trans.
		Neural Netw. Learn. Syst.}, vol.~33, pp. 494--514, 2022.
	
	\bibitem{Topos_and_Stacks'21}
	\BIBentryALTinterwordspacing
	J.-C. Belfiore and D.~Bennequin, ``Topos and stacks of deep neural networks,''
	2021. [Online]. Available: \url{https://arxiv.org/pdf/2106.14587.pdf}
	\BIBentrySTDinterwordspacing
	
	\bibitem{Tetlow_Toward_Semantic_Info_Theory'22}
	\BIBentryALTinterwordspacing
	P.~Tetlow, D.~Garg, L.~Chase, M.~Mattingley-Scott, N.~Bronn, K.~Naidoo, and
	E.~Reinert, ``Towards a semantic information theory (introducing quantum
	corollas),'' 2022. [Online]. Available:
	\url{https://arxiv.org/pdf/2201.05478.pdf}
	\BIBentrySTDinterwordspacing
	
	\bibitem{Tong_FL_ASC'21}
	H.~Tong, Z.~Yang, S.~Wang, Y.~Hu, O.~Semiari, W.~Saad, and C.~Yin, ``Federated
	learning for audio semantic communication,'' \emph{Front. Comms. Net.},
	vol.~2, 2021.
	
	\bibitem{Xie_DL-based_SemCom'21}
	H.~Xie, Z.~Qin, G.~Li, and B.-H. Juang, ``Deep learning enabled semantic
	communication systems,'' \emph{IEEE Trans. Signal Process.}, vol.~69, pp.
	2663--2675, Apr. 2021.
	
	\bibitem{Kalfa_Toward_GO_Semantic_Signal_Processing'21}
	M.~Kalfa, M.~Gok, A.~Atalik, B.~Tegin, T.~M. Duman, and O.~Arikan, ``Towards
	goal-oriented semantic signal processing: Applications and future
	challenges,'' \emph{Digit. Signal Process.}, vol. 119, pp. 103--134, Dec.
	2021.
	
	\bibitem{Zhou_WiCom_Letters'22}
	Q.~Zhou, R.~Li, Z.~Zhao, C.~Peng, and H.~Zhang, ``Semantic communication with
	adaptive universal transformer,'' \emph{IEEE Wirel. Commun. Lett.}, vol.~11,
	no.~3, pp. 453--457, 2022.
	
	\bibitem{SemCom_Net_Systems'21}
	\BIBentryALTinterwordspacing
	E.~Uysal, O.~Kaya, A.~Ephremides, J.~Gross, M.~Codreanu, P.~Popovski,
	M.~Assaad, G.~Liva, A.~Munari, T.~Soleymani, B.~Soret, and K.~H. Johansson,
	``Semantic communications in networked systems: A data significance
	perspective,'' 2021. [Online]. Available:
	\url{https://arxiv.org/pdf/2103.05391}
	\BIBentrySTDinterwordspacing
	
	\bibitem{Zhang_Goal-Oriented_Commun'22}
	\BIBentryALTinterwordspacing
	C.~Zhang, H.~Zou, S.~Lasaulce, W.~Saad, M.~Kountouris, and M.~Bennis,
	``Goal-oriented communications for the {IoT} and application to data
	compression,'' 2022. [Online]. Available:
	\url{https://arxiv.org/pdf/2211.05378.pdf}
	\BIBentrySTDinterwordspacing
	
	\bibitem{Sem_Empowered_Commun'22}
	\BIBentryALTinterwordspacing
	Z.~Lu, R.~Li, K.~Lu, X.~Chen, E.~Hossain, Z.~Zhao, and H.~Zhang,
	``Semantics-empowered communication: A tutorial-cum-survey,'' 2022. [Online].
	Available: \url{https://arxiv.org/pdf/2212.08487.pdf}
	\BIBentrySTDinterwordspacing
	
	\bibitem{Xie_Robust_IB'22}
	\BIBentryALTinterwordspacing
	S.~Xie, Y.~Wu, S.~Ma, M.~Ding, Y.~Shi, and M.~Tang, ``Robust information
	bottleneck for task-oriented communication with digital modulation,'' 2022.
	[Online]. Available: \url{https://arxiv.org/pdf/2209.10382.pdf}
	\BIBentrySTDinterwordspacing
	
	\bibitem{Chaccour_Building_NG_SemCom_Networks'22}
	\BIBentryALTinterwordspacing
	C.~Chaccour, W.~Saad, M.~Debbah, Z.~Han, and H.~V. Poor, ``Less data, more
	knowledge: Building next generation semantic communication networks,'' 2022.
	[Online]. Available: \url{https://arxiv.org/pdf/2211.14343.pdf}
	\BIBentrySTDinterwordspacing
	
	\bibitem{Qin_Sem_Com_Principles_Apps'22}
	\BIBentryALTinterwordspacing
	Z.~Qin, X.~Tao, J.~Lu, W.~Tong, and G.~Y. Li, ``Semantic communications:
	Principles and challenges,'' 2022. [Online]. Available:
	\url{https://arxiv.org/pdf/2201.01389v5.pdf}
	\BIBentrySTDinterwordspacing
	
	\bibitem{Luo_SemCom_Overview'22}
	X.~Luo, H.-H. Chen, and Q.~Guo, ``Semantic communications: Overview, open
	issues, and future research directions,'' \emph{IEEE Wirel. Commun.},
	vol.~29, no.~1, pp. 210--219, 2022.
	
	\bibitem{Rethinking_modern_com_Lu_2022}
	K.~Lu, Q.~Zhou, R.~Li, Z.~Zhao, X.~Chen, J.~Wu, and H.~Zhang, ``Rethinking
	modern communication from semantic coding to semantic communication,''
	\emph{IEEE Wirel. Commun.}, pp. 1--13, 2022.
	
	\bibitem{Niu_Towards_SemCom'22}
	\BIBentryALTinterwordspacing
	K.~Niu, J.~Dai, S.~Yao, S.~Wang, Z.~Si, X.~Qin, and P.~Zhang, ``Towards
	semantic communications: A paradigm shift,'' 2022. [Online]. Available:
	\url{https://arxiv.org/pdf/2203.06692.pdf}
	\BIBentrySTDinterwordspacing
	
	\bibitem{SemCom_for_6G_Future_Internet'22}
	\BIBentryALTinterwordspacing
	W.~Yang, H.~Du, Z.~Liew, W.~Y.~B. Lim, Z.~Xiong, D.~Niyato, X.~Chi, X.~S. Shen,
	and C.~Miao, ``Semantic communications for {6G} future internet:
	Fundamentals, applications, and challenges,'' 2022. [Online]. Available:
	\url{https://arxiv.org/pdf/2207.00427.pdf}
	\BIBentrySTDinterwordspacing
	
	\bibitem{Engineering_SemCom'22}
	\BIBentryALTinterwordspacing
	D.~Wheeler and B.~Natarajan, ``Engineering semantic communication: A survey,''
	2022. [Online]. Available: \url{https://arxiv.org/pdf/2208.06314.pdf}
	\BIBentrySTDinterwordspacing
	
	\bibitem{Zhang_a_New_Paradigm'22}
	Y.~Zhang, F.~Wang, W.~Xu, and C.~Liu, ``Semantic communications: A new paradigm
	for networked intelligence,'' in \emph{Proc. MLSP}, 2022, pp. 1--6.
	
	\bibitem{Jiang_Wireless_Semantic_Transmission'22}
	\BIBentryALTinterwordspacing
	P.~Jiang, C.-K. Wen, S.~Jin, and G.~Y. Li, ``Wireless semantic transmission via
	revising modules in conventional communications,'' 2022. [Online]. Available:
	\url{https://arxiv.org/pdf/2210.00473.pdf}
	\BIBentrySTDinterwordspacing
	
	\bibitem{Van_Meter_QNetworking'14}
	R.~Van~Meter, \emph{Quantum Networking}.\hskip 1em plus 0.5em minus 0.4em\relax
	Hoboken, NJ, USA: Wiley, 2014.
	
	\bibitem{Nielsen_Chuang_QC'10}
	{M. A. Nielsen and I. L. Chuang}, \emph{Quantum Computation and Quantum
		Information}, {10th Anniversary}~ed.\hskip 1em plus 0.5em minus 0.4em\relax
	New York, NY, USA: Cambridge Univ. Press, 2010.
	
	\bibitem{Wilde_QIT'2017}
	M.~M. Wilde, \emph{Quantum Information Theory}, 2nd~ed.\hskip 1em plus 0.5em
	minus 0.4em\relax Cambridge, UK: Cambridge Univ. Press, 2017.
	
	\bibitem{Quantum_Mech_for_Scien_Engineers'08}
	D.~A.~B. Miller, \emph{Quantum Mechanics for Scientists and Engineers}.\hskip
	1em plus 0.5em minus 0.4em\relax New York, NY, USA: Cambridge Univ. Press,
	2008.
	
	\bibitem{Wang_Qudits_HD_QC_2020}
	Y.~Wang, Z.~Hu, B.~C. Sanders, and S.~Kais, ``Qudits and high-dimensional
	quantum computing,'' \emph{Front. Phys.}, vol.~8, Nov. 2020.
	
	\bibitem{Cozzolino_HD_QC'19}
	D.~Cozzolino, B.~D. Lio, D.~Bacco, and L.~K. Oxenl{\o}we, ``High-dimensional
	quantum communication: Benefits, progress, and future challenges,''
	\emph{Adv. Quantum Technol.}, vol.~2, no.~12, p. 1900038, Oct. 2019.
	
	\bibitem{Wootters_Q_no_cloning'82}
	W.~K. Wootters and W.~H. Zurek, ``A single quantum cannot be cloned,''
	\emph{Nature}, vol. 299, pp. 802--803, 1982.
	
	\bibitem{Chehimi_Quantum_SemCom'22}
	\BIBentryALTinterwordspacing
	M.~Chehimi, C.~Chaccour, and W.~Saad, ``Quantum semantic communications: An
	unexplored avenue for contextual networking,'' 2022. [Online]. Available:
	\url{https://arxiv.org/pdf/2205.02422.pdf}
	\BIBentrySTDinterwordspacing
	
	\bibitem{Tong_Zhu__6G'21}
	{W. Tong and P. Zhu (Eds.)}, \emph{{6G}: The Next Horizon: From Connected
		People and Things to Connected Intelligence}.\hskip 1em plus 0.5em minus
	0.4em\relax Cambridge, UK: Cambridge Univ. Press, 2021.
	
	\bibitem{Russel_AI_Book_18}
	{S. Russel and P. Norvig}, \emph{Artificial Intelligence: A Modern Approach},
	3rd~ed.\hskip 1em plus 0.5em minus 0.4em\relax Englewood Cliffs, NJ, USA:
	Prentice Hall, 2018.
	
	\bibitem{Rebooting_AI'19}
	G.~Marcus and E.~Davis, \emph{Rebooting {AI}: Building Artificial Intelligence
		We Can Trust}.\hskip 1em plus 0.5em minus 0.4em\relax USA: New York, NY, USA,
	Pantheon Books, 2019.
	
	\bibitem{Dietterich_Toward_Robust_AI'17}
	T.~G. Dietterich, ``Steps toward robust artificial intelligence,'' \emph{AI
		Mag.}, vol.~38, pp. 3--24, 2017.
	
	\bibitem{Jordan_ML_Science'15}
	M.~Jordan and T.~Mitchell, ``Machine learning: Trends, perspectives, and
	prospects,'' \emph{Science}, vol. 349, pp. 255--60, Jul. 2015.
	
	\bibitem{MUWCM19}
	\BIBentryALTinterwordspacing
	M.~{Chen}, U.~{Challita}, W.~{Saad}, C.~{Yin}, and M.~{Debbah}, ``Artificial
	neural networks-based machine learning for wireless networks: A tutorial.''
	[Online]. Available: \url{https://arxiv.org/pdf/1710.02913.pdf}
	\BIBentrySTDinterwordspacing
	
	\bibitem{Ghahramani'2015_Probabilistic_ML}
	Z.~Ghahramani, ``Probabilistic machine learning and artificial intelligence,''
	\emph{Nature}, vol. 521, pp. 452--459, 2015.
	
	\bibitem{YYBGH_15}
	{Y. LeCun, Y. Bengio, and G. Hinton}, ``Deep learning,'' \emph{Nature}, vol.
	521, no. 436, pp. 436--444, 2015.
	
	\bibitem{DL_Revolution'18}
	{T. J. Sejnowski}, \emph{The Deep Learning Revolution}.\hskip 1em plus 0.5em
	minus 0.4em\relax Cambridge, MA, USA: The MIT Press, 2018.
	
	\bibitem{IGYAC16}
	I.~Goodfellow, Y.~Bengio, and A.~Courville, \emph{Deep Learning}.\hskip 1em
	plus 0.5em minus 0.4em\relax MIT Press, 2016.
	
	\bibitem{Scherer_Math_for_QC'19}
	W.~Scherer, \emph{Mathematics of Quantum Computing: An Introduction}.\hskip 1em
	plus 0.5em minus 0.4em\relax Cham, Switzerland: Springer, 2019.
	
	\bibitem{Preskill2018quantumcomputingin}
	J.~Preskill, ``Quantum computing in the {NISQ} era and beyond,''
	\emph{{Quantum}}, vol.~2, p.~79, Aug. 2018.
	
	\bibitem{Gisin_QCommun'07}
	N.~Gisin and R.~Thew, ``Quantum communication,'' \emph{Nat. Photon.}, vol.~1,
	no.~3, pp. 165--171, Mar. 2007.
	
	\bibitem{Imre_Advanced_QC'12}
	S.~Imre and L.~Gyongyosi, \emph{Advanced Quantum Communications: An Engineering
		Approach}.\hskip 1em plus 0.5em minus 0.4em\relax Hoboken, NJ, USA:
	Wiley-IEEE Press, 2012.
	
	\bibitem{Cariolaro2015QuantumC}
	G.~L. Cariolaro, \emph{Quantum Communications}.\hskip 1em plus 0.5em minus
	0.4em\relax Cham, Switzerland: Springer, 2015.
	
	\bibitem{Bassoli_QCNs'2021}
	R.~Bassoli, H.~Boche, C.~Deppe, R.~Ferrara, F.~H.~P. Fitzek, G.~Janssen, and
	S.~Saeedinaeeni, \emph{Quantum Communication Networks}, Cham, Switzerland,
	2021.
	
	\bibitem{Djordjevic_QC_QN_and_QS'22}
	I.~Djordjevic, \emph{Quantum Communication, Quantum Networks, and Quantum
		Sensing}.\hskip 1em plus 0.5em minus 0.4em\relax Cambridge, MA, USA:
	Elsevier, Jan. 2022.
	
	\bibitem{Universal_transformers'19}
	{M. Dehghani, S Gouws, O. Vinyals, J. Uszkoreit, and L. Kaiser}, ``Universal
	transformers,'' in \emph{Proc. Int. Conf. Learn. Represent. (ICLR)}, 2019.
	
	\bibitem{Liu_Swin_Transformer'21}
	Z.~Liu, Y.~Lin, Y.~Cao, H.~Hu, Y.~Wei, Z.~Zhang, S.~Lin, and B.~Guo, ``Swin
	transformer: Hierarchical vision transformer using shifted windows,'' in
	\emph{Proc. ICCV}, 2021, pp. 9992--10\,002.
	
	\bibitem{Wang_Transformer_empowered_6G'22}
	\BIBentryALTinterwordspacing
	Y.~Wang, Z.~Gao, D.~Zheng, S.~Chen, D.~Gündüz, and H.~V. Poor,
	``Transformer-empowered {6G} intelligent networks: From massive {MIMO}
	processing to semantic communication,'' 2022. [Online]. Available:
	\url{https://arxiv.org/pdf/2205.03770.pdf}
	\BIBentrySTDinterwordspacing
	
	\bibitem{Hu_Robust_SemCom_with_Masked_VQ-VAE'22}
	\BIBentryALTinterwordspacing
	Q.~Hu, G.~Zhang, Z.~Qin, Y.~Cai, G.~Yu, and G.~Y. Li, ``Robust semantic
	communications with masked {VQ-VAE} enabled codebook,'' 2022. [Online].
	Available: \url{https://arxiv.org/pdf/2206.04011.pdf}
	\BIBentrySTDinterwordspacing
	
	\bibitem{SemCom_Game'18}
	B.~Güler, A.~Yener, and A.~Swami, ``The semantic communication game,''
	\emph{IEEE Trans. Cogn. Commun. Netw.}, vol.~4, no.~4, pp. 787--802, 2018.
	
	\bibitem{Farsad_DL_JSCC'18}
	N.~Farsad, M.~Rao, and A.~Goldsmith, ``Deep learning for joint source-channel
	coding of text,'' in \emph{Proc. IEEE ICASSP}, 2018, pp. 2326--2330.
	
	\bibitem{Xie_Lite_distributed_SemCom'21}
	H.~Xie and Z.~Qin, ``A lite distributed semantic communication system for
	internet of things,'' \emph{IEEE J. Sel. Areas Commun.}, vol.~39, no.~1, pp.
	142--153, 2021.
	
	\bibitem{Peng_Robust_DL-Based_SemCom'22}
	\BIBentryALTinterwordspacing
	X.~Peng, Z.~Qin, D.~Huang, X.~Tao, J.~Lu, G.~Liu, and C.~Pan, ``A robust deep
	learning enabled semantic communication system for text,'' 2022. [Online].
	Available: \url{https://arxiv.org/abs/2206.02596}
	\BIBentrySTDinterwordspacing
	
	\bibitem{Yao_Semantic_Coding'22}
	S.~Yao, K.~Niu, S.~Wang, and J.~Dai, ``Semantic coding for text transmission:
	An iterative design,'' \emph{IEEE Trans. Cogn. Commun. Netw.}, pp. 1--1,
	2022.
	
	\bibitem{Lu_RL-powered_SemCom'21}
	\BIBentryALTinterwordspacing
	K.~Lu, R.~Li, X.~Chen, Z.~Zhao, and H.~Zhang, ``Reinforcement learning-powered
	semantic communication via semantic similarity,'' 2021. [Online]. Available:
	\url{https://arxiv.org/pdf/2108.12121.pdf}
	\BIBentrySTDinterwordspacing
	
	\bibitem{Luo_SemCom_with_relay'21}
	\BIBentryALTinterwordspacing
	X.~Luo, Z.~Chen, B.~Xia, and J.~Wang, ``Autoencoder-based semantic
	communication systems with relay channels,'' 2021. [Online]. Available:
	\url{https://arxiv.org/pdf/2111.10083.pdf}
	\BIBentrySTDinterwordspacing
	
	\bibitem{Jiang_Deep_Source-Channel_coding'22}
	P.~Jiang, C.-K. Wen, S.~Jin, and G.~Y. Li, ``Deep source-channel coding for
	sentence semantic transmission with {HARQ},'' \emph{IEEE Trans. Commun.},
	vol.~70, pp. 5225--5240, 2022.
	
	\bibitem{Liu_Context-Based_SemCom'22}
	Y.~Liu, S.~Jiang, Y.~Zhang, K.~Cao, L.~Zhou, B.-C. Seet, H.~Zhao, and J.~Wei,
	``Extended context-based semantic communication system for text
	transmission,'' \emph{Digit. Commun. Netw.}, Oct. 2022.
	
	\bibitem{Weng_SemCom_Sys_Speech_Trans'21}
	Z.~Weng and Z.~Qin, ``Semantic communication systems for speech transmission,''
	\emph{IEEE J. Sel. Areas Commun.}, vol.~39, no.~8, pp. 2434--2444, 2021.
	
	\bibitem{SemCom_for_speech_signals'20}
	\BIBentryALTinterwordspacing
	Z.~Weng, Z.~Qin, and G.~Y. Li, ``Semantic communications for speech signals,''
	2020. [Online]. Available: \url{https://arxiv.org/pdf/2012.05369.pdf}
	\BIBentrySTDinterwordspacing
	
	\bibitem{Weng_SemCom_Speech_Recognition'21}
	\BIBentryALTinterwordspacing
	{Z. Weng, Z. Qin, and G. Y. Li}, ``Semantic communications for speech
	recognition,'' 2021. [Online]. Available:
	\url{https://arxiv.org/pdf/2107.11190.pdf}
	\BIBentrySTDinterwordspacing
	
	\bibitem{Han_Semantic-aware_Speech2Text_Transmission'22}
	\BIBentryALTinterwordspacing
	T.~Han, Q.~Yang, Z.~Shi, S.~He, and Z.~Zhang, ``Semantic-aware speech to text
	transmission with redundancy removal,'' 2022. [Online]. Available:
	\url{https://arxiv.org/pdf/2202.03211.pdf}
	\BIBentrySTDinterwordspacing
	
	\bibitem{Weng_DL-enabled_SemCom'22}
	\BIBentryALTinterwordspacing
	Z.~Weng, Z.~Qin, X.~Tao, C.~Pan, G.~Liu, and G.~Y. Li, ``Deep learning enabled
	semantic communications with speech recognition and synthesis,'' 2022.
	[Online]. Available: \url{https://arxiv.org/pdf/2205.04603.pdf}
	\BIBentrySTDinterwordspacing
	
	\bibitem{Eirina_JSCC'19}
	E.~Bourtsoulatze, D.~Burth~Kurka, and D.~Gündüz, ``Deep joint source-channel
	coding for wireless image transmission,'' \emph{IEEE Trans. Cogn. Commun.},
	vol.~5, no.~3, pp. 567--579, 2019.
	
	\bibitem{Kurka_Deep_JSCC-f'20}
	D.~B. Kurka and D.~G{\"u}nd{\"u}z, ``{DeepJSCC-f}: Deep joint source-channel
	coding of images with feedback,'' \emph{IEEE J. Sel. Areas Inf. Theory},
	vol.~1, pp. 178--193, 2020.
	
	\bibitem{Bandwidth_Agile_Image_Transmission'21}
	D.~B. Kurka and D.~Gündüz, ``Bandwidth-agile image transmission with deep
	joint source-channel coding,'' \emph{IEEE Trans. Wirel. Commun.}, vol.~20,
	no.~12, pp. 8081--8095, 2021.
	
	\bibitem{Zhang_Wireless_Information_Transmission_of_Image'22}
	\BIBentryALTinterwordspacing
	Z.~Zhang, Q.~Yang, S.~He, M.~Sun, and J.~Chen, ``Wireless transmission of
	images with the assistance of multi-level semantic information,'' 2022.
	[Online]. Available: \url{https://arxiv.org/pdf/2202.04754}
	\BIBentrySTDinterwordspacing
	
	\bibitem{Xu_Wireless_Image_Transmission'22}
	J.~Xu, B.~Ai, W.~Chen, A.~Yang, P.~Sun, and M.~Rodrigues, ``Wireless image
	transmission using deep source channel coding with attention modules,''
	\emph{IEEE Trans. Circuits Syst. Video Technol.}, vol.~32, no.~4, pp.
	2315--2328, 2022.
	
	\bibitem{Pan_IM-SemCom'22}
	\BIBentryALTinterwordspacing
	Q.~Pan, H.~Tong, J.~Lv, T.~Luo, Z.~Zhang, C.~Yin, and J.~Li, ``Image
	segmentation semantic communication over internet of vehicles,'' 2022.
	[Online]. Available: \url{https://arxiv.org/pdf/2210.05321.pdf}
	\BIBentrySTDinterwordspacing
	
	\bibitem{Yang_WITT'22}
	\BIBentryALTinterwordspacing
	K.~Yang, S.~Wang, J.~Dai, K.~Tan, K.~Niu, and P.~Zhang, ``{WITT}: A wireless
	image transmission transformer for semantic communications,'' 2022. [Online].
	Available: \url{https://arxiv.org/pdf/2211.00937.pdf}
	\BIBentrySTDinterwordspacing
	
	\bibitem{Dai_NLT_SCC'22}
	J.~Dai, S.~Wang, K.~Tan, Z.~Si, X.~Qin, K.~Niu, and P.~Zhang, ``Nonlinear
	transform source-channel coding for semantic communications,'' \emph{IEEE J.
		Sel. Areas Commun.}, vol.~40, no.~8, pp. 2300--2316, 2022.
	
	\bibitem{Lee_Joint_Transmission_Recognition_for_IoTs'19}
	C.-H. Lee, J.-W. Lin, P.-H. Chen, and Y.-C. Chang, ``Deep learning-constructed
	joint transmission-recognition for internet of things,'' \emph{IEEE Access},
	vol.~7, pp. 76\,547--76\,561, 2019.
	
	\bibitem{Hu_Robust_SemCom'22}
	\BIBentryALTinterwordspacing
	{Q. Hu, G. Zhang, Guangyi, Z. Qin, Y. Cai, G. Yu, andG . Y. Li}, ``Robust
	semantic communications against semantic noise,'' 2022. [Online]. Available:
	\url{https://arxiv.org/pdf/2202.03338.pdf}
	\BIBentrySTDinterwordspacing
	
	\bibitem{Huang_Toward_SemCom'23}
	D.~Huang, F.~Gao, X.~Tao, Q.~Du, and J.~Lu, ``Toward semantic communications:
	Deep learning-based image semantic coding,'' \emph{IEEE J. Sel. Areas
		Commun.}, vol.~41, no.~1, pp. 55--71, 2023.
	
	\bibitem{Jiang_Wireless_SemCom'22}
	\BIBentryALTinterwordspacing
	P.~Jiang, C.-K. Wen, S.~Jin, and G.~Y. Li, ``Wireless semantic communications
	for video conferencing,'' 2022. [Online]. Available:
	\url{https://arxiv.org/pdf/2204.07790.pdf}
	\BIBentrySTDinterwordspacing
	
	\bibitem{Wang_Wireless_Deep_Video_Transmission}
	\BIBentryALTinterwordspacing
	S.~Wang, J.~Dai, Z.~Liang, K.~Niu, Z.~Si, C.~Dong, X.~Qin, and P.~Zhang,
	``Wireless deep video semantic transmission,'' 2022. [Online]. Available:
	\url{https://arxiv.org/pdf/2205.13129.pdf}
	\BIBentrySTDinterwordspacing
	
	\bibitem{Tung_DeepWiVe'21}
	\BIBentryALTinterwordspacing
	T.-Y. Tung and D.~Gündüz, ``{DeepWiVe}: Deep-learning-aided wireless video
	transmission,'' 2021. [Online]. Available:
	\url{https://arxiv.org/pdf/2111.13034.pdf}
	\BIBentrySTDinterwordspacing
	
	\bibitem{Huang_IS-SemCom'22}
	\BIBentryALTinterwordspacing
	Y.~Huang, B.~Bai, Y.~Zhu, X.~Qiao, X.~Su, and P.~Zhang, ``Iscom: Interest-aware
	semantic communication scheme for point cloud video streaming,'' 2022.
	[Online]. Available: \url{https://arxiv.org/pdf/2210.06808.pdf}
	\BIBentrySTDinterwordspacing
	
	\bibitem{WANG_Multimodal_SemCom'23}
	C.~Wang, X.~Yu, L.~Xu, Z.~Wang, and W.~Wang, ``Multimodal semantic
	communication accelerated bidirectional caching for {6G MEC},'' \emph{Future
		Gener. Comput. Syst.}, vol. 140, pp. 225--237, 2023.
	
	\bibitem{Li_Cross-Modal_SemCom'22}
	A.~Li, X.~Wei, D.~Wu, and L.~Zhou, ``Cross-modal semantic communications,''
	\emph{IEEE Wirel. Commun.}, pp. 1--8, 2022.
	
	\bibitem{Cognitive_SemCom_Systems'22}
	\BIBentryALTinterwordspacing
	F.~Zhou, Y.~Li, X.~Zhang, Q.~Wu, X.~Lei, and R.~Q. Hu, ``Cognitive semantic
	communication systems driven by knowledge graph,'' 2022. [Online]. Available:
	\url{https://arxiv.org/abs/2202.11958}
	\BIBentrySTDinterwordspacing
	
	\bibitem{Xiao_Reasoning_on_the_Air'22}
	\BIBentryALTinterwordspacing
	Y.~Xiao, Y.~Li, G.~Shi, and H.~V. Poor, ``Reasoning on the air: An implicit
	semantic communication architecture,'' 2022. [Online]. Available:
	\url{https://arxiv.org/pdf/2202.01950.pdf}
	\BIBentrySTDinterwordspacing
	
	\bibitem{Dai_Adaptive_SemCom'22}
	\BIBentryALTinterwordspacing
	J.~Dai, S.~Wang, K.~Yang, K.~Tan, X.~Qin, Z.~Si, K.~Niu, and P.~Zhang,
	``Adaptive semantic communications: Overfitting the source and channel for
	profit,'' 2022. [Online]. Available:
	\url{https://arxiv.org/pdf/2211.04339.pdf}
	\BIBentrySTDinterwordspacing
	
	\bibitem{zhang_Context-Based_SemCom'22}
	Y.~Zhang, H.~Zhao, J.~Wei, J.~Zhang, M.~F. Flanagan, and J.~Xiong,
	``Context-based semantic communication via dynamic programming,'' \emph{IEEE
		Trans. Cogn. Commun. Netw.}, pp. 1--1, 2022.
	
	\bibitem{Learning_Based_Digital_SemCom'22}
	\BIBentryALTinterwordspacing
	Y.~Bo, Y.~Duan, S.~Shao, and M.~Tao, ``Learning based joint coding-modulation
	for digital semantic communication systems,'' 2022. [Online]. Available:
	\url{https://arxiv.org/pdf/2208.05704.pdf}
	\BIBentrySTDinterwordspacing
	
	\bibitem{Fu_VQ_SemCom'22}
	\BIBentryALTinterwordspacing
	Q.~Fu, H.~Xie, Z.~Qin, G.~Slabaugh, and X.~Tao, ``Vector quantized semantic
	communication system,'' 2022. [Online]. Available:
	\url{https://arxiv.org/pdf/2209.11519.pdf}
	\BIBentrySTDinterwordspacing
	
	\bibitem{SemCom_with_Conceptual_Spaces'22}
	\BIBentryALTinterwordspacing
	D.~Wheeler, E.~E. Tripp, and B.~Natarajan, ``Semantic communication with
	conceptual spaces,'' 2022. [Online]. Available:
	\url{https://arxiv.org/pdf/2210.01629.pdf}
	\BIBentrySTDinterwordspacing
	
	\bibitem{Du_RIS-aided_Encoding'22}
	\BIBentryALTinterwordspacing
	H.~Du, J.~Wang, D.~Niyato, J.~Kang, Z.~Xiong, J.~Zhang, Xuemin, and {Shen},
	``Semantic communications for wireless sensing: {RIS}-aided encoding and
	self-supervised decoding,'' 2022. [Online]. Available:
	\url{https://arxiv.org/pdf/2211.12727.pdf}
	\BIBentrySTDinterwordspacing
	
	\bibitem{Hu_One-to-Many_SemCom'22}
	H.~Hu, X.~Zhu, F.~Zhou, W.~Wu, R.~Q. Hu, and H.~Zhu, ``One-to-many semantic
	communication systems: Design, implementation, performance evaluation,''
	\emph{IEEE Commun. Lett.}, vol.~26, no.~12, pp. 2959--2963, 2022.
	
	\bibitem{Xu_SemCom_for_IoV'22}
	\BIBentryALTinterwordspacing
	W.~Xu, Y.~Zhang, F.~Wang, Z.~Qin, C.~Liu, and P.~Zhang, ``Semantic
	communication for internet of vehicles: A multi-user cooperative approach,''
	2022. [Online]. Available: \url{https://arxiv.org/pdf/2212.03037.pdf}
	\BIBentrySTDinterwordspacing
	
	\bibitem{Xiao_RD_Theory_Strategic_SemCom'22}
	\BIBentryALTinterwordspacing
	Y.~Xiao, X.~Zhang, Y.~Li, and G.~Shi, ``Rate-distortion theory for strategic
	semantic communication,'' 2022. [Online]. Available:
	\url{https://arxiv.org/pdf/2202.03711.pdf}
	\BIBentrySTDinterwordspacing
	
	\bibitem{Luo_Encrypted_SemCom'22}
	\BIBentryALTinterwordspacing
	X.~Luo, Z.~Chen, M.~Tao, and F.~Yang, ``Encrypted semantic communication using
	adversarial training for privacy preserving,'' 2022. [Online]. Available:
	\url{https://arxiv.org/pdf/2209.09008.pdf}
	\BIBentrySTDinterwordspacing
	
	\bibitem{Qiao_What_is_SemCom'21}
	Q.~Lan, D.~Wen, Z.~Zhang, Q.~Zeng, X.~Chen, P.~Popovski, and K.~Huang, ``What
	is semantic communication? a view on conveying meaning in the era of machine
	intelligence,'' \emph{J. Commun. Inf. Netw.}, vol.~6, no.~4, pp. 336--371,
	Dec. 2021.
	
	\bibitem{Yang_SemCom_meets_Edge_Intelligence'22}
	\BIBentryALTinterwordspacing
	W.~Yang, Z.~Q. Liew, W.~Y.~B. Lim, Z.~Xiong, D.~Niyato, X.~Chi, X.~Cao, and
	K.~B. Letaief, ``Semantic communication meets edge intelligence,'' 2022.
	[Online]. Available: \url{https://arxiv.org/pdf/2202.06471.pdf}
	\BIBentrySTDinterwordspacing
	
	\bibitem{Strinati_Beyond_Shannon'20}
	\BIBentryALTinterwordspacing
	E.~C. Strinati and S.~Barbarossa, ``{6G} networks: Beyond {S}hannon towards
	semantic and goal-oriented communications,'' 2020. [Online]. Available:
	\url{https://arxiv.org/pdf/2011.14844.pdf}
	\BIBentrySTDinterwordspacing
	
	\bibitem{Zhang_Wisdom_Evolutionary_6G'21}
	P.~Zhang, W.~Xu, H.~Gao, K.~Niu, X.~Xu, X.~Qin, C.~Yuan, Z.~Qin, H.~Zhao,
	J.~Wei, and F.~Zhang, ``Toward wisdom-evolutionary and primitive-concise
	{6G}: A new paradigm of semantic communication networks,''
	\emph{Engineering}, vol.~8, Nov. 2021.
	
	\bibitem{Beck_SemCom_Info_Bottleneck_View'22}
	\BIBentryALTinterwordspacing
	E.~Beck, C.~Bockelmann, and A.~Dekorsy, ``Semantic communication: An
	information bottleneck view,'' 2022. [Online]. Available:
	\url{https://arxiv.org/pdf/2204.13366.pdf}
	\BIBentrySTDinterwordspacing
	
	\bibitem{Phopovski_SE_Filtering'19}
	\BIBentryALTinterwordspacing
	P.~Popovski, O.~Simeone, F.~Boccardi, D.~Gunduz, and O.~Sahin,
	``Semantic-effectiveness filtering and control for post-{5G} wireless
	connectivity,'' 2019. [Online]. Available:
	\url{https://arxiv.org/pdf/1907.02441.pdf}
	\BIBentrySTDinterwordspacing
	
	\bibitem{Shi_to_Semantic_Fidelity'21}
	\BIBentryALTinterwordspacing
	G.~Shi, D.~Gao, X.~Song, J.~Chai, M.~Yang, X.~Xie, L.~Li, and X.~Li, ``A new
	communication paradigm: from bit accuracy to semantic fidelity,'' 2021.
	[Online]. Available: \url{https://arxiv.org/pdf/2101.12649.pdf}
	\BIBentrySTDinterwordspacing
	
	\bibitem{Shi_From_SemCom_to_Sematic-aware_Networking'20}
	\BIBentryALTinterwordspacing
	G.~Shi, Y.~Xiao, Y.~Li, and X.~Xie, ``From semantic communication to
	semantic-aware networking: Model, architecture, and open problems,'' 2020.
	[Online]. Available: \url{https://arxiv.org/pdf/2012.15405.pdf}
	\BIBentrySTDinterwordspacing
	
	\bibitem{Commun_Beyond_Transmitting_Bits'22}
	\BIBentryALTinterwordspacing
	J.~Dai, P.~Zhang, K.~Niu, S.~Wang, Z.~Si, and X.~Qin, ``Communication beyond
	transmitting bits: Semantics-guided source and channel coding,'' 2022.
	[Online]. Available: \url{https://arxiv.org/pdf/2208.02481.pdf}
	\BIBentrySTDinterwordspacing
	
	\bibitem{Dong_Semantic_Cognitive_Intell'22}
	\BIBentryALTinterwordspacing
	P.~Dong, Q.~Wu, X.~Zhang, and G.~Ding, ``Edge semantic cognitive intelligence
	for {6G} networks: Novel theoretical models, enabling framework, and typical
	applications,'' 2022. [Online]. Available:
	\url{https://arxiv.org/pdf/2205.12073v2.pdf}
	\BIBentrySTDinterwordspacing
	
	\bibitem{Zhao_Semantic-Native_Communication'22}
	\BIBentryALTinterwordspacing
	Q.~Zhao, M.~Bennis, M.~Debbah, and D.~B. da~Costa, ``Semantic-native
	communication: A simplicial complex perspective,'' 2022. [Online]. Available:
	\url{https://arxiv.org/pdf/2210.16970.pdf}
	\BIBentrySTDinterwordspacing
	
	\bibitem{Seo_SemCom_Protocols'22}
	\BIBentryALTinterwordspacing
	S.~Seo, J.~Park, S.-W. Ko, J.~Choi, M.~Bennis, and S.-L. Kim, ``Towards
	semantic communication protocols: A probabilistic logic perspective,'' 2022.
	[Online]. Available: \url{https://arxiv.org/pdf/2207.03920.pdf}
	\BIBentrySTDinterwordspacing
	
	\bibitem{Pokhrel_UBT_SemCom'22}
	S.~R. Pokhrel and J.~Choi, ``Understand-before-talk ({UBT}): A semantic
	communication approach to {6G} networks,'' \emph{IEEE Trans. Veh. Technol.},
	pp. 1--13, 2022.
	
	\bibitem{Yu_Optical_SemCom'22}
	\BIBentryALTinterwordspacing
	Z.~Yu, H.~Huang, L.~Cheng, W.~Zhang, Y.~Mu, and K.~Xu, ``Semantic optical fiber
	communication system,'' 2022. [Online]. Available:
	\url{https://arxiv.org/pdf/2212.14739}
	\BIBentrySTDinterwordspacing
	
	\bibitem{JBPPNWS_17}
	J.~Biamonte, P.~Wittek, N.~Pancotti, P.~Rebentrost, N.~Wiebe, and S.~Lloyd,
	``\BIBforeignlanguage{English (US)}{Quantum machine learning},''
	\emph{\BIBforeignlanguage{English (US)}{Nature}}, vol. 549, no. 7671, pp.
	195--202, Sep. 2017.
	
	\bibitem{SSSMM19}
	S.~J. Nawaz, S.~K. Sharma, S.~Wyne, M.~N. Patwary, and M.~Asaduzzaman,
	``Quantum machine learning for {6G} communication networks: State-of-the-art
	and vision for the future,'' \emph{IEEE Access}, vol.~7, pp.
	46\,317--46\,350, Apr. 2019.
	
	\bibitem{NW_QDL_2019}
	\BIBentryALTinterwordspacing
	{N. Wiebe, A. Kapoor, and K. M. Svore}, ``Quantum deep learning,'' 22 May 2015.
	[Online]. Available: \url{https://arxiv.org/pdf/1412.3489.pdf}
	\BIBentrySTDinterwordspacing
	
	\bibitem{Schuld_PRL'2019}
	M.~Schuld and N.~Killoran, ``Quantum machine learning in feature {H}ilbert
	spaces,'' \emph{Phys. Rev. Lett.}, vol. 122, no.~4, Feb. 2019.
	
	\bibitem{Raussendorf_PRL_one-way_QC'01}
	R.~Raussendorf and H.~J. Briegel, ``A one-way quantum computer,'' \emph{Phys.
		Rev. Lett.}, vol.~86, pp. 5188--5191, May 2001.
	
	\bibitem{Adiabatic_QC_SIAM'07}
	D.~Aharonov, W.~{Van Dam}, J.~Kempe, Z.~Landau, S.~Lloyd, and O.~Regev,
	``Adiabatic quantum computation is equivalent to standard quantum
	computation,'' \emph{SIAM J. Comput.}, vol.~37, no.~1, pp. 166--194, 2007.
	
	\bibitem{Freedman_Topological'02}
	M.~H. Freedman, M.~Larsen, and Z.~Wang, ``A modular functor which is universal
	for quantum computation,'' \emph{Commun. Math. Phys.}, vol. 227, pp.
	605--622, 2002.
	
	\bibitem{Chaccour_Fellowshop_Com_Sensing'22}
	C.~Chaccour, M.~N. Soorki, W.~Saad, M.~Bennis, P.~Popovski, and M.~Debbah,
	``Seven defining features of terahertz ({THz}) wireless systems: A fellowship
	of communication and sensing,'' \emph{IEEE Commun. Surv. Tutor.}, vol.~24,
	no.~2, pp. 967--993, 2022.
	
	\bibitem{Pan_EntanglementPF_01}
	J.-W. Pan, C.~Simon, \v{C}aslav Brukner, and A.~Zeilinger, ``Entanglement
	purification for quantum communication,'' \emph{Nature}, vol. 410, no. 6832,
	pp. 1067--1070, Apr. 2001.
	
	\bibitem{Kaewpuang_Cooperative_Resource_Management'22}
	\BIBentryALTinterwordspacing
	R.~Kaewpuang, M.~Xu, W.~Y.~B. Lim, D.~Niyato, H.~Yu, J.~Kang, and X.~S. Shen,
	``Cooperative resource management in quantum key distribution ({QKD})
	networks for semantic communication,'' 2022. [Online]. Available:
	\url{https://arxiv.org/pdf/2209.11957.pdf}
	\BIBentrySTDinterwordspacing
	
	\bibitem{Cao_QKD_Networks_Survey'22}
	Y.~Cao, Y.~Zhao, Q.~Wang, J.~Zhang, S.~X. Ng, and L.~Hanzo, ``The evolution of
	quantum key distribution networks: On the road to the {Qinternet},''
	\emph{IEEE Commun. Surv. Tutor.}, vol.~24, no.~2, pp. 839--894, 2022.
	
	\bibitem{Kountouris_Semantics_EmpoweredCF'21}
	M.~Kountouris and N.~Pappas, ``Semantics-empowered communication for networked
	intelligent systems,'' \emph{IEEE Commun. Mag.}, vol.~59, pp. 96--102, 2021.
	
	\bibitem{Juba_Universal_SemCom'08}
	B.~A. Juba and M.~Sudan, ``Universal semantic communication {II}: A theory of
	goal-oriented communication,'' \emph{Electron. Colloquium Comput. Complex.},
	vol.~15, 2008.
	
	\bibitem{Yang_SemCom_with_AI_Tasks'21}
	\BIBentryALTinterwordspacing
	Y.~Yang, C.~Guo, F.~Liu, C.~Liu, L.~Sun, Q.~Sun, and J.~Chen, ``Semantic
	communications with {AI} tasks,'' 2021. [Online]. Available:
	\url{https://arxiv.org/pdf/2109.14170.pdf}
	\BIBentrySTDinterwordspacing
	
	\bibitem{Thomas_NeuroSymb_AI_SemCom'22}
	\BIBentryALTinterwordspacing
	{C. K. Thomas and W. Saad}, ``Neuro-symbolic artificial intelligence ({AI}) for
	intent based semantic communication,'' 2022. [Online]. Available:
	\url{https://arxiv.org/abs/2205.10768}
	\BIBentrySTDinterwordspacing
	
	\bibitem{Thomas_Neuro-Symbolic_Causal_Reasoning'22}
	\BIBentryALTinterwordspacing
	C.~K. Thomas and W.~Saad, ``Neuro-symbolic causal reasoning meets signaling
	game for emergent semantic communications,'' 2022. [Online]. Available:
	\url{https://arxiv.org/pdf/2210.12040.pdf}
	\BIBentrySTDinterwordspacing
	
	\bibitem{Xie_Task-Oriented_MU-SemCom'22}
	H.~Xie, Z.~Qin, X.~Tao, and K.~B. Letaief, ``Task-oriented multi-user semantic
	communications,'' \emph{IEEE J. Sel. Areas Commun.}, vol.~40, no.~9, pp.
	2584--2597, 2022.
	
	\bibitem{Tung_Effective_Commun'21}
	T.-Y. Tung, S.~Kobus, J.~P. Roig, and D.~Gündüz, ``Effective communications:
	A joint learning and communication framework for multi-agent reinforcement
	learning over noisy channels,'' \emph{IEEE J. Sel. Areas Commun.}, vol.~39,
	no.~8, pp. 2590--2603, 2021.
	
	\bibitem{Effective_Commun_for_6G'22}
	\BIBentryALTinterwordspacing
	E.~G. Soyak and O.~Ercetin, ``Effective communications for {6G}: Challenges and
	opportunities,'' 2022. [Online]. Available:
	\url{https://arxiv.org/pdf/2203.11695.pdf}
	\BIBentrySTDinterwordspacing
	
	\bibitem{Sana_Learning_Semantics'21}
	\BIBentryALTinterwordspacing
	M.~Sana and E.~C. Strinati, ``Learning semantics: An opportunity for effective
	{6G} communications,'' 2021. [Online]. Available:
	\url{https://arxiv.org/pdf/2110.08049.pdf}
	\BIBentrySTDinterwordspacing
	
	\bibitem{Goal-oriented_SemCom_Thesis}
	M.~Goek, ``Semantic and goal-oriented signal processing: semantic extraction,''
	Master's thesis, Bilkent University, Turkey, Aug. 2022.
	
	\bibitem{Resource_allocation_text_SemCom'22}
	\BIBentryALTinterwordspacing
	L.~Yan, Z.~Qin, R.~Zhang, Y.~Li, and G.~Y. Li, ``Resource allocation for text
	semantic communications,'' 2022. [Online]. Available:
	\url{https://arxiv.org/pdf/2201.06023.pdf}
	\BIBentrySTDinterwordspacing
	
	\bibitem{Peters_BERT_Paper'18}
	M.~E. Peters, M.~Neumann, M.~Iyyer, M.~Gardner, C.~Clark, K.~Lee, and
	L.~Zettlemoyer, ``Deep contextualized word representations,'' in \emph{Proc.
		North Amer. Chapter Assoc. Comput. Linguistics: Hum. Lang. Tech.}, New
	Orleans, LA, USA, 2018, pp. 2227--2237.
	
	\bibitem{Papineni-etal-bleu'02}
	K.~Papineni, S.~Roukos, T.~Ward, and W.-J. Zhu, ``{B}leu: a method for
	automatic evaluation of machine translation,'' in \emph{Proc. Annu. Meeting
		Assoc. Comput. Linguistics}, Jul. 2002, pp. 311--318.
	
	\bibitem{CIDEr_paper'15}
	R.~Vedantam, C.~L. Zitnick, and D.~Parikh, ``{CIDEr}: Consensus-based image
	description evaluation,'' in \emph{Proc. IEEE Conf. Comput. Vis. Pattern
		Recognit. (CVPR)}, 2015, pp. 4566--4575.
	
	\bibitem{Cosine_similarity}
	Wikipedia, ``Cosine similarity,'' [Online].
	https://en.wikipedia.org/wiki/Cosine\_similarity (accessed Oct. 2022).
	
	\bibitem{Jiang_Reliable_SemCom'22}
	S.~Jiang, Y.~Liu, Y.~Zhang, P.~Luo, K.~Cao, J.~Xiong, H.~Zhao, and J.~Wei,
	``Reliable semantic communication system enabled by knowledge graph,''
	\emph{Entropy}, vol.~24, no.~6, 2022.
	
	\bibitem{arXiv_Getu_DeepSC_Performance_Limits'23}
	\BIBentryALTinterwordspacing
	T.~M. Getu, W.~Saad, G.~Kaddoum, and M.~Bennis, ``Performance limits of a deep
	learning-enabled text semantic communication under interference,'' 2023.
	[Online]. Available: \url{https://arxiv.org/pdf/2302.14702.pdf}
	\BIBentrySTDinterwordspacing
	
	\bibitem{Poggio_Theo_Issues_Dnets_2020}
	T.~Poggio, A.~Banburski, and Q.~Liao, ``Theoretical issues in deep networks,''
	\emph{Proc. Natl. Acad. Sci. U.S.A.}, Jun. 2020.
	
	\bibitem{Toward_Science_of_Interpretable_ML'17}
	\BIBentryALTinterwordspacing
	F.~Doshi-Velez and B.~Kim, ``Towards a rigorous science of interpretable
	machine learning,'' 2017. [Online]. Available:
	\url{https://arxiv.org/pdf/1702.08608.pdf}
	\BIBentrySTDinterwordspacing
	
	\bibitem{Li-etal-2020-sentence_embeddings}
	B.~Li, H.~Zhou, J.~He, M.~Wang, Y.~Yang, and L.~Li, ``On the sentence
	embeddings from pre-trained language models,'' in \emph{Proc. Conf. on
		Empirical Methods in Natural Language Processing (EMNLP)}, Nov. 2020, pp.
	9119--9130.
	
	\bibitem{Lee_EQ2SEQ-SC'22}
	\BIBentryALTinterwordspacing
	J.-H. Lee, D.-H. Lee, E.~Sheen, T.~Choi, J.~Pujara, and J.~Kim, ``{Seq2Seq-SC}:
	End-to-end semantic communication systems with pre-trained language model,''
	2022. [Online]. Available: \url{https://arxiv.org/pdf/2210.15237.pdf}
	\BIBentrySTDinterwordspacing
	
	\bibitem{Reimers_SBERT'19}
	\BIBentryALTinterwordspacing
	N.~Reimers and I.~Gurevych, ``{Sentence-BERT}: Sentence embeddings using
	{Siamese BERT}-networks,'' 2019. [Online]. Available:
	\url{https://arxiv.org/pdf/1908.10084.pdf}
	\BIBentrySTDinterwordspacing
	
	\bibitem{METEOR_paper'05}
	S.~Banerjee and A.~Lavie, ``{METEOR}: An automatic metric for {MT} evaluation
	with improved correlation with human judgments,'' in \emph{Proceedings of the
		{ACL} Workshop on Intrinsic and Extrinsic Evaluation Measures for Machine
		Translation and/or Summarization}, Ann Arbor, Michigan, Jun. 2005, pp.
	65--72.
	
	\bibitem{Fellbaum'00_WordNetA}
	C.~D. Fellbaum, ``{WordNet}: an electronic lexical database,'' \emph{Language},
	vol.~76, pp. 706--708, 2000.
	
	\bibitem{Extending_METEOR'10}
	M.~Denkowski and A.~Lavie, ``Extending the {METEOR} machine translation
	evaluation metric to the phrase level,'' in \emph{Proc. Annual Conf. the
		North American Chapter of the Association for Computational Linguistics},
	2010, pp. 250--253.
	
	\bibitem{Chandrasekaran_Semantic_Similarity'22}
	\BIBentryALTinterwordspacing
	D.~Chandrasekaran and V.~Mago, ``Evolution of semantic similarity{\textemdash}a
	survey,'' \emph{{ACM} Comput. Surv.}, vol.~54, no.~2, pp. 1--37, 2022.
	[Online]. Available: \url{https://doi.org/10.1145%2F3440755}
	\BIBentrySTDinterwordspacing
	
	\bibitem{Semantic_Textual_Similarity_Methods'16}
	G.~Majumder, D.~P. Pakray, A.~Gelbukh, and D.~Pinto, ``Semantic textual
	similarity methods, tools, and applications: A survey,'' \emph{Computacion y
		Sistemas}, vol.~20, pp. 647--665, Dec. 2016.
	
	\bibitem{Han_Semantic-preserved_Com_System'22}
	\BIBentryALTinterwordspacing
	T.~Han, Q.~Yang, Z.~Shi, S.~He, and Z.~Zhang, ``Semantic-preserved
	communication system for highly efficient speech transmission,'' 2022.
	[Online]. Available: \url{https://arxiv.org/pdf/2205.12727.pdf}
	\BIBentrySTDinterwordspacing
	
	\bibitem{Vincent_Blind_Source_Separation'06}
	E.~Vincent, R.~Gribonval, and C.~Fevotte, ``Performance measurement in blind
	audio source separation,'' \emph{IEEE Trans. Audio, Speech, Language
		Process.}, vol.~14, no.~4, pp. 1462--1469, 2006.
	
	\bibitem{PESQ_paper'01}
	A.~Rix, J.~Beerends, M.~Hollier, and A.~Hekstra, ``Perceptual evaluation of
	speech quality ({PESQ})-a new method for speech quality assessment of
	telephone networks and codecs,'' in \emph{Proc. IEEE ICASSP}, vol.~2, 2001,
	pp. 749--752 vol.2.
	
	\bibitem{ITU-T_recommendation_P.862'01}
	\BIBentryALTinterwordspacing
	{ITU-T}, ``Perceptual evaluation of speech quality ({PESQ}): An objective
	method for end-to-end speech quality assessment of narrow-band telephone
	networks and speech codecs,'' {ITU-T Recommendation P.862}, Feb. 2001.
	[Online]. Available: \url{https://www.itu.int/rec/T-REC-P.862}
	\BIBentrySTDinterwordspacing
	
	\bibitem{High_Fidelity_Speech_Synthesis'19}
	\BIBentryALTinterwordspacing
	M.~Bińkowski, J.~Donahue, S.~Dieleman, A.~Clark, E.~Elsen, N.~Casagrande,
	L.~C. Cobo, and K.~Simonyan, ``High fidelity speech synthesis with
	adversarial networks,'' 2019. [Online]. Available:
	\url{https://arxiv.org/pdf/1909.11646.pdf}
	\BIBentrySTDinterwordspacing
	
	\bibitem{Liu_VQA_Survey'13}
	T.-J. Liu, J.~Y.-c. Lin, W.~Lin, and C.-C.~J. Kuo, ``Visual quality assessment:
	Recent developments, coding applications and future trends,'' \emph{APSIPA
		Trans. Signal Inf. Process.}, vol.~2, Jan. 2013.
	
	\bibitem{Wang_SSIM_Metric'04}
	Z.~Wang, A.~Bovik, H.~Sheikh, and E.~Simoncelli, ``Image quality assessment:
	from error visibility to structural similarity,'' \emph{IEEE Trans. Image
		Process.}, vol.~13, no.~4, pp. 600--612, 2004.
	
	\bibitem{Li2009_3-SSIM}
	C.~Li and A.~C. Bovik, ``Three-component weighted structural similarity
	index,'' in \emph{Proc. SPIE. Int. Soc. Opt. Eng.}, San Jose, CA, USA, 2009.
	
	\bibitem{Zhang_FSIM'11}
	L.~Zhang, L.~Zhang, X.~Mou, and D.~Zhang, ``{FSIM}: A feature similarity index
	for image quality assessment,'' \emph{IEEE Trans. Image Process.}, vol.~20,
	no.~8, pp. 2378--2386, 2011.
	
	\bibitem{Wang_MS-SSIM'03}
	Z.~Wang, E.~Simoncelli, and A.~Bovik, ``Multi-scale structural similarity for
	image quality assessment,'' in \emph{Proc. IEEE Asilomar Conf. Signals, Syst.
		\& Comput.}, vol.~2, 2003, pp. 1398--1402.
	
	\bibitem{Ding_Comparison'2021}
	\BIBentryALTinterwordspacing
	K.~Ding, K.~Ma, S.~Wang, and E.~P. Simoncelli, ``Comparison of full-reference
	image quality models for optimization of image processing systems,''
	\emph{Int. J. Comput. Vis.}, vol. 129, no.~4, pp. 1258--1281, Jan. 2021.
	[Online]. Available: \url{https://doi.org/10.1007%2Fs11263-020-01419-7}
	\BIBentrySTDinterwordspacing
	
	\bibitem{Wang_Perceptual_Learned'22}
	\BIBentryALTinterwordspacing
	J.~Wang, S.~Wang, J.~Dai, Z.~Si, D.~Zhou, and K.~Niu, ``Perceptual learned
	source-channel coding for high-fidelity image semantic transmission,'' 2022.
	[Online]. Available: \url{https://arxiv.org/pdf/2205.13120.pdf}
	\BIBentrySTDinterwordspacing
	
	\bibitem{Zhang_The_Unreasonable_Effectiveness'18}
	\BIBentryALTinterwordspacing
	R.~Zhang, P.~Isola, A.~A. Efros, E.~Shechtman, and O.~Wang, ``The unreasonable
	effectiveness of deep features as a perceptual metric,'' 2018. [Online].
	Available: \url{https://arxiv.org/pdf/1801.03924.pdf}
	\BIBentrySTDinterwordspacing
	
	\bibitem{Engman_Perceptual_Metric'20}
	J.~Engman and H.~Nilsson, ``{A Novel Perceptual Metric in Deep Learning: A
		Comparison of Loss Functions for Image Denoising and Reconstruction},''
	Master's thesis, Lund University, Lund, Sweden, 2020.
	
	\bibitem{Kaplanyan_DeepFovea'19}
	A.~S. Kaplanyan, A.~Sochenov, T.~Leimk\"{u}hler, M.~Okunev, T.~Goodall, and
	G.~Rufo, ``{DeepFovea}: Neural reconstruction for foveated rendering and
	video compression using learned statistics of natural videos,'' \emph{ACM
		Trans. Graph.}, vol.~38, no.~6, Nov. 2019.
	
	\bibitem{Zhang_Opt_in_Image_SemCom'23}
	\BIBentryALTinterwordspacing
	W.~Zhang, Y.~Wang, M.~Chen, T.~Luo, and D.~Niyato, ``Optimization of image
	transmission in a cooperative semantic communication networks,'' 2023.
	[Online]. Available: \url{https://arxiv.org/abs/2301.00433}
	\BIBentrySTDinterwordspacing
	
	\bibitem{Zhou_CW-SSIM'05}
	Z.~Wang and E.~Simoncelli, ``Translation insensitive image similarity in
	complex wavelet domain,'' in \emph{Proc. ICASSP}, vol.~2, 2005, pp.
	{II/573--II//576}.
	
	\bibitem{Chen_Fast_SSIM'11}
	M.-J. Chen and A.~C. Bovik, ``Fast structural similarity index algorithm,''
	\emph{J. Real-Time Image Process.}, vol.~6, no.~4, pp. 281--287, Dec. 2011.
	
	\bibitem{Wang_IW-SSIM'11}
	Z.~Wang and Q.~Li, ``Information content weighting for perceptual image quality
	assessment,'' \emph{IEEE Trans. Image Process.}, vol.~20, no.~5, pp.
	1185--1198, 2011.
	
	\bibitem{Sheikh_IFC'05}
	H.~Sheikh, A.~Bovik, and G.~de~Veciana, ``An information fidelity criterion for
	image quality assessment using natural scene statistics,'' \emph{IEEE Trans.
		Image Process.}, vol.~14, no.~12, pp. 2117--2128, 2005.
	
	\bibitem{Sheikh_VIF'06}
	H.~Sheikh and A.~Bovik, ``Image information and visual quality,'' \emph{IEEE
		Trans. Image Process.}, vol.~15, no.~2, pp. 430--444, 2006.
	
	\bibitem{Gao_MGA-based_IQA'09}
	X.~Gao, W.~Lu, D.~Tao, and X.~Li, ``Image quality assessment based on
	multiscale geometric analysis,'' \emph{IEEE Trans. Image Process.}, vol.~18,
	no.~7, pp. 1409--1423, 2009.
	
	\bibitem{Li_DLM'11}
	S.~Li, F.~Zhang, L.~Ma, and K.~N. Ngan, ``Image quality assessment by
	separately evaluating detail losses and additive impairments,'' \emph{IEEE
		Trans. Multimed.}, vol.~13, no.~5, pp. 935--949, 2011.
	
	\bibitem{Liu_MMF'13}
	T.-J. Liu, W.~Lin, and C.-C.~J. Kuo, ``Image quality assessment using
	multi-method fusion,'' \emph{IEEE Trans. Image Process.}, vol.~22, no.~5, pp.
	1793--1807, 2013.
	
	\bibitem{Larson_MAD'10}
	E.~Larson and D.~Chandler, ``Most apparent distortion: Full-reference image
	quality assessment and the role of strategy,'' \emph{J. Electron. Imaging},
	vol.~19, pp. 011\,006--1 -- 011\,006--21, Jan. 2010.
	
	\bibitem{Ponomarenko_Ponomarenko'07}
	N.~Ponomarenko, F.~Silvestri, K.~Egiazarian, M.~Carli, J.~Astola, and V.~Lukin,
	``On between-coefficient contrast masking of {DCT} basis functions,'' in
	\emph{Proc. Int. Workshop on Video Processing and Quality Metrics for
		Consumer Electronics}, Scottsdale, AZ, USA, 2007.
	
	\bibitem{Damera-Venkata_NQM'00}
	N.~Damera-Venkata, T.~Kite, W.~Geisler, B.~Evans, and A.~Bovik, ``Image quality
	assessment based on a degradation model,'' \emph{IEEE Trans. Image Process.},
	vol.~9, no.~4, pp. 636--650, 2000.
	
	\bibitem{Chandler_VSNR'07}
	D.~M. Chandler and S.~S. Hemami, ``{VSNR}: A wavelet-based visual
	signal-to-noise ratio for natural images,'' \emph{IEEE Trans. Image
		Process.}, vol.~16, no.~9, pp. 2284--2298, 2007.
	
	\bibitem{GANs_NIPS2014}
	{I. Goodfellow \textit{et al.}}, ``Generative adversarial nets,'' in
	\emph{Proc. NIPS}, 2014, pp. 2672--2680.
	
	\bibitem{Creswell_GANs_18}
	{A. {Creswell} \textit{et al.}}, ``Generative adversarial networks: An
	overview,'' \emph{IEEE Signal Process. Mag.}, vol.~35, no.~1, pp. 53--65,
	2018.
	
	\bibitem{Wang2019GenerativeAN}
	\BIBentryALTinterwordspacing
	{Z. Wang \textit{et al.}}, ``Generative adversarial networks in computer
	vision: A survey and taxonomy.'' [Online]. Available:
	\url{https://arxiv.org/pdf/1906.01529.pdf}
	\BIBentrySTDinterwordspacing
	
	\bibitem{Tim_Improved_Training_for_GANs'16}
	\BIBentryALTinterwordspacing
	T.~Salimans, I.~Goodfellow, W.~Zaremba, V.~Cheung, A.~Radford, and X.~Chen,
	``Improved techniques for training {GANs},'' 2016. [Online]. Available:
	\url{https://arxiv.org/pdf/1606.03498.pdf}
	\BIBentrySTDinterwordspacing
	
	\bibitem{Heusel_FID_metric'17}
	\BIBentryALTinterwordspacing
	M.~Heusel, H.~Ramsauer, T.~Unterthiner, B.~Nessler, and S.~Hochreiter, ``{GANs}
	trained by a two time-scale update rule converge to a local {Nash}
	equilibrium,'' 2017. [Online]. Available:
	\url{https://arxiv.org/pdf/1706.08500.pdf}
	\BIBentrySTDinterwordspacing
	
	\bibitem{MMD_GANs'18}
	\BIBentryALTinterwordspacing
	M.~Bińkowski, D.~J. Sutherland, M.~Arbel, and A.~Gretton, ``Demystifying {MMD
		GANs},'' 2018. [Online]. Available:
	\url{https://arxiv.org/pdf/1801.01401.pdf}
	\BIBentrySTDinterwordspacing
	
	\bibitem{Wang_MSE_SPM'09}
	Z.~Wang and A.~C. Bovik, ``Mean squared error: Love it or leave it? a new look
	at signal fidelity measures,'' \emph{IEEE Signal Process. Mag.}, vol.~26,
	no.~1, pp. 98--117, 2009.
	
	\bibitem{Seshadrinathan_MOVIE'10}
	K.~Seshadrinathan and A.~C. Bovik, ``Motion tuned spatio-temporal quality
	assessment of natural videos,'' \emph{IEEE Trans. Image Process.}, vol.~19,
	no.~2, pp. 335--350, 2010.
	
	\bibitem{Lin_FVQA'14}
	J.~Y. Lin, T.-J. Liu, E.~C.-H. Wu, and C.-C.~J. Kuo, ``A fusion-based video
	quality assessment ({FVQA}) index,'' in \emph{Proc. APSIPA Annual Summit and
		Conf.}, 2014, pp. 1--5.
	
	\bibitem{Pinson_VQM'04}
	M.~Pinson and S.~Wolf, ``A new standardized method for objectively measuring
	video quality,'' \emph{IEEE Trans. Broadcast.}, vol.~50, no.~3, pp. 312--322,
	2004.
	
	\bibitem{Wolf_VQM_VFD_NTIA'11}
	\BIBentryALTinterwordspacing
	{S. Wolf and M. H. Pinson}, ``Video quality model for variable frame delay
	({VQM$\_$VFD}),'' U.S. Dept. Commer., Nat. Telecommun. Inf. Admin. (NTIA)
	Tech. Memo. TM-11–482, Boulder, CO, USA, 2011. [Online]. Available:
	\url{https://its.ntia.gov/umbraco/surface/download/publication?reportNumber=11-482.pdf}
	\BIBentrySTDinterwordspacing
	
	\bibitem{Rassool_VMAF'17}
	R.~Rassool, ``{VMAF} reproducibility: Validating a perceptual practical video
	quality metric,'' in \emph{Proc. IEEE Int. Symposium on Broadband Multimedia
		Systems and Broadcasting}, 2017, pp. 1--2.
	
	\bibitem{NetFlix_VMAF_Blog'16}
	\BIBentryALTinterwordspacing
	Z.~Li, A.~Aaron, I.~Katsavounidis, A.~Moorthy, and M.~Manohara, ``Toward a
	practical perceptual video quality metric,'' Netflix Technology Blog, Jun.
	2016. [Online]. Available:
	\url{https://netflixtechblog.com/toward-a-practical-perceptual-video-quality-metric-653f208b9652}
	\BIBentrySTDinterwordspacing
	
	\bibitem{ITU-T_recommendation_P.910'99}
	\BIBentryALTinterwordspacing
	{ITU-T}, ``Subjective video quality assessment methods for multimedia
	applications,'' {ITU-T Recommendation P.910}, Sep. 1999. [Online]. Available:
	\url{https://handle.itu.int/11.1002/1000/4751}
	\BIBentrySTDinterwordspacing
	
	\bibitem{Lin_MCL-V'15}
	J.~Y.-c. Lin, R.~Song, C.-H. Wu, T.-J. Liu, H.~Wang, and C.-C.~J. Kuo,
	``{MCL-V}: A streaming video quality assessment database,'' \emph{J. Vis.
		Commun. Image Represent.}, vol.~30, pp. 1--9, Jul. 2015.
	
	\bibitem{Wang_Speed-SSIM'07}
	Z.~Wang and Q.~Li, ``Video quality assessment using a statistical model of
	human visual speed perception,'' \emph{J. Opt. Soc. Am. A}, vol.~24, no.~12,
	pp. B61--B69, Dec. 2007.
	
	\bibitem{Watson_DVQ'01}
	A.~B. Watson, J.~Hu, and J.~F. McGowan, ``Digital video quality metric based on
	human vision,'' \emph{J. Electron. Imaging}, vol.~10, pp. 20--29, 2001.
	
	\bibitem{Yoshikazu_CVQ'08}
	Y.~Kawayoke and Y.~Horita, ``{NR} objective continuous video quality assessment
	model based on frame quality measure,'' in \emph{Proc. ICIP}, 2008, pp.
	385--388.
	
	\bibitem{Barkowsky_TetraVQM'09}
	M.~Barkowsky, J.~Bialkowski, B.~Eskofier, R.~Bitto, and A.~Kaup, ``Temporal
	trajectory aware video quality measure,'' \emph{IEEE J. Sel. Top. Signal
		Process.}, vol.~3, pp. 266--279, May 2009.
	
	\bibitem{Winkler_V-Factor'08}
	S.~Winkler and P.~Mohandas, ``The evolution of video quality measurement: From
	{PSNR} to hybrid metrics,'' \emph{IEEE Trans. Broadcast.}, vol.~54, no.~3,
	pp. 660--668, 2008.
	
	\bibitem{Amirshahi_STAQ'11}
	S.~A. Amirshahi and M.-C. Larabi, ``Spatial-temporal video quality metric based
	on an estimation of {QoE},'' in \emph{Int. Workshop on Quality of Multimedia
		Experience}, 2011, pp. 84--89.
	
	\bibitem{Vu_ST-MAD'11}
	P.~V. Vu, C.~T. Vu, and D.~M. Chandler, ``A spatiotemporal
	most-apparent-distortion model for video quality assessment,'' in \emph{Proc.
		ICIP}, 2011, pp. 2505--2508.
	
	\bibitem{Ionescu_Human3.6M'14}
	C.~Ionescu, D.~Papava, V.~Olaru, and C.~Sminchisescu, ``{Human3.6M}: Large
	scale datasets and predictive methods for {3D} human sensing in natural
	environments,'' \emph{IEEE Trans. Pattern Anal. Mach. Intell.}, vol.~36,
	no.~7, pp. 1325--1339, 2014.
	
	\bibitem{Zhang_Semantic_Sensing_and_Commun'22}
	\BIBentryALTinterwordspacing
	B.~Zhang, Z.~Qin, Y.~Guo, and G.~Y. Li, ``Semantic sensing and communications
	for ultimate extended reality,'' 2022. [Online]. Available:
	\url{https://arxiv.org/pdf/2212.08533.pdf}
	\BIBentrySTDinterwordspacing
	
	\bibitem{AoI_Foundations_and_Trends'17}
	A.~Kosta, N.~Pappas, and V.~Angelakis, ``Age of information: A new concept,
	metric, and tool,'' \emph{Found. Trends Netw.}, vol.~12, no.~3, pp. 162--259,
	2017.
	
	\bibitem{AoI_Survey_Introduction'21}
	R.~D. Yates, Y.~Sun, D.~R. Brown, S.~K. Kaul, E.~H. Modiano, and S.~Ulukus,
	``Age of information: An introduction and survey,'' \emph{IEEE J. Sel. Areas
		Commun.}, vol.~39, pp. 1183--1210, 2021.
	
	\bibitem{Uysal_AoI_in_Practice'21}
	\BIBentryALTinterwordspacing
	E.~Uysal, O.~Kaya, S.~Baghaee, and H.~B. Beytur, ``Age of information in
	practice,'' 2021. [Online]. Available:
	\url{https://arxiv.org/pdf/2106.02491.pdf}
	\BIBentrySTDinterwordspacing
	
	\bibitem{Maatouk_AoII'20}
	\BIBentryALTinterwordspacing
	A.~Maatouk, M.~Assaad, and A.~Ephremides, ``The age of incorrect information:
	an enabler of semantics-empowered communication,'' 2020. [Online]. Available:
	\url{https://arxiv.org/pdf/2012.13214.pdf}
	\BIBentrySTDinterwordspacing
	
	\bibitem{Maatouk_Opt_Wireless_Networks'20}
	\BIBentryALTinterwordspacing
	A.~Maatouk, ``{Optimization of Wireless Networks: Freshness in
		Communications},'' Ph.D. dissertation, {Universit{\'e} Paris-Saclay}, Nov.
	2020. [Online]. Available:
	\url{https://tel.archives-ouvertes.fr/tel-03028195}
	\BIBentrySTDinterwordspacing
	
	\bibitem{Optimizing_AoI'20}
	\BIBentryALTinterwordspacing
	J.~Holm, A.~E. Kalør, F.~Chiariotti, B.~Soret, S.~K. Jensen, T.~B. Pedersen,
	and P.~Popovski, ``Freshness on demand: Optimizing age of information for the
	query process,'' 2020. [Online]. Available:
	\url{https://arxiv.org/abs/2011.00917}
	\BIBentrySTDinterwordspacing
	
	\bibitem{Kaul_Rreal_time_status'12}
	S.~Kaul, R.~Yates, and M.~Gruteser, ``Real-time status: How often should one
	update?'' in \emph{Proc. IEEE INFOCOM}, 2012, pp. 2731--2735.
	
	\bibitem{Minimizing_AoI_TMC'21}
	I.~Kadota and E.~Modiano, ``Minimizing the age of information in wireless
	networks with stochastic arrivals,'' \emph{IEEE Trans. Mob. Comput.},
	vol.~20, no.~3, pp. 1173--1185, 2021.
	
	\bibitem{Perspectives_on_time'22}
	P.~Popovski, F.~Chiariotti, K.~Huang, A.~E. Kalør, M.~Kountouris, N.~Pappas,
	and B.~Soret, ``A perspective on time toward wireless {6G},'' \emph{Proc.
		IEEE}, vol. 110, no.~8, pp. 1116--1146, 2022.
	
	\bibitem{Sun_AoI_Book'19}
	{Y. Sun, I. Kadota, R. Talak, and E. Modiano}, \emph{Age of Information: A New
		Metric for Information Freshness}.\hskip 1em plus 0.5em minus 0.4em\relax San
	Rafael, CA, USA: Morgan $\&$ Claypool Publishers, 2019.
	
	\bibitem{Costa_PAoI_paper'14}
	M.~Costa, M.~Codreanu, and A.~Ephremides, ``Age of information with packet
	management,'' in \emph{Proc. IEEE ISIT}, 2014, pp. 1583--1587.
	
	\bibitem{Costa_PAoI_journal'16}
	{M. Costa, M. Codreanu, and A. Ephremides}, ``On the age of information in
	status update systems with packet management,'' \emph{IEEE Trans. Inf.
		Theory}, vol.~62, no.~4, pp. 1897--1910, 2016.
	
	\bibitem{Zou_rAoI'19}
	P.~Zou, O.~Ozel, and S.~Subramaniam, ``Relative age of information: A new
	metric for status update systems,'' in \emph{Proc. ITW}, 2019, pp. 1--5.
	
	\bibitem{AoII_metric_journal'20}
	A.~Maatouk, S.~Kriouile, M.~Assaad, and A.~Ephremides, ``The age of incorrect
	information: A new performance metric for status updates,'' \emph{IEEE/ACM
		Trans. Netw.}, vol.~28, no.~5, p. 2215–2228, Oct. 2020.
	
	\bibitem{Sun_Sampling_Wiener_Process'20}
	Y.~Sun, Y.~Polyanskiy, and E.~Uysal, ``Sampling of the wiener process for
	remote estimation over a channel with random delay,'' \emph{IEEE Trans. Inf.
		Theory}, vol.~66, no.~2, pp. 1118--1135, 2020.
	
	\bibitem{Soleymani_dissertation_19}
	T.~Soleymani, ``Value of information analysis in feedback control,'' {Ph.D.}
	dissertation, {Technical University of Munich, Germany}, 2019.
	
	\bibitem{Howard_Info_Value_Theory'1966}
	R.~A. Howard, ``Information value theory,'' \emph{IEEE Trans. Syst. Sci.
		Cybern.}, vol.~2, pp. 22--26, 1966.
	
	\bibitem{Ayan_AoI_vs_VoI'19}
	O.~Ayan, M.~Vilgelm, M.~Kl\"{u}gel, S.~Hirche, and W.~Kellerer,
	``Age-of-information vs. value-of-information scheduling for cellular
	networked control systems,'' in \emph{Proc. ACM/IEEE Int. Conf.
		Cyber-Physical Systems}, New York, NY, USA, 2019, p. 109–117.
	
	\bibitem{Molin_Estimators_Based_on_VoI'19}
	A.~Molin, H.~Esen, and K.~H. Johansson, ``Scheduling networked state estimators
	based on value of information,'' \emph{Automatica}, vol. 110, no.~C, Dec.
	2019.
	
	\bibitem{Wang_SemCom_Per_Optimization'21}
	Y.~Wang, M.~Chen, W.~Saad, T.~Luo, S.~Cui, and H.~V. Poor, ``Performance
	optimization for semantic communications: An attention-based learning
	approach,'' in \emph{Proc. IEEE GLOBECOM}, 2021, pp. 1--6.
	
	\bibitem{Wang_JSAC_Performance_Optimi_SemCom'22}
	Y.~Wang, M.~Chen, T.~Luo, W.~Saad, D.~Niyato, H.~V. Poor, and S.~Cui,
	``Performance optimization for semantic communications: An attention-based
	reinforcement learning approach,'' \emph{IEEE J. Sel. Areas Commun.},
	vol.~40, no.~9, pp. 2598--2613, 2022.
	
	\bibitem{Yan_QoE_Aware_Resource_Allocation_SemCom'22}
	\BIBentryALTinterwordspacing
	{L. Yan, Z. Qin, R. Zhang, Y. Li, and G. Y. Li}, ``{QoE}-aware resource
	allocation for semantic communication networks,'' 2022. [Online]. Available:
	\url{https://arxiv.org/pdf/2205.14530.pdf}
	\BIBentrySTDinterwordspacing
	
	\bibitem{Xia_Resource_Management_SemCom'22}
	\BIBentryALTinterwordspacing
	L.~Xia, Y.~Sun, X.~Li, G.~Feng, and M.~A. Imran, ``Wireless resource management
	in intelligent semantic communication networks,'' 2022. [Online]. Available:
	\url{https://arxiv.org/pdf/2202.07632.pdf}
	\BIBentrySTDinterwordspacing
	
	\bibitem{Dong_Innovative_SemCom'22}
	\BIBentryALTinterwordspacing
	C.~Dong, H.~Liang, X.~Xu, S.~Han, B.~Wang, and P.~Zhang, ``Innovative semantic
	communication system,'' 2022. [Online]. Available:
	\url{https://arxiv.org/pdf/2202.09595.pdf}
	\BIBentrySTDinterwordspacing
	
	\bibitem{Kang_Personalized_Saliency'23}
	J.~Kang, H.~Du, Z.~Li, Z.~Xiong, S.~Ma, D.~Niyato, and Y.~Li, ``Personalized
	saliency in task-oriented semantic communications: Image transmission and
	performance analysis,'' \emph{IEEE J. Sel. Areas Commun.}, vol.~41, no.~1,
	pp. 186--201, 2023.
	
	\bibitem{Zhang_New_Results'18}
	J.~Zhang, W.~Zeng, X.~Li, Q.~Sun, and K.~P. Peppas, ``New results on the
	fluctuating two-ray model with arbitrary fading parameters and its
	applications,'' \emph{IEEE Trans. Veh. Technol.}, vol.~67, no.~3, pp.
	2766--2770, 2018.
	
	\bibitem{Sun_semantic-assisted'22}
	\BIBentryALTinterwordspacing
	Q.~Sun, C.~Guo, Y.~Yang, J.~Chen, and X.~Xue, ``Semantic-assisted image
	compression,'' 2022. [Online]. Available:
	\url{https://arxiv.org/pdf/2201.12599.pdf}
	\BIBentrySTDinterwordspacing
	
	\bibitem{Cheng_CLUB'20}
	\BIBentryALTinterwordspacing
	P.~Cheng, W.~Hao, S.~Dai, J.~Liu, Z.~Gan, and L.~Carin, ``{CLUB}: A contrastive
	log-ratio upper bound of mutual information,'' 2020. [Online]. Available:
	\url{https://arxiv.org/pdf/2006.12013.pdf}
	\BIBentrySTDinterwordspacing
	
	\bibitem{Watrous_QIT'2018}
	J.~Watrous, \emph{The Theory of Quantum Information}.\hskip 1em plus 0.5em
	minus 0.4em\relax Cambridge, UK: Cambridge Univ. Press, 2018.
	
	\bibitem{Wilde_QIT_arXiv'19}
	\BIBentryALTinterwordspacing
	M.~M. Wilde, ``From classical to quantum shannon theory,'' 2019. [Online].
	Available: \url{https://arxiv.org/pdf/1106.1445.pdf}
	\BIBentrySTDinterwordspacing
	
	\bibitem{Fonseca_HD_quantum_teleportation'19}
	A.~Fonseca, ``High-dimensional quantum teleportation under noisy
	environments,'' \emph{Phys. Rev. A}, vol. 100, no.~6, Dec. 2019.
	
	\bibitem{Pappas_Goal-oriented_Commun'21}
	N.~Pappas and M.~Kountouris, ``Goal-oriented communication for real-time
	tracking in autonomous systems,'' in \emph{Proc. IEEE Int. Conf. Autonomous
		Systems}, 2021, pp. 1--5.
	
	\bibitem{Shao_TO_Commun'22}
	\BIBentryALTinterwordspacing
	J.~Shao, X.~Zhang, and J.~Zhang, ``Task-oriented communication for edge video
	analytics,'' 2022. [Online]. Available:
	\url{https://arxiv.org/pdf/2211.14049.pdf}
	\BIBentrySTDinterwordspacing
	
	\bibitem{asturi_Performance_Evaluation_Framework'09}
	R.~Kasturi, D.~Goldgof, P.~Soundararajan, V.~Manohar, J.~Garofolo, R.~Bowers,
	M.~Boonstra, V.~Korzhova, and J.~Zhang, ``Framework for performance
	evaluation of face, text, and vehicle detection and tracking in video: Data,
	metrics, and protocol,'' \emph{IEEE Trans. Pattern Anal. Mach. Intell.},
	vol.~31, no.~2, pp. 319--336, 2009.
	
\end{thebibliography}
\end{document}